\documentclass[12pt]{iopart}
\usepackage{graphicx,amssymb,epsf}
\usepackage[]{natbib}

\def\be{\begin{equation}}
\def\ee{\end{equation}}
\def\bea{\begin{eqnarray}}
\def\eea{\end{eqnarray}}
\def\nn{\nonumber }

\def\eg{{\it e.g.}}
\def\ie{{\it i.e.}}

\def\ho{\mbox{H$_\circ$} }

\def\boltz     {{k_{\rm B}}}
\def\te        {{T_{\rm e}}}
\def\tcmb      {\mbox{${T_{\rm cmb}}$} }
\def\p		   {{^\prime}}
\def\o		{\over }
\newbox\simlessbox \newbox\simgreatbox

\def\ARAA{{\it Ann. Rev. of Astron. Astrop.}}
\def\AaA{{\it Astron. Astrop.}}
\def\AaAS{{\it Astron. Astrop. Supp.}}
\def\ApJ{{\it Astrophys. J.}}

\def\ApJS{{\it Astrophys. J. Supp.}}
\def\MNRAS{{\it Mon. Not. Roy. Ast. Soc.}}
\def\Nat{{\it Nature}}

\def\PhRvD{{\it Physical. Rev. D}}
\def\PhRvL{{\it Physical. Rev. Lett.}}
 
\def\PASJ{{\it Proc. of the Astrop. Soc. of Japan}}

\begin{document}
\pagenumbering{roman}

\review[Secondary anisotropies of the CMB]{Secondary anisotropies of the CMB}

\author{Nabila Aghanim$^1$, Subhabrata Majumdar$^2$ and Joseph Silk$^3$}

\address{$^1$ Institut d'Astrophysique Spatiale (IAS), CNRS,
  B$\hat{\rm{a}}$t. 121, Universit\'e Paris-Sud, F-91405, Orsay, France}

\address{$^2$ Department of Astronomy \& Astrophysics, Tata Institute of Fundamental Research (TIFR), Homi Bhabha Road, Mumbai, India}

\address{$^3$ Denys Wilkinson Building, University of Oxford, Keble Road, Oxford, OX1 3RH, UK }

\eads{\mailto{nabila.aghanim@ias.u-psud.fr}, \mailto{subha@tifr.res.in}}
\begin{abstract}

The Cosmic Microwave Background fluctuations provide a powerful probe
of the dark ages of the universe through the imprint of the secondary
anisotropies associated with the reionisation of the universe and the
growth of structure. We review the relation between the secondary
anisotropies and and the primary anisotropies that are directly
generated by quantum fluctuations in the very early universe. The
physics of secondary fluctuations is described, with emphasis on the
ionisation history and the evolution of structure. We discuss the
different signatures arising from the secondary effects in terms of
their induced temperature fluctuations, polarisation and
statistics. The secondary anisotropies are being actively pursued at
present, and we review the future and current observational status.

\end{abstract}
\maketitle

\newpage
\pagenumbering{arabic}
\setcounter{page}{1}

\section{Introduction}
In the post WMAP era for Cosmic Microwave Background (CMB)
measurements and in preparation of Planck and the post-Planck era,
attention is now shifting towards small angular scales of the order of
a few arc-minutes or even smaller. At these scales, CMB temperature
and polarisation fluctuations are no longer dominated by primary
effects at the surface of last scattering but rather by the so-called
secondary effects induced by the interaction of CMB photons with the
matter in the line of sight.

Current and future CMB experiments have two main goals: i) measuring
small angular scale temperature fluctuations (below a few arc minutes),
and ii) measuring the CMB polarisation power spectrum. These goals are
fundamental for our understanding of the universe.  The small-scale
anisotropies are directly related to the presence of structures in the
universe whereas the two types of polarisation ($E$ and $B$-modes, which we
discuss later) probe both the reionisation of the universe,
i.e. the formation of the first emitting objects, and the inflationary
potential. In this review, we focus on the end of the dark ages and the
astrophysical probes of reionisation.

\begin{figure} 
\epsfxsize=12cm
\epsfysize=10cm
\hspace{2cm}
\epsffile{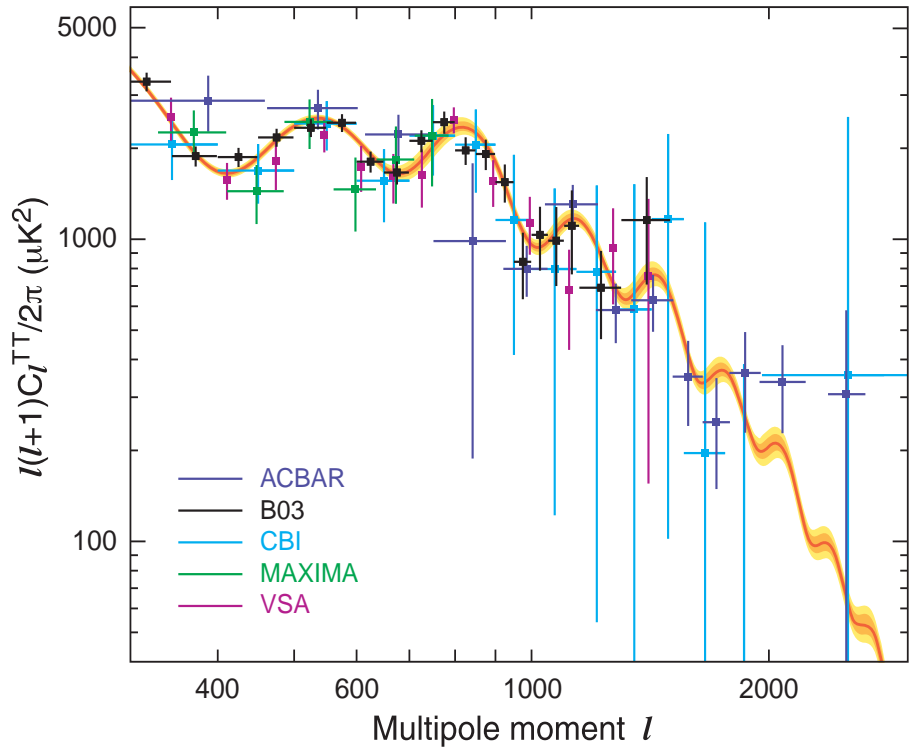}
\caption{From Spergel et al. (2006): The compilation of the small
scale CMB measurements from ground-based and balloon experiments (Ruhl
et al.  2003, Abroe et al. 2004, Kuo et al. 2004, Readhead et al. 2004,
Dickinson et al. 2004).  The red, dark orange and light orange lines
represent the predictions from the $\Lambda$CDM model fit to the WMAP
data for the best fit, the 68\% and 95\% confidence levels
respectively. Excess of power is seen at the largest $l$ values.}
\label{fig:high-l}
\end{figure}

There have been rapid and important advances in the recent past. We
already have on the one hand measurements, by ACBAR, CBI, BIMA, VSA,
of the temperature power spectrum for $2000<\ell<4000$ with CBI and
BIMA data showing an excess of power as compared with the predicted
damping tail of the CMB (Figure ~\ref{fig:high-l}). On the other hand,
DASI, Archeops, Boomerang, Maxipol, CBI, QUaD and WMAP have direct
measurements of the $E$-mode polarisation. The situation will change
even more in the near future with anticipated results from experiments
currently taking data or in preparation (QUaD, BICEP, EBEX, CLOVER,
QUIET, SPIDER, Planck).

All of this experimental activity is motivated by what now amounts to
the standard model of cosmology. The CMB temperature fluctuations
which are generated prior to decoupling are measured on scales from 90
degrees to several arc minutes. This has led to a model of precision
cosmology. The basic infrastructure is the Friedmann-Lemaitre model
with zero curvature, a cosmological constant (or dark energy), a
baryonic content and non-baryonic dominant cold dark matter component
(with fractions given by the recent WMAP data (Spergel et
al. 2007)). Superimposed on the cosmological background are the
primordial adiabatic density fluctuations, described by a nearly
scale-invariant power spectrum $|\delta_k|^2 \propto k^{n-1},$ at
horizon crossing (in the comoving gauge), that generated the
large-scale structure via gravitational instability of the cold matter
component. However it has become increasingly apparent that to further
refine these parameters, and to face the more intriguing challenge of
establishing possible deviations from the concordance model one has to
address the degeneracies between cosmological parameters with those
from the secondary anisotropies as well as the extragalactic
astrophysical foregrounds.

The primary CMB anisotropies are due to the gravitational redshift at
large angular scales (Sachs \& Wolfe 1967) and to the evolution of the
primordial photon-baryon fluid evolution under gravity and Compton
scattering at lower scales (Silk 1967, Peebles \& Yu 1970, Sunyaev \&
Zel'dovich 1970) to which one adds photon diffusion damping at small
scales (Silk 1967). Primary fluctuations have provided us with an
unparalleled probe of the primordial density fluctuations that seeded
large-scale structure formation.  Indeed on large angular scales,
greater than the angular scale subtended by the sound horizon at
recombination, one can directly view the approximately scale-invariant
spectrum of primordial quantum fluctuations.

On their way towards us, the photons interact with cosmic structures
and their frequency, energy or direction of propagation are
affected. These are the secondary effects that involve the
density and velocity fields and incorporate Compton scattering off
electrons. This review is devoted to a study of these secondary
anisotropies.

\begin{figure} 
\hspace{-2cm}
\epsfxsize=16.cm
\epsfysize=9cm
\hspace{2.0cm}
\epsffile{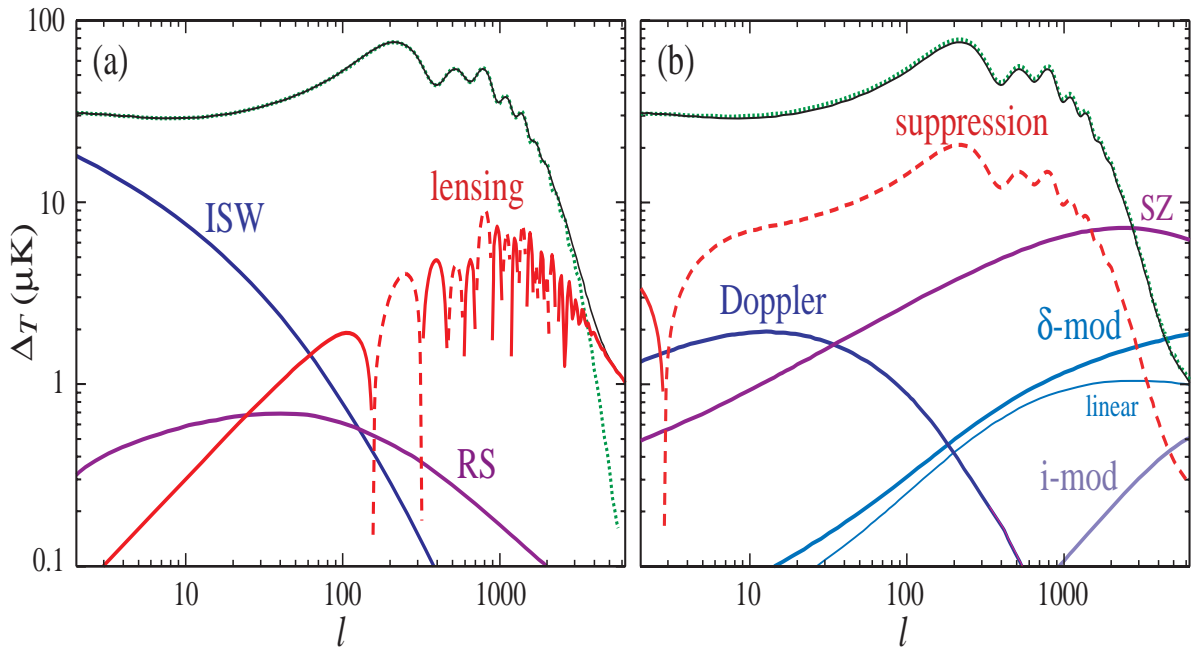}
\caption{From Hu \& Dodelson (2002): The power spectrum of the
  secondary temperature anisotropies arising from gravitational
  effects (panel a) and scattering effects (panel b). The power
  spectrum of primary anisotropies is shown for comparison. The
  calculations use a flat universe with $\Omega_\Lambda = 0.67,
  \Omega_{\rm b}h^2 = 0.02, \Omega_{\rm m}h^2 = 0.16, n=1$. Acronyms
  are defined in the text. $\delta$- and i-mod refer to density and
  ionisation fraction modulation respectively
  (Sect. \ref{sec:reion}). ``Suppression'' and ``Doppler'' refer to
  the damping and anisotropy generation at reionisation
  (Sect. \ref{sec:reion}). }
\label{fig:sectot}
\end{figure}

The CMB photons we observe today have traversed the universe from the
last scattering surface to us and have thus interacted with matter
along their path through the universe. These interactions generate the
secondary anisotropies that arise from two major families of
interactions. The first family includes the gravitational effects
(Figure \ref{fig:sectot} panel a), including gravitational lensing, the
Rees-Sciama effect (RS), moving lenses and decaying potentials usually
referred to as the integrated Sachs-Wolfe effect (ISW). These
anisotropies arise from the interactions of the photons with
gravitational potential wells. The second family incorporates the
effects of scattering between CMB photons and free electrons
(Figure \ref{fig:sectot} panel b) such as inverse Compton interaction
(the Sunyaev-Zel'dovich (SZ) effect) and velocity-induced scatterings
such as the Ostriker-Vishniac (OV) effect and inhomogeneous
reionisation.

We define secondary anisotropies in the CMB to include all temperature
fluctuations generated since the epoch of matter-radiation decoupling
at $z\sim 1100.$ The following contributions may be distinguished.

\begin{enumerate}
\item {} The integrated Sachs-Wolfe (ISW) effect is due to CMB photons
traversing a time-varying linear gravitational potential. The relevant
scale is the curvature scale freeze-out  in concordance cosmology: the
horizon at $1+z\sim (\Omega_\Lambda/\Omega_{\rm m})^{1/3}$. This corresponds
to an angular scale of about 10$^\circ$.
\item {} The Rees-Sciama (RS) effect is due to CMB photons
traversing a non-linear gravitational potential,  usually
associated with   gravitational collapse. The relevant scales are
those of galaxy clusters and superclusters, corresponding to  angular
scales of 5-10 arc minutes.
\item {} Gravitational lensing of the CMB by intervening large-scale
structure does not change the total power in fluctuations, but power
is redistributed preferentially towards smaller scales. The effects
are significant only below a few arc minutes. Its effects may be
significant on large scales when the observable of interest is the
$B$-mode power spectrum.
\item {} The Sunyaev-Zel'dovich (SZ) effect from hot gas in clusters is due
to the first order correction for energy transfer in Thomson
scattering.  It is on the scale of galaxy clusters and superclusters,
although it may be produced on very small scales by the first stars in
the universe.  There is a spectral distortion, energy being
transferred from photons in the Rayleigh-Jeans tail of the cosmic
blackbody radiation to the Wien tail.
\item {} The kinetic Sunyaev-Zel'dovich effect is the Doppler effect
due to the motion of hot gas in clusters that scatters the CMB. It causes
no spectral distortion.
\item {} The Ostriker-Vishniac (OV linear) effect is also due to Doppler
boosting. It is the linear version of the kinetic Sunyaev-Zel'dovich
effect. It is proportional to the product of $\Delta n_e$ and
$\Delta v,$. This is effective on the scale of order 1 arc minute.
\item {} Discrete sources provide an appreciable
foreground, especially at lower frequencies for radio sources and high
frequencies for infra-red and submillimetre sources.  
\item {} Polarisation is primarily a secondary phenomenon.  The
primary effect from last scattering is induced by out-of-phase
velocity perturbations and provides evidence for the acausal nature of
the fluctuations. The secondary polarisation is associated with the
reionisation of the universe and is on large scales corresponding to
the horizon at reionisation. Inhomogeneous reionisation and scattering
at the galaxy cluster scale leads to smaller scale polarisation. The
reionisation signal is weak, amounting to no more than 10 percent of
the primary signal.
\item {} $B$-mode polarisation can be induced by shear
perturbations. One source is gravitational lensing of primary CMB
fluctuations.  A second is relic gravity waves from inflation. These
are pure $B$-modes, and fall off rapidly on scales smaller than the
horizon at recombination, corresponding to about half a degree.
Mixing by Faraday rotation in the intracluster medium also contributes
to $B$-mode generation on small angular scales. The $B$-mode
polarisation amplitude only amounts to about a percent of the primary
signal, and its discovery will pose the major challenge for future
experiments.

\end{enumerate}

\section{Reionisation}\label{sec:reion}

\subsection{Basics of Physics}\label{sec:reion-bas}

In dealing with secondary CMB anisotropies at reionisation or arising
from ionised structure like the hot gas in galaxy clusters, we are
concerned with the scattering of the CMB photons by the
plasma. A specific example of this is the Sunyaev-Zel'dovich effect,
which is discussed in detail in section $\ref{sec_SZ}$, where the
intra-cluster gas up-scatters the cold microwave photons.

Since the secondary anisotropies are distortions of the CMB, which is
the radiation field, we start by looking at the properties of an
isotropic and thermal radiation background.  The
distribution function, $f_\alpha(\mbox{{\bf r,p}}_\nu,$t), of any
radiation field is defined such that $f _\alpha d^3rd^3p_\nu$ is the
number of photons in the real space volume $d^3r$ about {\bf r} and
the momentum space volume $d^3p_\nu$ about {\bf p}$_\nu$ ($\nu$ being
the frequency) at time $t$ with polarisation $\alpha=1,2$. This
distribution can be related to the photon occupation number,
$n_\alpha(\mbox{{\bf r,p}}_\nu,$t), by 
\be n_\alpha(\mbox{{\bf
r,p}}_\nu,\mbox{t}) = h^3\,f_\alpha(\mbox{{\bf r,p}}_\nu,\mbox{t}).
\ee 
For polarisation a description in
terms of the pure polarisation states pre-supposes fully
polarised radiation.
For CMB radiation, the occupation number has a Planckian
distribution given by 
\be n_\alpha = \left( e^{h_{\rm pl} \nu / \boltz \tcmb}
-1 \right)^{-1} \quad \rm for \ \alpha=1,2~~~, \label{eq:planckdist}
\ee 
where \tcmb is the temperature of the CMB photons. The specific
intensity of radiation is related to the distribution function by 
\be
I_\nu({\bf \hat k},{\bf r},t) = \sum_{\alpha=1}^2 \, \left( { h_{\rm pl}^4 \,
\nu^3 \over c^2 } \right) \, f_\alpha({\bf r},{\bf p}_\nu,t) ~~~.  \ee

Commonly, the specific intensity is described in units of brightness
temperature, $T_{\rm R-J}$, which is defined as the temperature of the
thermal radiation field which in the Rayleigh-Jeans (R-J) limit (\ie, low
frequency) would have the same brightness as the radiation that is
being described. In the R-J limit, the specific
intensity reduces to $I_\nu = 2 \boltz \tcmb \nu^2 / c^2$, so that 
\be
T_{\rm R-J}(\nu) = {c^2 I_\nu \over 2 \boltz \nu^2} \quad .  
\ee

Now let us consider the scattering between two species (namely photons
and electrons).  For an ensemble of particles, if the motion of one
particle is completely independent of all other particles, then to
describe the state of the particles, one can specify the single
particle distribution function given by $f ({\bf r,
p,}t)\,d^3r\,d^3p$, which is the probability of finding a single
particle in the phase space volume $d^3r\,d^3p$ around the point ({\bf
r, p}) at time $t$.  If there are no interactions between the
particles and if they are non-relativistic, then the distribution
obeys the Liouville equation \be {{ df}\o{dt}}\,=\,{{\partial
f}\o{\partial t}}\,+\,{{\bf p}\o{m}}.{{\partial f}\o{\partial {\bf
r}}} \,+\, F\,({\bf r,p},t).{{\partial f}\o{\partial {\bf p}}} 
\,\, ,
\label{eq:liouville}
\ee
where ${\bf F}$ is any force that may be present, and $m$ is the mass of a
 particle, assumed to be the same for all particles.

In the case of inter-particle interactions being random and statistical
in nature, one cannot describe the system by a mean force ${\bf F}$,
but one has to consider instantaneous collisions between the particles
(this is the case for photon - electron interactions). These
collisions will remove particles from (or add particles to) a cell in
phase-space. If one carefully balances these changes of particles in
each cell, then for non-relativistic elastic collisions, one ends up
with the Boltzmann equation
$$
{{ df}\o{dt}}\,=\,{{\partial f}\o{\partial t}}\,+\,{{\bf
p}\o{m}}.{{\partial f}\o{\partial {\bf r}}} \,+\, F\,({\bf r,p},t).{{\partial
f}\o{\partial {\bf p}}} \,=\, \nn $$
\be
\int d^3{\bf p}_1\,{{|{\bf p -
p}_1|}\o{m}}\,{{d\sigma}\o{d\Omega}}\,d\Omega
\left[f({\bf p\p}_1)\,f({\bf p\p}) \,-\,f({\bf p}_1)\,f({\bf p}) \right],
\label{eq:boltzman}
\ee 
where the scattering solid angle $d\Omega$ is determined by the
conservation of momentum and energy and $d\sigma$ is the scattering
cross section.  Moreover, the collisions take place between particles
with momenta ${\bf p}$ and ${\bf p}_1$ and produced particles with
momenta ${\bf p\p}$ and ${\bf p\p}_1$. The Boltzmann equation, being
integro-differential, is difficult to solve analytically. However, it
can be tackled under some approximations which can be made when ${\bf
p}$ is close to ${\bf p\p}$ and ${\bf p}_1$ is close to ${\bf
p\p}_1$. It is then possible to expand the right hand side of Equation
(\ref{eq:boltzman}) in powers of $\Delta\,{\bf p}\,=\, {\bf p\p - p}$
and carry out the integral. The result can be expressed in terms
of a Taylor series to give the Fokker-Planck
equation. A simplification of the Fokker-Planck equation yields the
Kompaneets equation, whose solution for the case of photon-electron
collisions in astrophysical situations gives the Sunyaev-Zel'dovich
distortion (Section $\ref{sec_SZ}$).  \par\bigskip

At matter-radiation decoupling, the free electrons are
non-relativistic and the scattering between them and the photons is
simply Thomson scattering. The incident electromagnetic radiation
with linear polarisation ${\mathbf \epsilon_{\rm i}}$ is scattered by an
electron at rest in a radiation field of polarisation ${\mathbf
  \epsilon_{\rm e}}$
into a solid angle ${\mathrm d}\Omega$ with a probability:
$$
{\mathrm d}\sigma=\frac{3\sigma_T}{8\pi}|{\mathbf \epsilon_{\rm i}
\cdot\epsilon_{\rm e}}|^2\,{\mathrm d}\Omega .
$$ 
In the plane perpendicular to the scattering direction there is no
variation of the polarisation. In the scattering plane, however, there
is a net polarisation.  As a consequence, if the incident radiation
propagating along the $z$ axis comes from two orthogonal directions
there will be no polarisation transmitted along the $z$ axis. 
Isotropic non-polarised incident radiation will induce the same
identical polarisation along $x$ and $y$ axis. If the incident
radiation is anisotropic and quadrupolar the scattered radiation shows
an excess of energy and thus a non-zero polarisation oriented
according to the quadrupole orientation. As a result, the Thomson
scattering induces a linear polarisation under the condition that the
incident radiation has at least a quadrupolar geometry.  In the
cosmological context, anisotropies are induced by density
perturbations and the velocity gradients are responsible for the
quadrupole moment. We therefore expect a Thomson scattering-induced
polarisation for the primary anisotropies.

The polarisation intensity is
governed by the Boltzmann equation (Peebles \& Yu 1970, Sunyaev \&
Zel'dovich 1972,  Bond \& Efstathiou 1984, Ma \&
Bertschinger 1995, Hu \& White 1997). This yields:
\begin{equation}
\dot{\Delta}_{Q\pm iU}+ {\bf \hat{n}}_i \partial_i \Delta_{Q\pm iU}=
n_{\rm e}\sigma_{\rm T}a(\eta)
\left(-\Delta_{Q\pm iU}+\sqrt{\frac{6\pi}{5}}\sum_{m=-2}^{m=2}\ _{\pm2}
Y_2^m({\bf \hat n}) \Pi^{(m)} \right),
\label{eq:bolpol}
\end{equation}
where $Q$ and $U$ are the two Stokes parameters, $\eta \equiv \int dt
/ a $ is the conformal time, $a$ is the expansion factor, and ${\bf
\hat{n}}$ the direction of photon propagation.  $\ _sY_\ell^m$ are the
spherical harmonics with spin-weight $s$, and $\Pi^{(m)}$ is defined
in terms of the quadrupole components of the temperature
($\Delta_{T2}^{(m)}({\bf r},\eta)$) and polarisation perturbations
\begin{equation}
\Pi^{(m)}({\bf r},\eta) \equiv \Delta_{T2}^{(m)}({\bf r},\eta)+
12\sqrt{6}\Delta_{+,2}^{(m)}({\bf r},\eta)+12\sqrt{6}\Delta_{-,2}^
{(m)}({\bf r},\eta),
\end{equation} 
${\bf r}$ is the comoving coordinate.  In equation \ref{eq:bolpol},
the dot stands for the time derivative, $\sigma_{\rm T}$ is the
Thomson cross section, and $n_{\rm e}$ is the free electron number
density which can be written as $n_{\rm e}({\bf r},\eta)=\bar{n}_{\rm
e}(\eta)[1+\delta_{\rm e}( {\bf r},\eta)]$, with $\delta_{\rm e}$ and
$\bar{n_{\rm e}}$ the fluctuation and the background of the electron
number density, respectively. The electron density fluctuations can be
due to matter density perturbations or to spatial variations of the ionisation
fraction. Replacing $n_{\rm e}({\bf r}, \eta)$  in
equation \ref{eq:bolpol} by its full expression allows us to separate first order
effects (proportional to $\bar{n_{\rm e}}$) from second order effects
(proportional to
$\delta_{\rm e}$). Finally, the polarisation perturbations at present
can be obtained by integrating the Boltzmann equation
 (Equation (\ref{eq:bolpol})) along the line of sight.

Assuming that primary temperature fluctuations dominate over
polarisation perturbations, the polarisation at reionisation is due to
coupling between the electron density and the quadrupole moment. The
solution for a single Fourier mode, $\Delta_{Q\pm iU}$, of the
Boltzmann equation Eq. (\ref{eq:bolpol}) is then given (e.g. Ng \& Ng
1996) by:
\begin{equation}
\Delta_{Q\pm iU}({\bf k}, {\bf \hat{n}}, \eta_0)=\sqrt{\frac{6\pi}{5}}
\int^{\eta_0}_0 d\eta e^{i k(\eta_0-\eta)\mu}g(\eta)\sum_m\ _{\pm 2}
Y_2^m({\bf \hat n}) X^{(m)}({\bf k},\eta),
\label{sol2}
\end{equation}
where $X^{(m)}({\bf k},\eta)$ equals $\Pi^{(0)}({\bf k},\eta)$ for the
first order contribution and $S^{(m)}({\bf k},\eta)=
\delta_{\rm e}({\bf k}, \tau)Q(\eta)$ for the second order
contribution, with $Q(\eta)$ being the radiation quadrupole.
The visibility function $g(\eta)$:
\begin{equation}
g(\eta)\equiv -\frac{{\rm d}\tau}{{\rm d}\eta}{\rm e}^{-\tau(\eta)},
\label{eq:vis}
\end{equation}
provides us with the probability that a photon had its last scattering
at $\eta$ and reached the observer at the present time, $\eta_0$. In
equation (\ref{eq:vis}), $\tau(\eta) \equiv \int_\eta^{\eta_0}d\eta'
a(\eta) n_{\rm e} \sigma_{\rm T}$ is the optical depth and $\mu={\bf
k}\cdot {\bf \hat{n}}$.

\subsection{Constraints on reionisation}

As the CMB radiation possesses an rms primary quadrupole moment $Q_{\rm
rms}$, Thomson scattering between the CMB photons and free electrons
generates linear polarisation. This is the case at recombination but
in particular it is true at reionisation. Re-scattering of the CMB
photons at reionisation generates a new polarisation anisotropy at
large angular scale because the horizon has grown to a much larger
size by that epoch (Ng \& Ng 1996). The location of the anisotropy (a
bump), $\ell_{\rm peak}$, relates to the horizon size at the new
``last scattering'' and thus depends on the ionisation redshift
$z_{\rm ion}$. A fitting formula was given by Liu et al (2001):
\begin{equation}
\ell_{\rm peak}=0.74(1+z_{\rm ion})^{0.73}\Omega_0^{0.11}.
\end{equation}
The height of the bump relates to the optical depth or in other words
to the duration of the last scattering. Such a signature (bump at
large scales) has first been observed by WMAP (Kogut et al. 2003,
Spergel et al. 2003) by correlating the temperature and the
polarisation power spectra. The first year WMAP observations
constrained the optical depth at reionisation to a high value
$\tau\sim 0.17$ and provided a simple model for the reionisation, the
ionisation redshift was found to be $z_\mathrm{ion}\sim 17$. The
optical depth is degenerate with the tilt of the primordial power
spectrum.

The WMAP first year result came as a surprise, in the context of
earlier studies of the Gunn-Peterson effect inferred to be present in
the most distant quasars at $z\sim 6$ (\eg Fan et al. 2003) and of the
high temperature of the intergalactic medium at $z\sim 3$ (Theuns et
al. 2002).  The situation was soon rectified with the WMAP 3 year data
release (Spergel et al. 2007). The improved data included an $E$-mode
polarisation map. The power spectrum is proportional to $\tau^2$ and
the new constraints on polarisation yielded an optical depth
$\tau=0.09\pm0.03$.  Together with a better understanding of
polarisation foregrounds, the improved measurements enabled the
degeneracy with the tilt to be reduced. The new tilt value of
$n=0.95\pm 0.02$ lowers the small-scale power.  Despite the reduced
WMAP 3 year normalisation, $\sigma_8=0.74\pm 0.06,$ the lower optical
depth implies that the constraints on the possible sources of
reionisation remain essentially unchanged (Alvarez et al. 2006).
Precise measurements (cosmic variance-limited) of the $E$-mode
polarisation power spectrum will eventually allow us to
phenomenologically reconstruct the reionisation history (e.g. Hu \&
Okamoto 2004). This will help constrain the reionisation models and
enable us to explore the transition between partial and total
reionisation (e.g. Holder et al. 2003).

Reionisation must have occurred before $z\sim 6$ and the universe is
now generally considered to have become reionised at a redshift
between 7 and 20. The major question now is to identify the sources
responsible for the reionisation of the universe.  The ionising
sources cannot be a population of normal galaxies or known quasars.
Optical studies of the bright quasar luminosity function (Haiman, Abel
\& Madau 2001, Wyithe \& Loeb 2003), as well the associated X-ray
background (Djikstra, Haiman \& Loeb 2004) rule out the known quasar
population as a reionisation source.  However miniquasars with
correspondingly softer spectra could evade this constraint.  Recourse
must therefore be had to Population III stars or to miniquasars, both
of which represent hypothetical but plausible populations of the first
objects in the universe that are significant sources of ionising
photons.

We discuss theoretical issues in Section \ref{reio-sourc}. Here we ask
whether one can observationally distinguish between the alternative
hypotheses of stellar versus miniquasar ionisation sources.  The most
promising techniques for probing reionisation include 21~cm emission
and absorption, Lyman-alpha absorption against high redshift quasars,
and the statistics of Lyman-alpha emitters.  One distinguishing
feature is the intrinsic source spectrum, which is thermal for stars
but with a cut-off at a few times the Lyman limit frequency, whereas
it is a power-law for miniquasars with a spectrum that extends to
higher energies with nearly equal logarithmic increments in energy per
decade of frequency.  One can also explore the evolution of the
intergalactic medium during reionisation through the study of the
redshifted 21~cm hyperfine triplet-singlet level transition of the
ground state of neutral hydrogen (HI).  This line allows the detection
of the HI gas in the early universe. It thus represents a unique way
to map the spatial distribution of intergalactic hydrogen (e.g. Madau,
Meiksin \& Reese 1997, Ciardi \& Madau 2003). Therefore it permits, in
principle, a reconstruction of the reionisation history as governed by
the first luminous sources.  The size of the ionised structures that
could be detected depends on the design of future radio
telescopes. The forthcoming radio telescope, in the frequency range
80-180 MHz, LOw Frequency ARray (LOFAR)\footnote{www.lofar.org} should
have the sensitivity and resolution ($\sim3$ arc minutes) needed.
Using cosmological radiative transfer numerical computations with an
idealised LOFAR array, Valdes et al. (2006) have simulated
observations of the reionisation signal for both early and late
reionisation scenarios.  They show that if reionisation occurs late,
LOFAR will be able to detect individual HI structures on arc minute
scales, emitting at a brightness temperature of $\approx 35$~mK as a
3-$\sigma$ signal in about 1000 hours of observing time.  Zaroubi \&
Silk (2005) showed that we could even distinguish between stars and
miniquasars as sources of reionisation since there is a dramatic
difference between these two cases in the widths of the ionisation
fronts. Only the miniquasar model translates to scale-dependent 21~cm
brightness temperature fluctuations that should be measurable by
forthcoming LOFAR studies of the 21~cm angular correlation function
(Zaroubi et al. 2007).  A hitherto undetected population of Lyman
alpha-emitting galaxies is a possible reionisation source and may be
visible during the pre-reionisation era. One can hope to detect such
objects to $z\sim 10$ relative to the damping wing of the
Gunn-Peterson absorption from the neutral intergalactic medium outside
their HII regions (Gnedin \& Prada 2004).

\subsection{Secondary anisotropies from reionisation} 
When reionisation is completed, the scattering between CMB photons and
electrons moving along the line of sight generates secondary
anisotropies through the Doppler effect. The amplitude of the fluctuations
is given by:
\begin{equation}
\frac{\Delta T}{T}({\mathbf \theta})=\int{\mathrm d}\eta\,
a(\eta)g(\eta)v_{\rm r}({\mathbf \theta},\eta)= -\int{\mathrm d}t\,
\sigma_{\rm T}{\rm e}^{-\tau({\mathbf \theta},t)}n_{\rm e}({\mathbf \theta},t)
v_{\rm r}({\mathbf \theta},t)
\label{eq:dop}
\end{equation}
with $v_{\rm r}({\mathbf \theta},t)$ the velocity along the line of sight
(i.e. radial velocity). The electron 
density can be written as $n_{\rm e}({\mathbf \theta},t)=
n({\mathbf \theta},t)\times \chi_{\rm e}({\mathbf \theta},t)$ the product
of the matter density $n({\mathbf \theta},t)$ and the ionisation
fraction $\chi_e({\mathbf \theta},t)$. Both quantities vary around 
their average values. We can finally write the electron 
density as $n_{\rm e}({\mathbf \theta},t)={\bar n_{\rm e}({\mathbf
\theta},t)}[1+\delta+\delta_{\chi_{\rm e}}]$, with ${\bar n_{\rm e}({\mathbf
\theta},t)}$ the average number of electrons and $\delta$ and 
$\delta_{\chi_{\rm e}}$ the fluctuations of density field and ionisation 
fraction respectively.

By replacing the electron density expression in equation \ref{eq:dop},
we can see that there is a first order effect which suffers from
cancellations, and two second order effects which affect the
probability of scattering of the CMB photons
(e.g. Dodelson \& Jubas 1995). They both generate secondary
anisotropies. They are sometimes referred to as modulations of the
Doppler effect (i.e. the velocity field) by density and ionisation
spatial variations.

\subsubsection{Density-induced anisotropies}

\begin{figure} 
\epsfxsize=8.cm
\epsffile{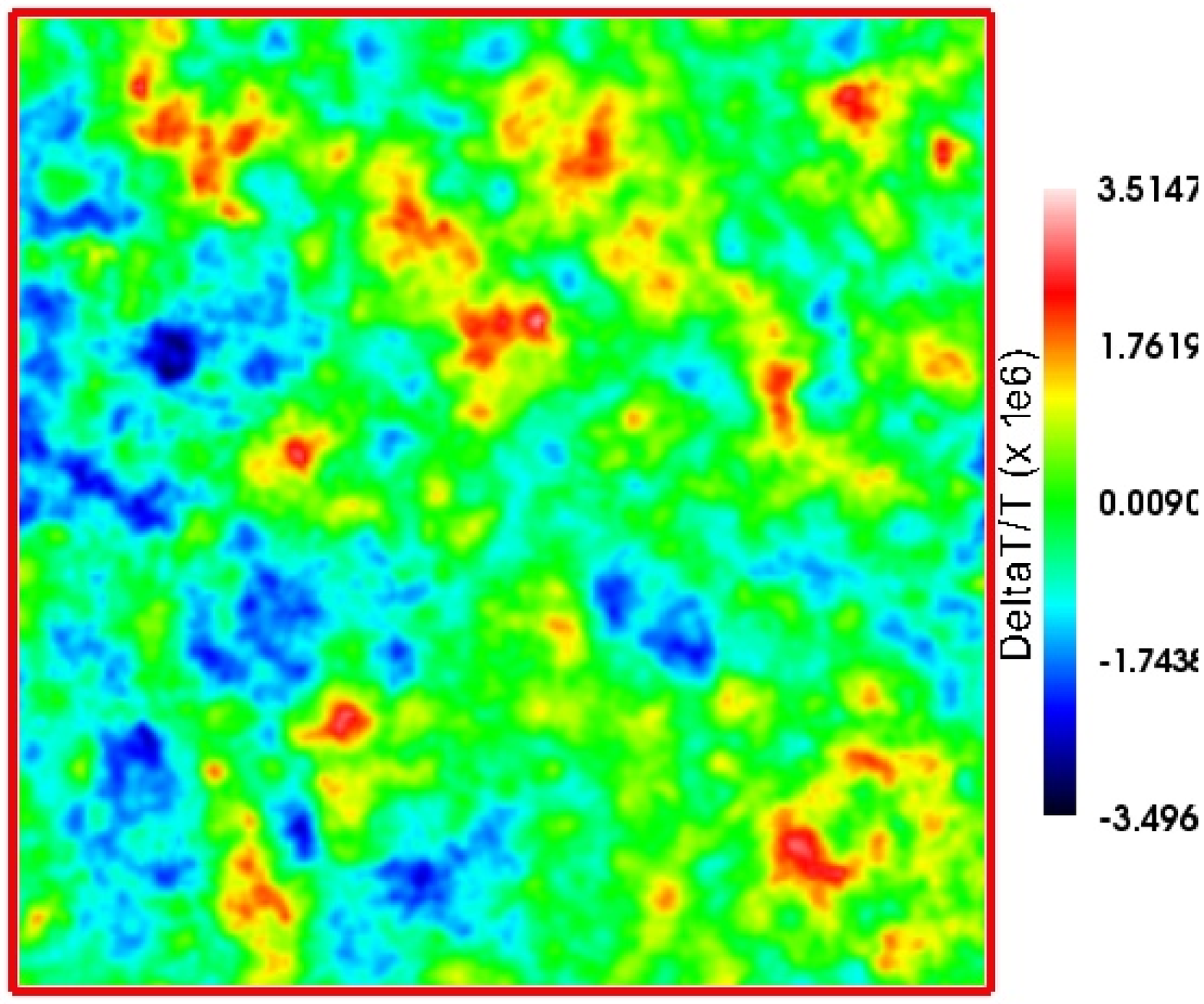}
\epsfxsize=8.cm
\epsffile{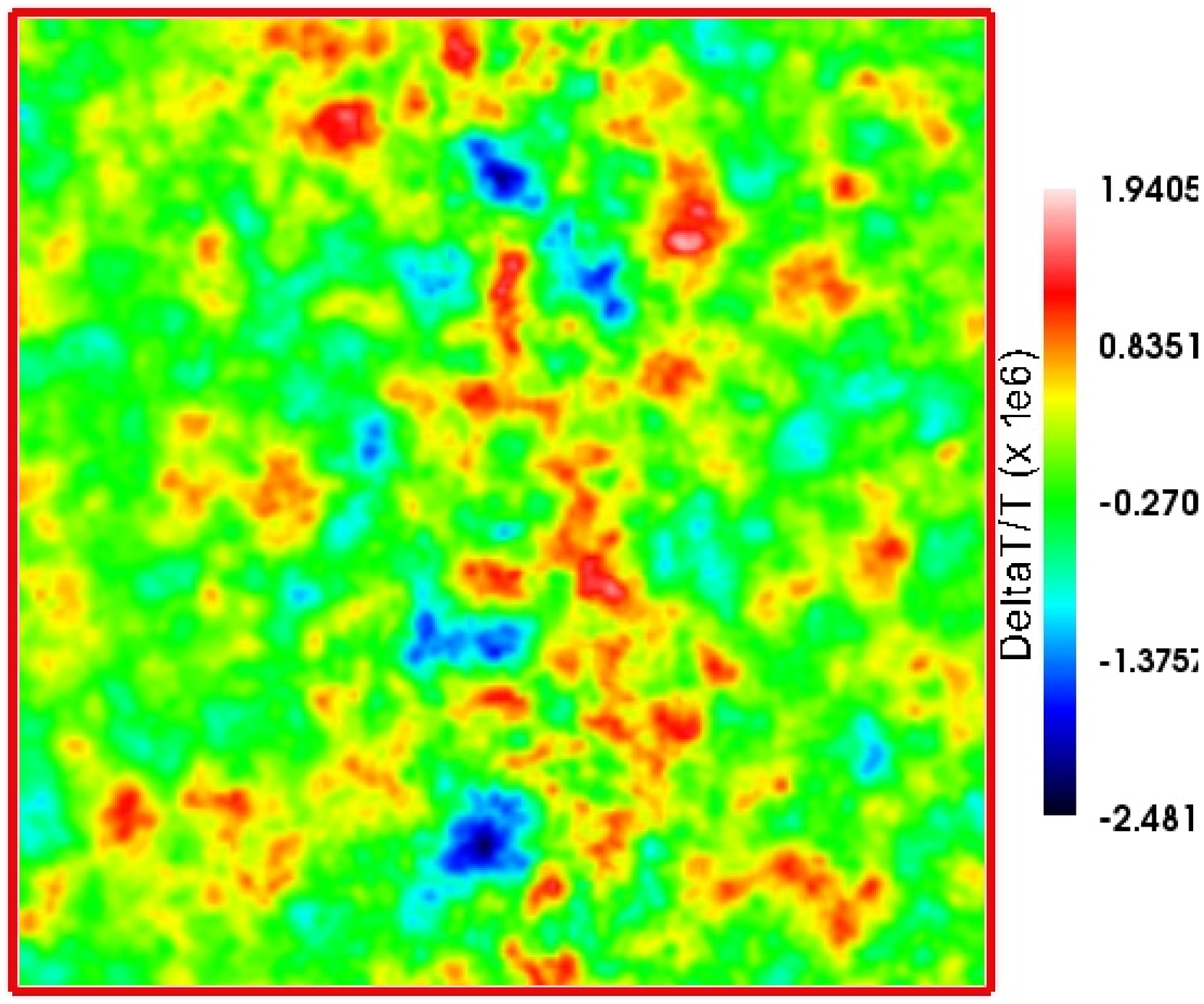}
\caption{From Iliev et al. (2007b): Doppler effect induced temperature
fluctuation maps from numerical simulations including radiative
transfer (right panel). The left panel shows the result after
correcting for the missing large-scale velocities.}
\label{fig:rion-im}
\end{figure}

These are produced when the ionisation fraction is homogeneous,
i.e. reionisation is completed, and when the Doppler effect is
modulated by spatial variations of the density field. The computation
in the linear regime first appeared in Sunyaev \& Zel'dovich (1970),
was revisited by Vishniac and Ostriker (Ostriker \& Vishniac 1986,
Vishniac 1987), and is known as the Ostriker-Vishniac (OV) effect (see
also Dodelson \& Jubas (1995), Hu \& White (1996), Jaffe \&
Kamionkowski (1998), Scannapieco (2000), Castro (2003)).  The OV
effect is a second order effect which weights as density squared
($\propto \delta^2$) and peaks at small angular scales (arc minutes) with
an {\it rms} amplitude of the order of $\mu$K.  The computation of the
density-induced anisotropies can be generalised to mildly non-linear
and non-linear regimes. Because these regimes are difficult to
describe analytically, a more appropriate tool is numerical
simulations (e.g. Gnedin \& Jaffe 2001, Zhang, Pen \& Trac 2004), see
also Figure \ref{fig:rion-im}.
However, one can also use the halo model (see review by Cooray \&
Sheth 2002) to model analytically the mildly non-linear regime as done
for example by Santos et al. (2003) or Ma \& Fry (2000, 2002).
\begin{figure} 
\epsfxsize=8.cm
\epsffile{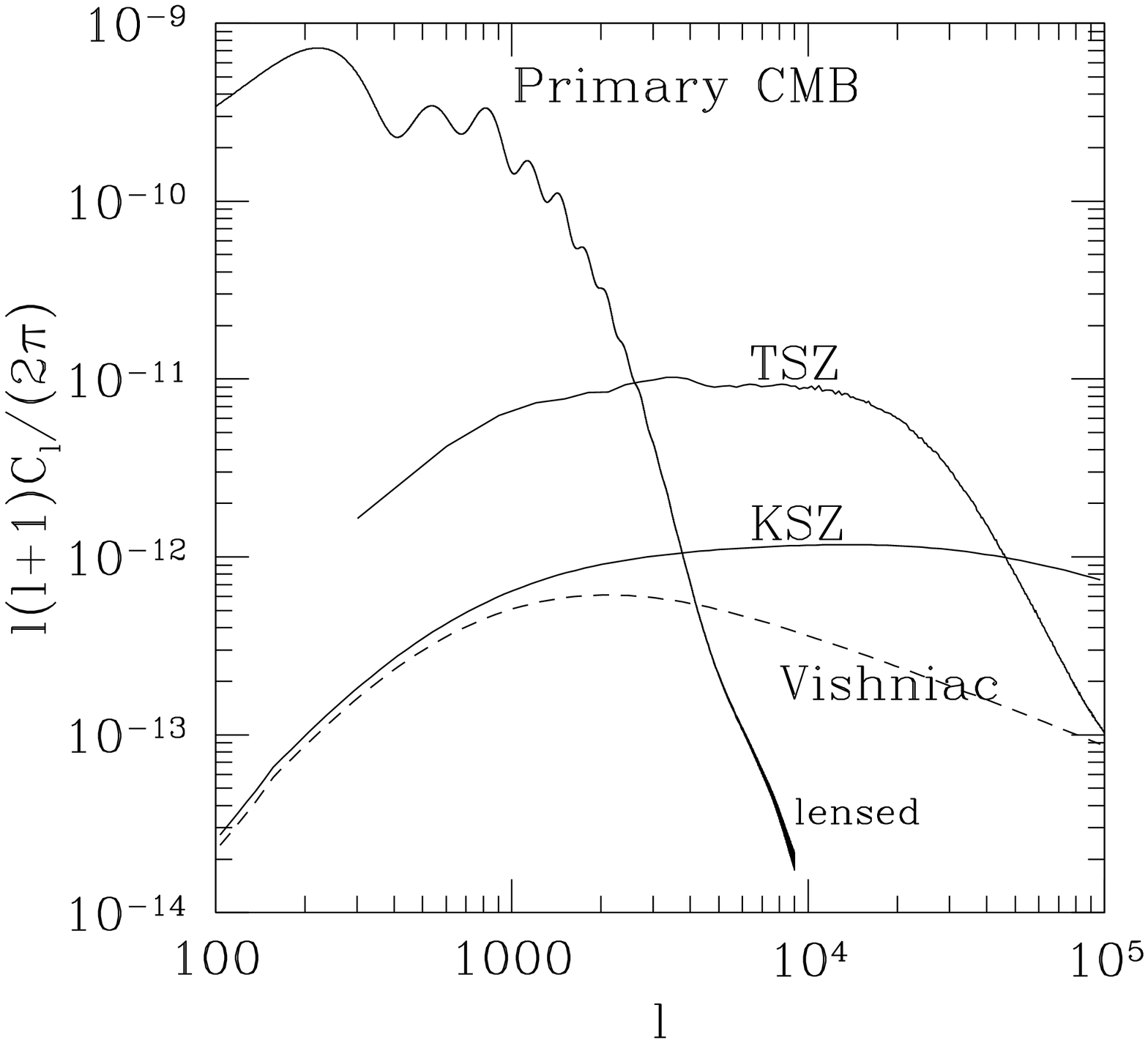}
\epsfxsize=8.3cm
\epsffile{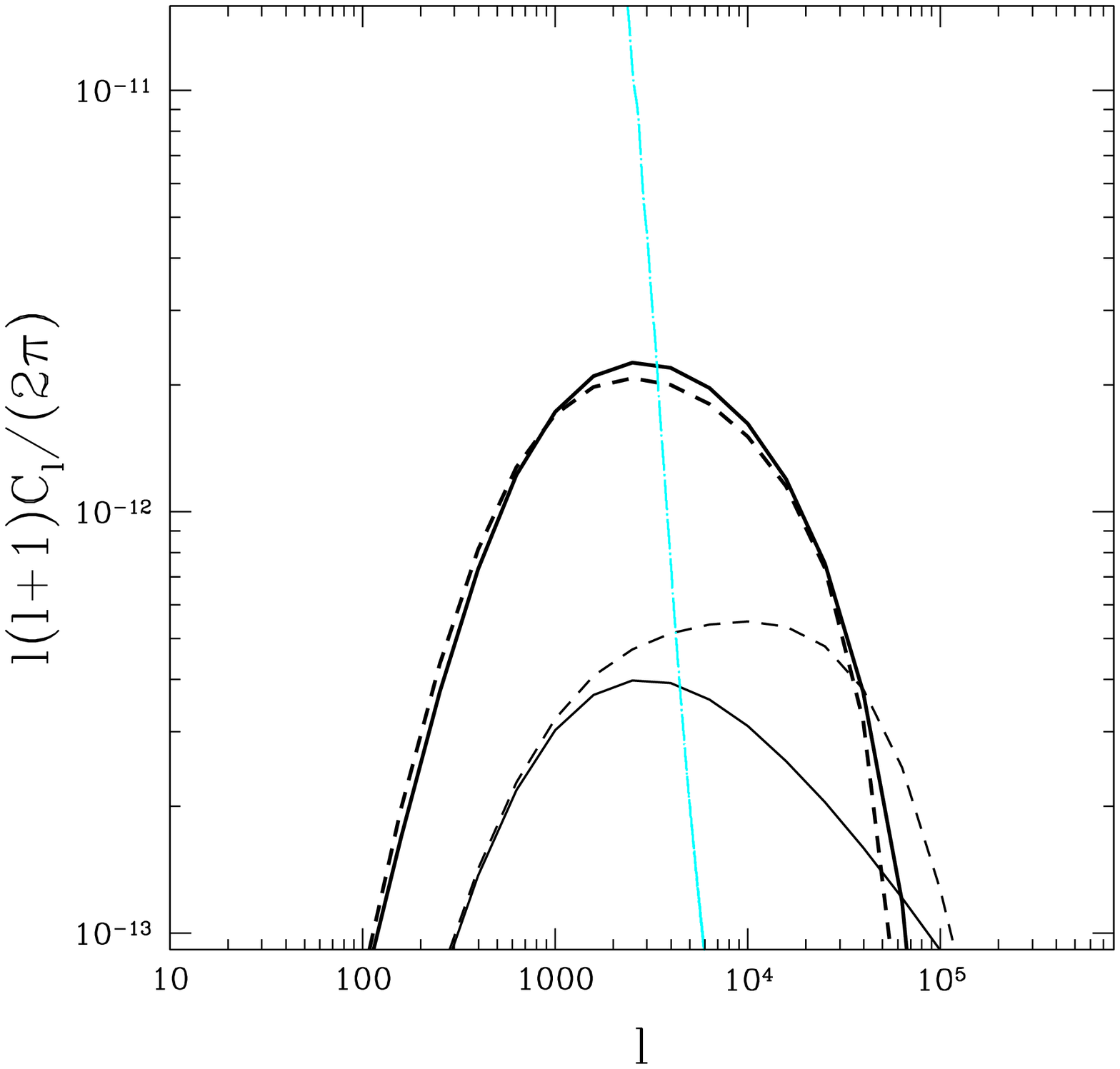}
\caption{Left panel, from Zhang, Pen \& Trac (2004): The Doppler
  effect induced temperature anisotropies (kinetic SZ) from numerical
  simulations. The results include non-linear regime and are obtained
  by assuming universe was reionised at $z=16.5$ and remained ionised
  after that. The contribution from the linear regime, OV effect,
  (dashed line) is plotted for comparison, together with primary power
  spectrum and thermal SZ spectrum in the R-J region (see
  Sect. \ref{sec_SZ}). Right panel, from Santos et al. (2003):
  Analytic computation of the secondary anisotropies produced by
  reionisation. Top thick lines are for the inhomogeneous
  reionisation-induced fluctuations. Bottom lines are for
  density-induced fluctuations where the solid thin line is for the
  linear OV effect and the dashed for the non-linear contribution to
  OV.}
\label{fig:reion-t}
\end{figure}
These studies showed that reionisation-induced anisotropies are
dominated by the OV effect at large angular scales. The contribution
from non-linear effects only intervenes at smaller scales with
amplitudes of $\sim$ a few $\mu K$ at $\ell>1000$
(Figure ~\ref{fig:reion-t}). The non-linear contributions from collapsed
and fully virialised structures such as galaxy clusters is
historically known as the kinetic Sunyaev-Zel'dovich effect and will
be discussed separately in section \ref{sec_SZ}.

\subsubsection{Sources of patchy reionisation}\label{reio-sourc}

Before reionisation is completed, ionised and neutral regions of the
universe co-exist. This is called the inhomogeneous reionisation (IHR)
regime.  In that case, the Doppler effect is modulated by variations
of the ionisation fraction $\chi_e$. Aghanim et al. (1996) computed the
first estimate of the power spectrum of secondary anisotropies induced
by early QSOs ionising the universe from $z=12$ to complete
reionisation at $z\sim 6$. They predicted a large contribution from
such fluctuations whose amplitude and distribution depended on the
number density of sources, their luminosities and their lifetimes.
The model was revisited by Gruzinov \& Hu (1998) and Knox, Scoccimarro
\& Dodelson (1998) who added the effect of spatial correlations
between sources.  The effect of an IHR on the CMB has been recently
revisited in the context of a reionisation scenario compatible with
WMAP data. In this work, Santos et al. (2003) found that secondary
fluctuations from IHR dominates over density-modulated (OV)
anisotropies.  IHR is intimately linked to the nature of the ionising
sources, to their formation and evolution history and to their spatial
distribution. As a result, predictions of the IHR effect span a large
range of amplitudes and angular scales.  A precise forecast of the
effects of IHR on the CMB anisotropies requires a precise treatment of
the reionisation history of the universe together with the formation
of the first ionising sources including radiative transfer (e.g. Iliev
et al. 2007a).

Stellar ionising sources have been studied by many authors (e.g.  Cen
2003, Ciardi, Ferrara \& White 2003, Haiman \& Holder 2003, Wyithe \&
Loeb 2003, Sokasian et al. 2003, Somerville \& Livio 2003).  The first
cosmological 3D simulations incorporating radiative transfer of
inhomogeneous reionisation by protogalaxies were performed by Gnedin
(2000).  He found that reionisation by protogalaxies spans the
redshift range from $z\sim 15$ until $z\sim 5$.  HII regions gradually
expand into the low-density intergalactic medium, leaving behind
neutral high-density protrusions, and within the next 10\% of the
Hubble time, the HII regions merge as the ionising background rises by
a large factor. The remaining dense neutral regions are gradually
ionised.  Sources as luminous as protogalaxies are too rare at these
redshifts and recourse must be had to a population of galactic
building blocks that are plausibly associated with dwarf galaxies or
miniquasars.

Recent studies find in general that in order to provide enough
ionising flux at, or before, $z=15$, for the usual scale-invariant
primordial density perturbation power spectrum, one needs Population
III stars, which provide about 20 times more ionising photons per
baryon than Population II (Schaerer 2002, Bromm, Kudritzki \& Loeb
2001), or an IMF that is initially dominated by high mass stars
(Daigne et al. 2004). This is in agreement with recent numerical
simulations of the formation of the first stars from primordial
molecular clouds suggesting that the first metal-free stars were
predominantly very massive, $m_{\rm star}\ge 100 M_\odot$ (Abel, Bryan
\& Norman 2000, 2002, Bromm, Coppi \& Larson 2002).  In general,
possibly unrealistically high ionising photon escape fractions are
required for a stellar reionisation source (Sokasian et al. 2004).

Miniquasars have also been considered as a significant ionising source
(e.g.  Ricotti \& Ostriker 2004, Ricotti, Ostriker \& Gnedin 2005,
Madau et al. 2004, Oh 2001, Dijkstra, Haiman \& Loeb 2004).  In
view of the correlation between central black hole mass and spheroid
velocity dispersion (Ferrarese \& Merritt 2000, Gebhardt et al. 2000),
miniquasars are as plausible ionisation sources as are Population III
stars, whose nucleosynthetic traces have not yet been seen even in the
most metal-poor halo stars nor in the high $z$ Lyman alpha forest. The
observed correlation suggests that seed black holes must have been
present before spheroid formation. Recent observations of a quasar
host galaxy at $z=6.42 $ (Walter et al. 2004) (and other AGN) suggest
that supermassive black holes were in place and predated the formation
of the spheroid. Theory suggests that the seeds from which the Super
Massive Black-Holes formed amounted to at least $1000 \rm M_\odot$ and
were in place before $z\sim 10$ (Islam, Taylor \& Silk 2003, Madau \&
Reese 2001, Volonteri, Haardt \& Madau 2003).

Decaying particles remain an option for reionisation that is difficult
to exclude. One recent example is provided by a decaying sterile
neutrino whose decay products, relativistic electrons, result in
partial ionisation of the smooth gas (Hansen \& Haiman 2004). A
neutrino with a mass of $\sim$ 200 MeV and a decay time of $\sim
10^8$yrs can account for an electron scattering optical depth as high
as 0.16 without violating existing astrophysical limits on the cosmic
microwave and gamma-ray backgrounds. In this scenario, reionisation is
completed by subsequent star formation at lower redshifts. Dark matter
annihilation during hydrogen recombination (at $z \sim 1000$) can
modify the recombination history of the Universe (Padmanabhan \&
Finkbeiner 2005). The residual ionization after recombination is
enhanced.  The surface of last scattering is broadened, partially
suppressing the small-scale primary temperature fluctuations and
enhancing the polarization fluctuations. In addition, the extended
recombination phase weakens some of the cosmological parameter
constraints, most notably on the scalar spectral index (Bean,
Melchiorri \& Silk 2007).

\subsection{Second order Polarisation at reionisation}
\label{sec:polsec}

In this section, we focus on the polarisation signal at small scales
induced at reionisation by the coupling between primary quadrupole and
fluctuations in the electron density at the new last scattering
surface. These electron density fluctuations can again have two
origins: They are either due to density fluctuations in a
homogeneously ionised universe (Seshadri \& Subrahmanian 1998, Hu 2000), or they can be
associated with fluctuations of the ionising fraction in an
inhomogeneously ionised universe
(Hu 2000, Mortonson \& Hu 2007). Additional polarisation fluctuations
from collapsed and virialised structures, such as galaxy clusters,
will be treated separately in Sect. \ref{sec:polclus}.
\begin{figure} 
\epsfxsize=8.cm
\epsffile{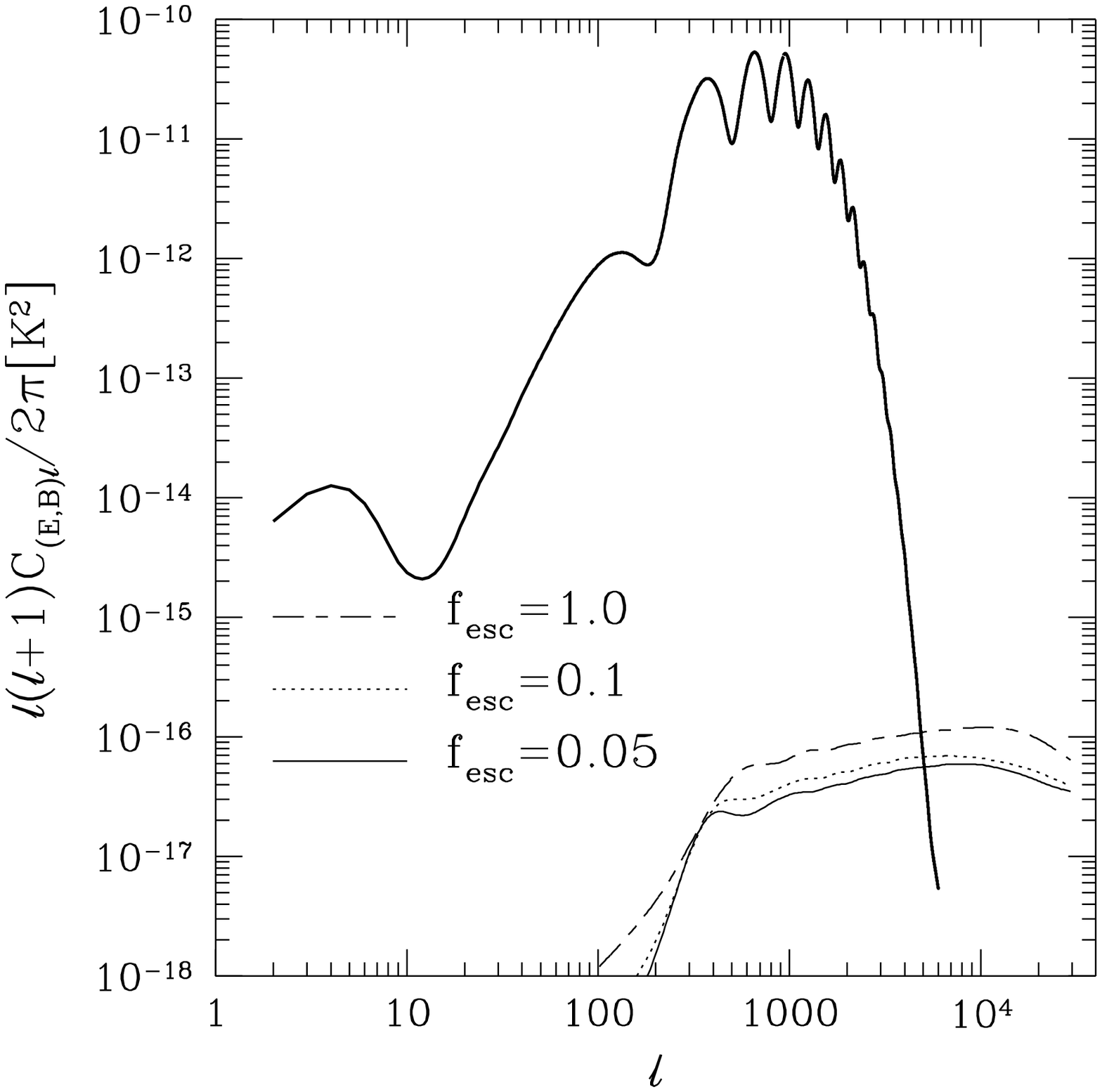}
\epsfxsize=8.cm
\epsffile{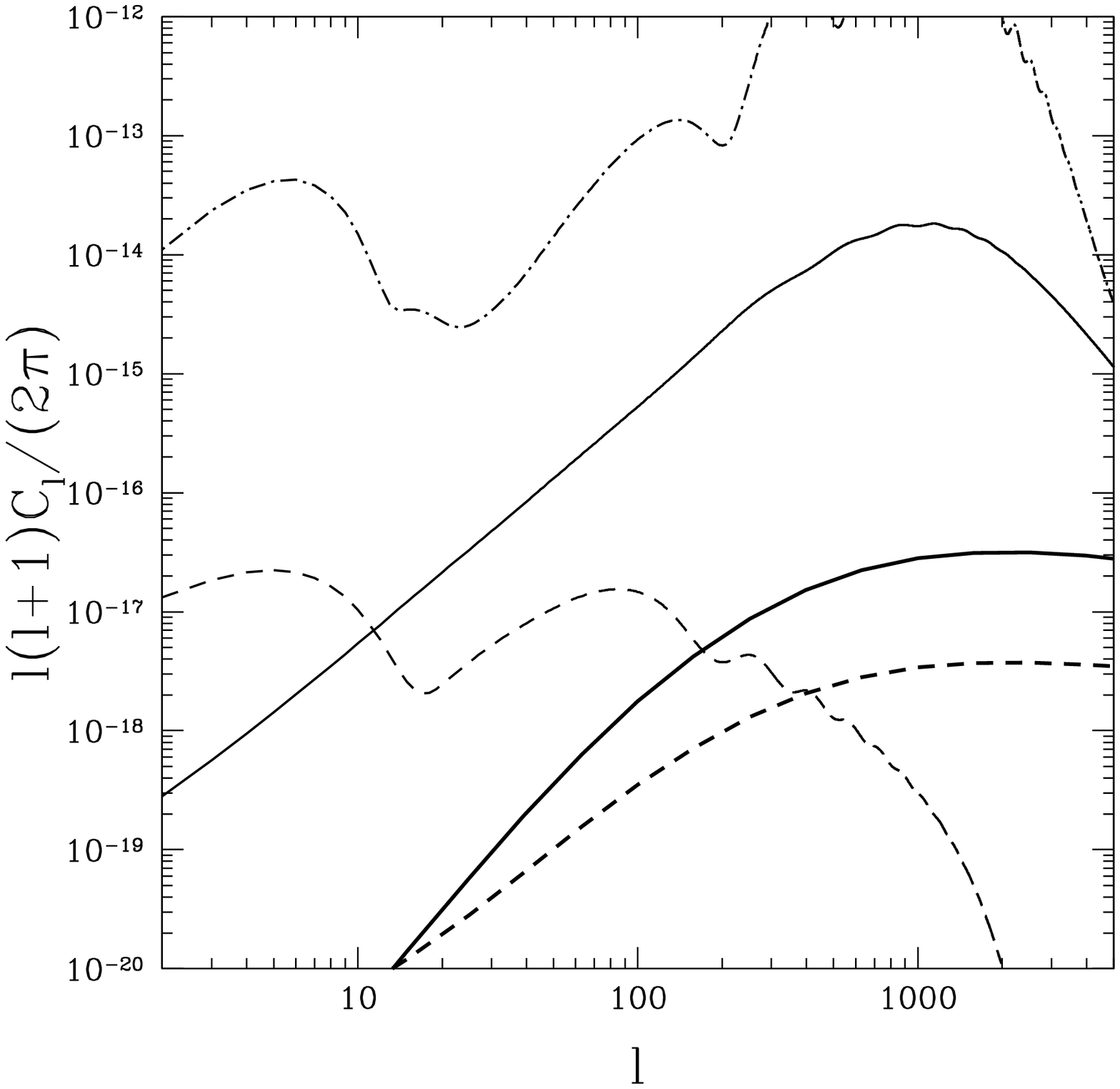}
\caption{Left panel, from Liu et al. (2001): Reionisation-induced
  polarisation (dashed, dotted and thin solid lines) with the
  first-order $E$-mode power spectrum (thick solid line).  The second
  order reionisation-induced polarisation is computed from numerical
  simulations with different escape fractions of ionising photons
  $f_{\rm esc}$. Right panel, from Santos et al. (2003): $B$-mode
  polarisation with contributions from lensing (thin solid line) and
  tensor modes (thin-dashed). The contribution, at reionisation, from
  density (thick dashed) and ionisation (thick solid) modulated
  scattering is also shown. The density modulated contribution uses
  the halo model for non-linear corrections. Also shown for comparison
  is the first-order $E$-mode power spectrum (dot-dashed line).}
\label{fig:polarsec}
\end{figure}

The dominant second order polarisation fluctuations are due to
coupling between primary quadrupole anisotropy $Q_{\rm rms}$ and electron 
density fluctuations $\delta_{\rm e}$ and are given by:
\begin{equation}
\Delta_{Q\pm iU}\propto \int d\tau g(\tau)Q_{\rm rms}\delta_{\rm e}
\propto\kappa Q_{\rm rms}\delta_{\rm e}.
\end{equation}
The quadrupole considered for generating polarisation through Thomson
scattering is in general the primary quadrupole. However in the rest
frame of the scattering electrons, a quadrupole moment is also
generated from quadratic Doppler effect (Sunyaev \& Zel'dovich 1980).
The amplitude of the polarisation induced by coupling with electron
density fluctuations in this case is smaller than those produced by
the primary quadrupole as discussed by (Hu 2000).

In all cases, the polarisation signal from secondary anisotropies
takes place at small angular scales, and has quite a small amplitude
(Figure \ref{fig:polarsec}).  Liu et al. (2001) found a typical amplitude
of $\sim10^{-2}\,\mu$K in a pre-WMAP reionisation model using
numerical simulations to describe reionisation
(Figure~\ref{fig:polarsec} left panel). More recently, this result was
confirmed by Dor\'e et al. (2007) who also used numerical simulation
compatible with current cosmological constraints.  In a model
reproducing the high optical depth suggested by 1st year WMAP
observations, Santos et al. (2003) generalised the computations to the
non-linear regime using the halo model. They conclude that the
modulation by ionising fraction inhomogeneities, i.e. patchy
reionisation, dominates over the modulation by density fluctuations
but the amplitudes remain small (Figure~\ref{fig:polarsec}, right
panel).

\section{Secondary effects from large-scale structure}\label{grav}

\subsection{The ISW effect }

After decoupling, as the universe continues to expand, seeds of cosmic
structures that scattered the CMB at the last scattering surface grow
due to gravitational instability giving rise to large scale structure.
The gravitational potential evolves with evolution of the structure
and the CMB photons are influenced once again by the change in the
gravitational potential which they traverse. One can subdivide the
gravitational secondaries broadly into two classes, one arising from
the time-variable metric perturbations and the other due to
gravitational lensing. The former is generally known as the integrated
Sachs-Wolfe effect (Sachs \& Wolfe 1967) in the linear region and goes
by the names of Rees-Sciama effect and moving-halo effect (sometimes
called the proper-motion effect) in the non-linear regime. The
integrated Sachs-Wolfe (ISW) effect is further divided in the
literature into an early ISW effect and a late ISW effect. The early
ISW effect is only important around recombination when anisotropies
can start growing and the radiation energy density is still
dynamically important. The final anisotropy for these gravitational
secondaries depends on the parameters of the background cosmology and
is also tightly coupled to the clustering and the spatial and temporal
evolution of the intervening structure.

In general, the temperature anisotropies, along any direction ${\bf
n}$, associated with the gravitational potential and proper motions
can be written in the form (Sachs \& Wolfe 1967, Hu, Scott \& Silk 1994,  see
Martinez-Gonz\'alez, Sanz \& Silk 1990, for a simple derivation)

\begin{equation}
\frac{\Delta T({\bf n})}{T} = (\phi_{\rm rec} - \phi_0)
+\int_{\eta_{\rm rec}}^{\eta_0}
2\dot{\phi}d\eta ,
\label{eqn:SachWolfe}
\end{equation}
where $\eta_{\rm rec}$ is the recombination time, $\eta_0$ the present
time and $\phi$ is the gravitational potential.  The first term
represents the Sachs-Wolfe effect due to different gravitational
potentials at recombination and present. The second term is the
integrated ISW effect and depends on the time derivative of $\phi$
with respect to the conformal time.  The numerical factors multiplying each
term in the equation depends on the choice of gauge and hence differ
among various authors.  A point to note is that the temperature change
due to the gravitational redshifting of photons is frequency
independent (in contrast to the SZ effect) and cannot be separated from
the primary anisotropies using spectral information only.

The origin of the late ISW effect lies in the decay of the
gravitational potential (Kofman \& Starobinsky 1985, Mukhanov, Feldman
\& Brandenberger 1992, Kamionkowski \& Spergel 1994). When the CMB
photons pass through structures they are blue and red-shifted when
they respectively enter and exit the gravitational potential wells of
the cosmic structures. The net effect is zero except in the case of a
non-static universe. This can happen naturally in a low matter density
universe and at the onset of dark energy (or spatial curvature)
domination typically occurring at late times.  The increased rate of
expansion of the universe reduces the amplitude of gravitational
potential.  The differential redshift of the photons climbing in and
out of the potential gives rise to a net temperature anisotropy. There
is one qualitative difference between the early ISW and the late ISW
effects. For the late ISW effect, the potential decays over a much
longer time (of the order of the present day Hubble time).  Thus the
photons have to travel through multiple peaks and troughs of the
perturbations and the chances of cancellation of the coherence in
gravitational redshifts becomes greater leaving, little net
perturbation to the photon temperature (Tuluie, Laguna \& Anninos
1996).

To study the amplitude of the late ISW effect, we start by
constructing its power spectrum. We expand the potential time
derivative, $\dot{\phi}$, in spherical basis to get the expression for
the power spectrum in a flat universe as

\begin{equation}
C_\ell=(4\pi)^2\int k^2P_\phi(k)dk \left[\int_0^{\eta_0} 2\dot{F}(\eta)
j_\ell(kr)d\eta\right ]^2\, ,
\label{cl}
\label{eqn:Cl_ISW1}
\end{equation}
where $j_\ell(x)$ is the spherical Bessel function and $r$ is the
comoving distance between the photon at a conformal time $\eta$ and
the observer. $F(k,\eta)=D/a$ is the growth rate of potential, where
$a$ is the expansion factor normalised to have $a_0 = 1$ and $D$ is
the linear growth factor. $D(z)$ governs the growth of amplitude of
density perturbation with time. It is simply equal to unity for
$\Omega_{\rm m}=1$ flat universe.  For universe with both matter and vacuum
energy (i.e $\Omega_\Lambda$), one has accurate fitting formulae for
the growth function (Carroll, Press \& Turner 1992).  The power spectrum of the
potential, $P_\phi$, is given by

\begin{equation}
\langle \phi(\vec{k}) \phi(\vec{k}^\prime) \rangle = (2\pi)^3 \delta_D (\vec{k} + \vec{k}^\prime) P_\phi(k).
\end{equation}

The main assumption in writing equation (\ref{eqn:Cl_ISW1}) is that in
the linear regime the mode does not change in phase and so the change
in its amplitude with time is simply described through the growth
factor. The equation also ignores gravitational lensing to be discussed
later.

In the small angular scale limit and under the assumptions that
the correlations at a distance $k^{-1}$ are slowly changing on a
timescale $(ck)^{-1}$, the radial integral in equation
(\ref{eqn:Cl_ISW1}) can be broken into a product of the spherical Bessel
function $j_\ell(kr)$ and a slowly changing function of time. Taking out
the slowly varying part outside the radial integral and using the
large $\ell$ approximation for the Bessel function, we can use the
Limber approximation to get

\begin{equation}
C_\ell =
32\pi^3\int_0^{\eta_0}\frac{\dot{F}^2(k = \ell/r,\eta)P_\phi(k = \ell/r) d\eta}{r^2},
\label{eqn:Cl_ISW2}
\end{equation}

From the above equation, we can define the power spectrum of the
potential time derivative as
$P_{\dot{\phi}}(k,\eta)=\dot{F}^2(k,\eta)P_\phi(k)$. Note that in the
non-linear regime, the growth factor depends on the wavenumber
$k$. However, equation (\ref{eqn:Cl_ISW2}) is still valid due to the
slow time dependence of $P_{\dot{\phi}}$.
 
To calculate $\dot{\phi}$ as a function of time and scale, we relate the
potential to the matter density via the Poisson equation. In
$k$-space, this can be written as

\begin{equation}
\phi + {3 \over 2}\frac{\Omega_{\rm m}}{a} \left(\frac{H_0}{k}\right)^2\delta = 0 ,
\label{eqn:kspacePoss}
\end{equation}
where $\Omega_{\rm m}$ is the present day matter density parameter and
$\delta$ is the matter density perturbation.

\begin{figure}  
\epsfxsize=12.cm
\epsfysize=10.cm
\hspace{1.5cm}
\epsffile{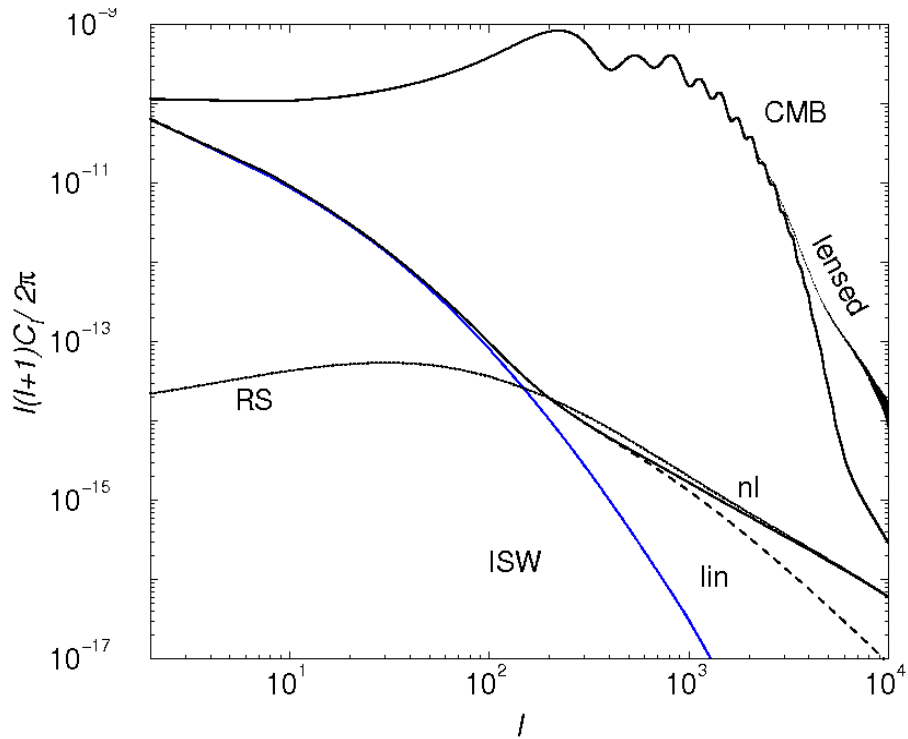}
\caption{From Cooray (2002a): The power spectrum of the ISW
effect, including non-linear contribution. The Rees-Sciama effect
shows the non-linear extension. The curve labeled ``nl'' is the
non-linear contribution while the curve labeled ``lin'' is the
contribution from the momentum field under the second order
perturbation theory. The primary anisotropy power spectrum accounting
for the lensing effect is shown for comparison. }
\label{fig:gravsecondaries}
\end{figure}

At this point it is straightforward to calculate the late ISW effect
once we put in an appropriate expression for $F(\eta)$. After we do this,
the first thing to notice is that for a flat matter-dominated
$\Omega_{\rm m}
= 1$ universe, $D(\eta) \propto a(\eta)$, and so in the linear regime
there is no ISW effect. Until non-linear effects are considered, the late
ISW effect occurs only in open and lambda dominated universes. The
linear ISW effect, the non-linear ISW effect and gravitational lensing
effect are shown in figure \ref{fig:gravsecondaries}.

\begin{figure}
\begin{center}
\resizebox{4.3in}{!}{\includegraphics{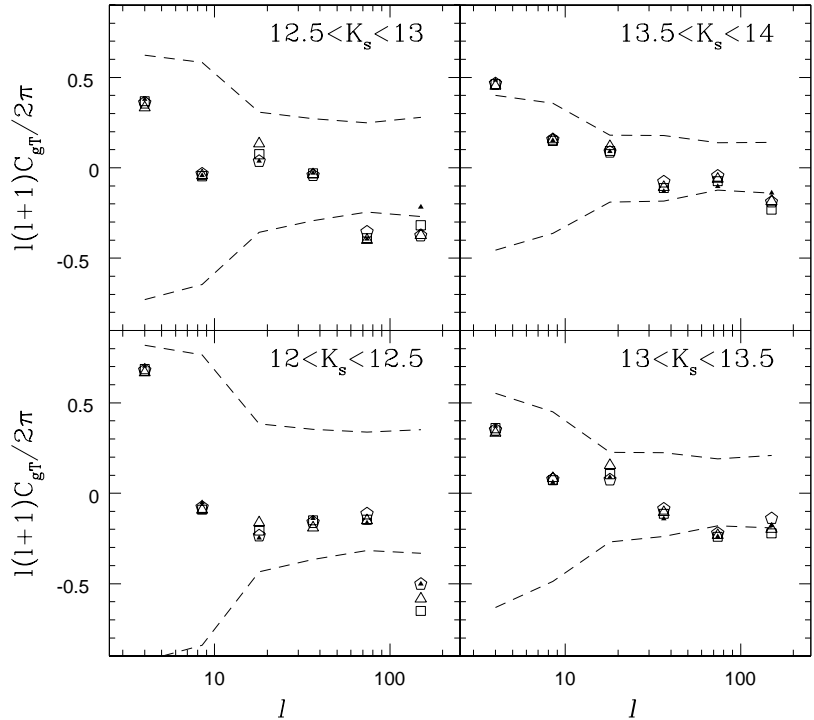}}
\end{center}
\caption{From Rassat et al.(2006): Results of the cross-correlation
  $C_{gT}(\mu K)$ for the Internal Linear Combination (small triangle),
  Q (open triangle), V (open square), and W (open pentagon) WMAP
  maps with different magnitude bins of the 2MASS galaxy survey.  The
  dashed lines are 1$\sigma$ error bars about the null hypothesis. An
  ISW effect is expected to be achromatic, which is observed, but the
  null hypothesis is not ruled out.}
\label{fig:isw-meas}
\end{figure}

The ISW effect is seen mainly in the lowest $\ell$-values in the power
spectrum (Tuluie, Laguna \& Anninos 1996). Its importance comes from
the fact that it is very sensitive to the amount, equation of state
and clustering properties of the dark energy. Detection of such a
signal is, however, limited by cosmic variance. The time evolution of
the potential that gives rise to the ISW effect may also be probed by
observations of large scale structure. One can thus expect the ISW to
be correlated with tracers of large scale structure. This idea was
first proposed by Crittenden \& Turok (1996) and has been widely
discussed in the literature (Kamionkowski 1996, Kinkhabwala \&
Kamionkowski 1999, Cooray 2002b, Afshordi 2004, Hu \& Scranton
2004). The ISW detection was attempted using the COBE data and radio
sources or the X-ray background (Boughn, Crittenden \& Turok 1998,
Boughn \& Crittenden 2002) without much success. The recent WMAP data
(Spergel et al. 2003, 2007) provide for the first time all-sky high
quality CMB measurements at large scales.  Those data were used
recently in combination with many large scale structure tracers to
detect the ISW signal.  The correlations are presently performed
mainly using galaxy surveys (2MASS, SDSS, NVSS, SDSS, APM, HEAO), see
Figure~\ref{fig:isw-meas} for a recent result. However, despite numerous
attempts both in real space (Diego, Hansen \& Silk 2003, Boughn \&
Crittenden 2004, Fosalba \& Gaztanaga 2004, Hernandez-Monteagudo \&
Rubiono-Martin 2004, Nolta et al. 2004, Afshordi, Lin \& Sanderson
2005,Padmanabhan et al. 2005,Gaztanaga, Maneram \& Multamaki
2006,Rassat et al. 2006) or in the wavelet domain (e.g. Vielva,
Martinez-Gonzalez \& Tucci 2006), there is very weak (or null) detection of the ISW effect
through correlations. The
ISW effect provides and offers a promising new way of inferring
cosmological constraints (e.g. Corasaniti, Gianantonio \& Melchiorri
2005, Pogosian 2006).

\subsection{The Rees-Sciama and the moving halo effects}\label{RS}

As mentioned in the previous section, the ISW is linear in first order
perturbation theory. Cancellations of the ISW on small spatial scales
leave second order and non-linear effects. In hierarchical structure
formation, the collapse of a structure can present a changing
gravitational potential to passing photons. If the photon crossing
time is a non-negligible fraction of the evolution time-scale, the net
effect of the blue and redshift is different from zero and the path
through the structures leaves a signature on the CMB.  This was first
pointed out by Reese \& Sciama (1968) for evolving density profiles of
any individual large scale structures (see also Dyer 1976). This goes
by the name of the Rees-Sciama (RS) effect. Subsequently, there have
been many studies of the RS effect from isolated structures using the
`Swiss-Cheese' model (Kaiser 1982, Thompson \& Vishniac 1987,
Martinez-Gonzalez, Sanz \& Silk 1990, Chodorowski 1992,1994),
Tolman-Bondi solutions (Panek 1992, Lasenby et al.1999) and from
clustering (Fang \& Wu 1993). Calculations have also been done for
non-linear regimes, both analytically (e.g. Cooray 2002a) and using
numerical simulation (e.g. Seljak 1996a, Dabrowski et al.1999). Much
of this work was concerned with the possible contamination of primary
anisotropies by the RS effect, since both are present at similar
angular scales and cannot be distinguished using multi-frequency
observations. As we shall see below, the RS effect is negligibly small
at all angular scales (Figure~\ref{fig:gravsecondaries}). The non-linear
evolution of primordial scalar fields generates some vector and tensor
modes, inducing, in turn, $B$ mode polarisation anisotropies
(Mollerach, Harary \& Matarrese 2004). This secondary signal although
smaller than the one associated with gravitational lensing effects
(see Section \ref{sec:lens}) might constitute a limiting background for future CMB
polarisation experiments.

For an isolated collapsed structure, there can be a change in the
gravitational potential along the line of sight due to its bulk motion
across the line of sight. For clusters of galaxies, this was first
shown by Birkinshaw \& Gull (1983); (see also Birkinshaw (1989) for a
correction to the original results) as a way to measure their
transverse velocities and is known as the ``moving-halo'' effect. At
the same time, these authors pointed out to the fact that CMB
anisotropies should be gravitationally lensed by such moving halos.  A
similar proposal for temperature anisotropies due to the presence of
cosmic string wakes was proposed by Kaisser \& Stebbins (1984) (see
also Stebbins 1988). The CMB photons entering ahead of a moving
structure (galaxy cluster or super cluster) traversing the line of
sight will be redshifted, while those entering the structure wake are
blueshifted. The transverse motion induces a bipolar imprint in CMB
whose amplitude is proportional to the velocity $v_{\rm t}$ and to the
depth of the potential well $M_{\rm tot}$ and aligned with the
direction of motion. The effect of moving local mass concentrations
like the Great Attractor or the Shapley concentration was recently
investigated Cooray \& Seto (2005) (see also Tomita 2005, 2006) to
explain the quadrupole and octopole alignment in the WMAP first year
data. This effect was found to be much smaller than that required for
explaining the low multipole anomalies (but see Vale (2005)).
In general, the bulk motion of
dark matter halos of all masses would contribute to this effect and is
found to be negligible for all angular scales (Aghanim et al.1998,
Molnar \& Birkinshaw 2000).  Lensing by moving massive clusters only
induce a few $\mu$K temperature distortion (Dodelson 2004, Holder \&
Kosowsky 2004).

We can try to combine the temperature anisotropy due to the RS and
moving-halo effects to make a simple estimate `non-linear ISW'
effect. For an isolated structure, the anisotropy can be written as

\begin{equation}
\frac{\Delta T}{T} \sim \frac{\phi}{t_{\rm c}}\delta t + \frac{v_{\rm
t}}{d} \delta t,
\label{eqn:panek1}
\end{equation}
where $t_{\rm c}$ is the characteristic dynamical time namely the
free-fall time, $\delta t$ is
the photon crossing time, $d \sim c\,\delta t$ is the physical
size. The potential $\phi$ can be determined from $\phi \sim M_{\rm
tot}/d$. The matter crossing time $d/ v_{\rm t}$ is taken as the
evolution time $t_{\rm c}$. From energy balance arguments, we get
$\phi \sim v^2_{\rm t}$. Thus, we have $t_{\rm c} \sim
d/\phi^{1/2}$. Putting all these together in equation
(\ref{eqn:panek1}), we can write

\begin{equation}
\frac{\Delta T}{T} \sim \phi^{3/2} + v_{\rm t}\phi . 
\label{eqn:panek2}
\end{equation}
The first term is the RS term and the second is the moving-halo
term. Finally, from linear perturbation theory, we have $v_{\rm t} \sim \phi
(1+z)^{-3/2} (dH_0)^{-1}$ and $\delta \sim \phi(1+z)^{-3}(dH_0)^{-2}$,
so we can rewrite equation (\ref{eqn:panek2}) as

\begin{equation}
\frac{\Delta T}{T} \sim 10^{-7} \left(\delta^{3/2} + \delta^{2}\right)
\left(\frac{d}{14 h^{-1}{\rm Mpc}} \right)^3 \left(1+z\right)^{9/2}.
\end{equation}

The above estimate is rather crude since we have used linear
perturbation theory to describe non-linear regions. Moreover, it only
applies to an isolated structure (for which the RS effect is
independently treated from the velocity effect) and a proper
justification can be only done using simulations where the phase
dependence of the growth factor is naturally taken into account. The
non-linear ISW effect can also be calculated using the halo model
which allows us to describe both the density and velocity fields of
the large scale structure in a coherent way (for details see
Cooray \& Sheth 2002). In such an approach, the basic idea is to take the time
derivative of the Poisson equation (i.e. equation
(\ref{eqn:kspacePoss})) and using the continuity equation in $k$-space
given by

\begin{equation}
\dot{\delta}+i\vec{k}\cdot \vec{p}=0,
\end{equation}
where the momentum density field $\vec{p}(\vec{r}) = (1+\delta)\vec{v}(\vec{r})$; 
one then obtains the following expression:
\begin{equation}
\dot{\phi}={3 \over 2} \frac{\Omega_{\rm m}}{a} \left( {H_0 \over k} \right)^2(
\frac{\dot{a}}{a}\delta+i\vec{k} \cdot \vec{v}),
\label{eqn:kspacephidot}.
\end{equation}

This relation connects the potential time derivative to the density
and the momentum density. One can now obtain the power spectrum of
$\dot{\phi}$ by averaging over all the $k$-modes. It is easy to see
from equation \ref{eqn:kspacephidot} that the power spectrum will
involve correlation between density fields and time derivatives of
density fields, as well as cross-correlation between density and
momentum fields. Thus the general result has information about both
the classical RS effect as well as the moving-halo effect. Numerical
simulations capture an important point that is often missed in
analytical perturbation theory calculations which is that in the
strongly non-linear regime the power spectrum of $\dot{\phi}$ is
dominated by the momentum density. 

The angular power spectrum including the non linear ISW effect is
shown in figure \ref{fig:gravsecondaries}. For all cases, the
temperature anisotropy $\frac{\Delta T}{T}$ is between $10^{-6}$ and
$10^{-7}$. The amplitude of the power spectrum goes as the
normalisation parameter
$\sigma^4_8$. Moreover for a given $\sigma_8$, change in $\Omega_{\rm m}h$
significantly affects the power spectrum at the low
$\ell$-values. Depending upon the background cosmology, the power
spectrum peaks at $\ell$ between $100$ and $300$ and is always $2-3$
orders of magnitude less than primary CMB power spectrum. The non
linear ISW effect becomes equal to the primary anisotropy at $\ell
\sim 5000$. However, well before this equality is reached, it is
overtaken by other sources of secondary anisotropies such as the
thermal SZ effect.

\section{Lensing of the CMB}\label{sec:lens}

\subsection{Lensing by large scale structure} 

\begin{figure} 
\epsfxsize=13.cm
\epsfysize=10cm
\hspace{2cm}
\epsffile{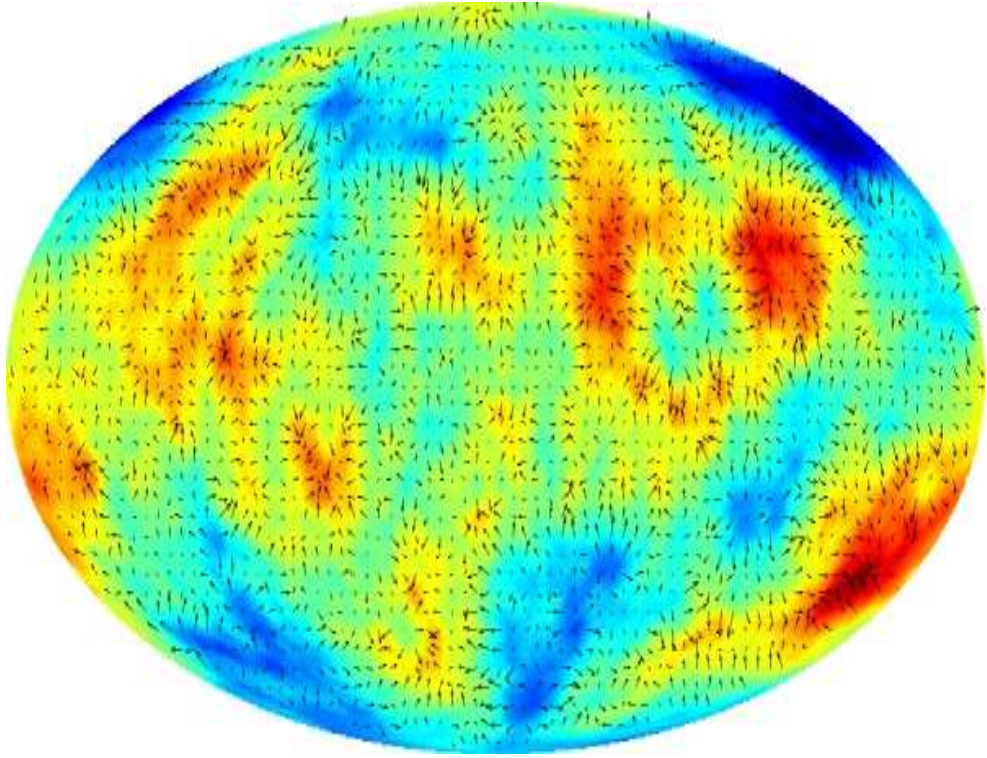}
\caption{From http://cosmologist.info/lenspix/: All-sky maps of the
  lensed CMB temperature and polarisation anisotropies performed using
  the formalism developped in Lewis (2005).}
\label{fig:cmblens}
\end{figure}

As the CMB photons propagate from the last scattering surface, the
intervening large scale structure can not only generate new secondary
anisotropies (as shown in the last section, Sect.~\ref{grav}) but can
also gravitationally lens the primary anisotropies (Blanchard \&
Schneider 1987,Kashlinsky 1988,Linder 1988, Cayon, Martinez-Gonzalez
\& Sanz 1993, Seljak 1996b, Metcalf \& Silk 1997, Hu 2000).  For a
detailed description of the process we refer the reader to a recent
and thorough review by Lewis \& Challinor (2006). Formally, lensing
does not generate any new temperature anisotropies. There are indeed
no new anisotropies generated if the gravitational potential is not
evolving (see previous section for this case), whereas lensing occurs
whenever there is a gravitational potential.  Since lensing conserves
surface brightness, the effect of gravitational lensing of the primary
CMB can only be observed if the latter has anisotropies. In this case
lensing magnifies certain patches in the sky and demagnifies others
(Figure \ref{fig:cmblens}). If the primary CMB were completely
isotropic, one would not be able to differentiate between the
different (de)magnifications.

For gravitational lensing, the absolute value of the light deflection
does not matter. What matters is the relative deflection of close-by
light rays. If all the adjacent CMB photons are isotropically
deflected, there would only be a coherent shift relative to the actual
pattern. However, if they are not isotropically deflected, then the
net dispersion of the deflection angles would change the intrinsic
anisotropies at the relevant angular scales. The net result of
gravitational lensing is to transfer power from larger scales (thus
smoothing the initial peaks in the CMB power spectrum) to smaller
scales. In the following, we detail the lensing effects on both the
temperature anisotropies as well as on the polarised signal.

In order to understand the effects of gravitational lensing on the CMB
power spectrum we have to write its effect on a single temperature and
polarisation anisotropy. Gravitational lensing modifies the CMB
anisotropies, which are then measured as an angular displacement in
the following way
\begin{equation}
T_{\rm obs}({\mathbf \theta})=T({\mathbf \theta+\xi(\theta)}),
\end{equation} 
where $\theta$ is the original undistorted angle. However, it is
inaccurate to approximate the observed temperature by a truncated
expansion in the deflection angle. This is only a good approximation
on scales where the CMB is very over the relevant lensing deflection,
i.e. on large scales, or very small (see Challinor \& Lewis 2005).
In the weak lensing
limit, the regime of interest for CMB studies, we can use the
perturbative approach and write the lensed CMB anisotropies as:
\begin{equation}
T_{\rm obs}({\mathbf \theta})\sim T({\mathbf \theta})+\xi^i({\mathbf\theta})
\cdot T_{,i}+\frac{1}{2}\xi^i({\mathbf\theta})\xi^j({\mathbf\theta})\cdot
T_{,ij}
\end{equation}
with $\xi$ given by:
\begin{equation}
\xi_i({\mathbf \theta})=\frac{-3}{2}\Omega_0\int\frac{dz'}{H(z')}\frac{1}{a}
\frac{D_0(z')D_0(z,z')}{D_0(z)}\varphi_{,i}^{(1)}({\mathbf \theta},z),
\end{equation}
$D_0(z,z')$ is the angular diameter distance between redshifts $z$ and
$z'$ and $\varphi_{,i}^{(1)}({\mathbf \theta}, z)$ is the
perpendicular gradient of the Newtonian potential in the direction ${\mathbf
\theta}$.
In the same way, the modified polarisation anisotropy is written as
$P_{\rm obs}(\mathbf{\theta})=P(\mathbf{\theta}_{\rm
obs})=P(\mathbf{\theta}+\mathbf{\xi})$ which similarly gives second
order:
\begin{equation}
P_{\rm obs}(\mathbf{\theta})\sim
P(\mathbf{\theta})+\xi^i(\mathbf{\theta})\cdot
{P_{,i}}+\frac{1}{2}\xi^i(\mathbf{\theta})\xi^j(\mathbf{\theta})\cdot
{P_{,ij}}
\label{eq:pollen}
\end{equation}


\subsubsection{Lensed CMB power spectrum} 
In order to have an idea of the effect of gravitational lensing on the
CMB power spectrum, we can consider its counterpart the two point
correlation function. In the small angle approximation, we can use the
perturbative approach to second order and obtain:
\[\langle T_{\rm obs}({0})T_{\rm obs}(\mathbf{\theta})\rangle=\langle 
T({0+\xi(0)})T(\mathbf{\theta+\xi(\theta)})\rangle\]
\[=\langle T({0})T(\mathbf{\theta})\rangle+\langle\xi^i({0})
\xi^j(\mathbf{\theta})\rangle\langle T_{,i}({0})T_{,j}(\mathbf
{\theta})\rangle \]
\[+\frac{1}{2}\langle\xi^i({0})\xi^j({0})
\rangle\langle T_{,ij}({0})T(\mathbf{\theta})\rangle 
 +\frac{1}{2}
\langle\xi^i(\mathbf{\theta})\xi^j(\mathbf{\theta})\rangle\langle
T({0})T_{,ij}(\mathbf{\theta})\rangle \] 

\begin{figure} 
\epsfxsize=11.cm
\epsfysize=10cm
\hspace{2cm}
\epsffile{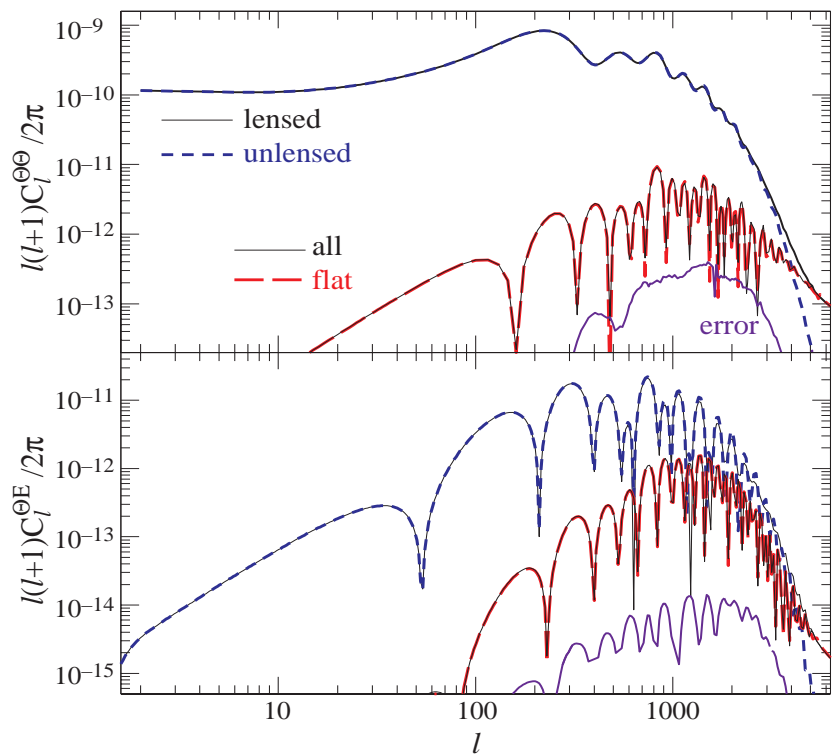}
\caption{From Hu (2000): The lensed and unlensed power spectra. The
error due to the flat sky approximation with respect to the all sky
computation is also shown. The corrections of the all sky computations
can be even larger (see Challinor \& Lewis 2005)}
\label{fig:cllense}
\end{figure}

The lensed power spectrum $C_\ell^{\rm obs}$ as a function of the
unaltered power spectrum $C_\ell$ is obtained after Fourier
transformation, which is directly associated with multipole
decomposition. It is given by

\begin{equation}
C_\ell^{\rm obs}=C_\ell\left[1-\int\frac{d^2k}{(2\pi)^2}
\frac{(\mathbf{\ell\cdot k})^2}{k^4}\bar{P}(k)\right]
 +\int\frac{d^2k}{(2\pi)^2} \frac{(\mathbf{\ell\cdot
 k})^2-k^4}{k^4}\bar{P}(k)C_{|\mathbf{\ell-k}|},
\end{equation}

where $\bar{P}$ is the projected power spectrum of the of the lensing
convergence.  A generalisation of the computation (Hu 2000) shows that
the errors introduced by the flat sky approximation are negligible as
shown in figure \ref{fig:cllense}.

The expression of $C_\ell^{\rm obs}$ clearly shows the effect of the
gravitation lensing on the CMB:
\begin{itemize}
\item The first term is a renormalisation due to the second order
effect introduced by the lenses in the perturbative formulae. 
\item The second term is a mode coupling due to the convolution of the
unperturbed spectrum by the projected power spectrum $\bar{P}$. Both
cause the smoothing of the acoustic peaks at small scales.
\end{itemize}

Weak lensing does not introduce any characteristic scale in the CMB.
Its effects are mostly noticeable at small scales where they modify
the CMB damping tail through power transfer from large to small
scales. This increase in power at large $\ell$s is significantly smaller
than the modifications due to scattering effects (e.g. SZ effect). To
identify the effects of gravitational lensing on the CMB it is
necessary to explore not only the power spectrum but also higher order
moments that possibly reveal the induced non-Gaussian signatures left
by the non-linear coupling (Bernardeau 1997, 1998, Zaldarriaga 2000,
Cooray 2002c, Kesden, Cooray \& Kamionkowski 2003).  The projected
mass distribution from $z\sim 1000$ to present and hence the lensing
effect can be reconstructed in principle via maximum likelihood
estimators or quadratic statistics in the temperature and polarisation
(e.g.  Goldberg \& Spergel 1999, Hu 2001, Okamoto \& Hu 2003, Cooray
\& Kesden 2003, Hirata \& Seljak 2003).  However, as shown for example
in Amblard, Vale \& White (2004), lensing reconstruction is affected
by other secondary effects indistinguishable from lensing such as the
KSZ effect or residual foreground contaminations. In addition to
providing the projected mass density, the weak lensing effect on the
CMB is a potentially powerful tool to probe the neutrino mass and dark
energy equation of state (e.g. Kaplinghat, Knox \& Song 2003,
Lesgourgues et al. 2006).

\subsubsection{Effects of lensing on CMB Polarisation} 
A curl-free vector field does not remain scalar if it is distorted.
Consequently in the case of CMB polarisation vector field, we expect
that gravitational lensing will mix the $E$ and $B$ components of the
polarisation. Computing equation \ref{eq:pollen} for $E$ and $B$
components implies second derivatives of a distorted field
(e.g. Benabed, Berbardeau \& van Waerbeke 2001) and gives:
\begin{equation}
\Delta E_{\rm obs}=(1-2\kappa)\Delta E+\mathbf{\xi\cdot\nabla}(\Delta E)
-2\delta^{ij}(\gamma_i\Delta
P_j+\mathbf{\nabla\gamma_i\cdot\nabla}P_j)
\end{equation}
and
\begin{equation}
\Delta B_{\rm obs}=(1-2\kappa)\Delta B+\mathbf{\xi\cdot\nabla}(\Delta B)
-2\epsilon^{ij}(\gamma_i\Delta
P_j+\mathbf{\nabla\gamma_i\cdot\nabla}P_j),
\end{equation}
where $\Delta$ denotes the Laplacian, $\kappa$ and $\gamma$ are the
convergence and shear of the gravitational field, and $\delta$ and
$\epsilon$ the identity and the anti-symmetric tensors.

These two expressions already show the three major effects of
gravitational lensing on polarisation:
\begin{itemize}
\item A displacement shown by the term
$(1+\mathbf{\xi\cdot\nabla}(\Delta E/B))$. 
\item An amplification expressed by $-2\kappa(\Delta E/B)$ and
controlled by the convergence of the lensing.
\item A mixing term representing the coupling between the shear of
the gravitational lensing and its gradient, with the polarisation
vector $\mathbf{P}$.
\end{itemize}

From the previous set of equations we immediately note that if
gravitational waves are negligible as it is the case for scalar
density perturbations the equation for the $B$ modes is written as:
\[\Delta B=-2\epsilon^{ij}(\gamma_i\Delta P_j+\mathbf{\nabla}\gamma_i
\mathbf{\cdot\nabla}P_j).\] 
This means that the convolution of the
primary polarisation, of initially scalar type (from Thomson
scattering) with the shear of the gravitational lensing generates a
$B$ mode polarisation.

The observed or lensed power spectrum of the CMB polarisation can be
computed in the flat sky approximation
(e.g. Zaldarriaga \& Seljak 1997, 1998). It gives:
\[(C_\ell^E)_{\rm obs}=C_\ell^E[1-l^2\sigma]+\int\frac{d^2k}{(2\pi)^2}
\frac{(\mathbf{\ell\cdot k})^2-k^4}{2k^4}\bar{P}(k)\]
\[\times\left[\left(C_{|\mathbf{\ell-k}|}^E+C_{|\mathbf{\ell-k}|}^B\right)
+\cos(4\phi_{\mathbf{\ell-k}})\left(C_{|\mathbf{\ell-k}|}^E-
C_{|\mathbf{\ell-k}|}^B\right)\right]\]
and
\[(C_\ell^B)_{\rm obs}=C_\ell^B[1-l^2\sigma]+\int\frac{d^2k}{(2\pi)^2}
\frac{(\mathbf{\ell\cdot k})^2-k^4}{2k^4}\bar{P}(k)\]
\[\times\left[\left(C_{|\mathbf{\ell-k}|}^E+C_{|\mathbf{\ell-k}|}^B\right)
-\cos(4\phi_{\mathbf{\ell-k}})\left(C_{|\mathbf{\ell-k}|}^E-
C_{|\mathbf{\ell-k}|}^B\right)\right],\] 

where $\sigma=\int\frac{d^2k}{(2\pi)^2}\frac{(\mathbf{\ell\cdot
k})^2}{k^4} \bar{P}(k)$. In the case of no or negligible primary $B$
mode polarisation the first term in the $(C_\ell^B)_{\rm obs}$ is
neglected and we are left with the coupling term. A comparison of the
full sky approach (Hu 2000) and a flat sky computation shows that the
error introduced by the simplification are negligible. 

\begin{figure} 
\epsfxsize=11.cm
\epsfysize=10cm
\hspace{2.cm}
\epsffile{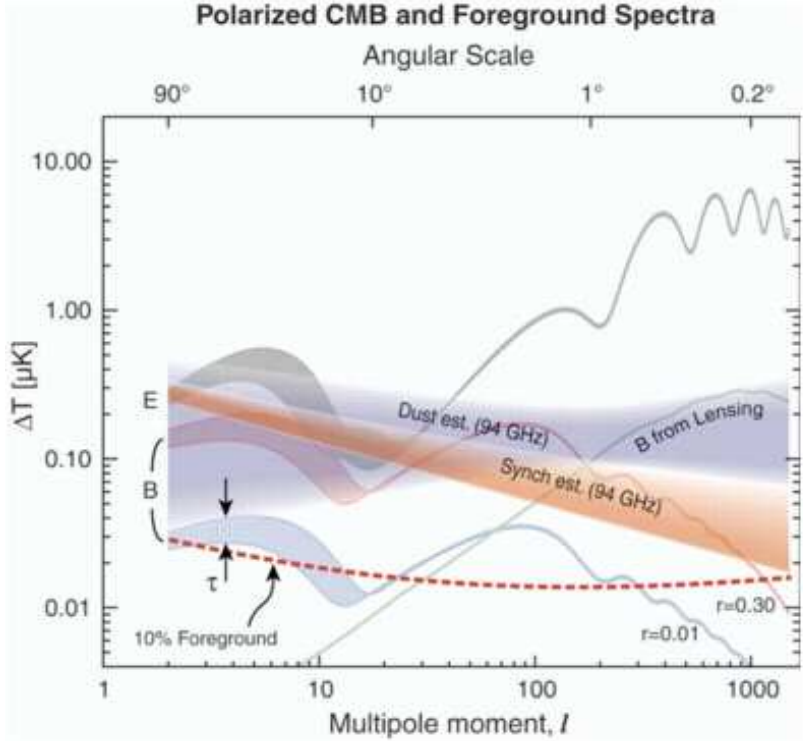}
\caption{From Bock et al.(2006): CMB polarisation power spectra of the
primary $E$-modes (grey line), the primary $B$-modes for two values of
the scalar to tensor ratio $r$ (light red and light blue lines) and
the lensing-induced $B$-modes (light green curve). Overplotted are the
current estimates of the polarised galactic foregrounds and their
uncertainty. The red dashed line is an estimate of the residual
foreground contamination using multi-frequency observations. }
\label{fig:polar-for}
\end{figure}

Weak lensing induced-$B$ mode polarisation, in addition to galactic
emission, is one of the major contamination for the future post-Planck
polarisation-devoted CMB experiments (see Figure \ref{fig:polar-for}),
whose main scientific goal will be to detect inflation-generated
gravitational waves. If the inflaton potential is such that $V\le
4.\times 10^{15}$ GeV, cleaning for lensing-induced polarisation is a
requirement.  However for larger potentials, deep integrations of
moderately large patches of the sky at low resolution should suffice
to account for the noise induced by lensing (this is the case for,
e.g., Planck, QUAD, BICEP, B-POL).  However, if the inflaton potential
is much smaller lensing-induced polarisation will be the dominant
foreground in the range $\ell\sim50$ to 100, once the galactic
contamination is removed (Figure ~\ref{fig:polar-for}). The lensing
signal can be separated from gravitational wave-$B$ modes using high
order statistics as is the case for the temperature anisotropies
(e.g. Hu \& Okamoto 2002, Kesden, Cooray \& Kamionkowski 2003,
Kaplinghat, Knox \& Song 2003). The separation between primordial $B$
modes and lensing-induced $B$ polarisation depends on the
reconstruction of the lensing signal. For the secondary polarisation
signal to be reduced by a factor 10 in power spectrum amplitude, a
full sky measure of temperature and polarisation with a resolution of
a few arc minutes and a noise of 1$\mu$K-arc minutes is needed.


\section{The Sunyaev-Zel'dovich effect} 
\label{sec_SZ}

The best known and most studied secondary contribution due to cosmic
structure is definitively the Sunyaev-Zel'dovich (SZ) effect
(Zel'dovich \& Sunyaev 1972, 1980; see also Rephaeli 1995, Birkinshaw
1999, Carlstrom, Holder \& Reese 2002). It is caused by the inverse
Compton interaction between the CMB photons and the free electrons of
a hot ionised gas along the line of sight. The SZ effect can be
broadly subdivided into: the thermal SZ (TSZ) effect where the photons
are scattered by the random motion of the thermal electrons and the
kinetic SZ (KSZ) effect which is due to the bulk motion of the
electrons. In the former case, the resultant CMB photons have a unique
spectral dependence, whereas the final spectrum remains Planckian in
the case of KSZ effect since it only Doppler shifts the incident
spectrum.

\subsection{The thermal SZ effect}
The TSZ effect describes comptonization, the process by which electron
scattering brings a photon gas to equilibrium.  The term
Comptonization is used if the electrons are in thermal equilibrium at
some temperature $T_{\rm e}$, and if both $k_{\rm B} T_{\rm e}<<m_{\rm
e} c^2$ and $h_{\rm pl} \nu << m_{\rm e} c^2$, where $\nu$ is the
frequency of the photon. This defines the non-relativistic nature of
the problem.  Comptonization becomes important when the temperature of
a low density electron gas becomes higher than the temperature of a
Planck function with the same energy density, and the absorption
optical depth is low enough that the photon spectrum falls below a
Planck function at the same electron temperature. This is typically
the situation in ionized regions (like the ICM, accretion flows around
compact objects, ionized bubbles around high redshift quasars, etc).
This makes the absorption process, proportional to the square of the
density, negligible compared to Compton scattering. Since low energy
photons are available, there can be energy transfer from the electrons
to the photons.  This is what is seen in the SZ effect observed for clusters
of galaxies and predicted for other astrophysical sources of hot
plasma.

In the non-relativistic limit, a differential Fokker-Planck equation,
the Kompaneets equation, can be written down to describe
the time evolution of the photon occupation number $n(\nu)$, which is
assumed to be isotropic.  Due to scattering, there is net energy
transfer from the electrons to the photons (or vice versa). However,
the total photon number is conserved.  
In the fully relativistic case, discussed in the next section, a
Fokker-Planck type equation cannot be given because the change in the
photon frequency $\Delta\nu$ due to scattering is not negligible
compared to its incident frequency $\nu$.

To derive the Kompaneets equation, one has to start the scattering
process between an electron and a photon (see
Sect. \ref{sec:reion-bas}).  This is given by the Boltzmann equation
(see Eq.  (\ref{eq:boltzman})), describing the evolution of the photon
occupation number. The electrons are assumed to have a Maxwellian
distribution with temperature $T_{\rm e}$.  Keeping in mind that
Compton scattering conserves the total number of photons, one ends up
with the Kompaneets equation given by

\be {{\partial n}\o{\partial y}} \; = \; {{1}\o{x^2}}
{{\partial}\o{\partial x}} \left[\, x^4 \, \left( \, {{\partial
n}\o{\partial x}} \, + \, n \, + \, n^2 \right) \right] .
\label{eq: kompaneets}	
\ee where $x = \frac{h_{\rm pl} \nu}{k_{\rm B} T_e}$ and one defines a
dimensionless scaled variable $y$ (called the ``Comptonization
parameter'' or ``Compton y-parameter'') given by

\be
y \; \equiv \; t \, c\, {{k_{\rm B} T_{\rm e}}\o{m_{\rm e} c^2}} \,
n_{\rm e} \, \sigma_{\rm T}  
\label{eq:y_var}
\ee

Note, that equation (\ref{eq:y_var}) can also be written as the
integral of the gas pressure, $p_{\rm e}$, along the line of sight through
the extent of the plasma, i.e.  \be y \;= \, \frac{\sigma_{\rm T}}{m_{\rm e}
c^2} \int p_{\rm e} dl \,.  \ee

The stationary solution of the Kompaneets equation is given by the
Bose-Einstein equilibrium distribution. Under the assumption that $
h_{\rm pl}\nu << k_{\rm B} T_{\rm e}$, we have $x << 1$, one can neglect
the $n$ and $n^2$ terms.

The resulting form of the
Kompaneets equation has a solution of the form

\be n(x,y) = \mbox{exp}\, \left[\,{{y}\o{x^2}}\,{{\partial}\o{\partial
x}} \,x^4\,{{\partial}\o{\partial x}} \right]\, n(x,0),
\label{eq:reduced_Kompaneets} 
\ee 

where $n(x,0)= {(e^{x} - 1)}^{-1}$, since in the absence of
distortions (\ie,~$y=0$), the photon spectrum is a black body.  If
$x^2\,y < 1$, then one can expand the exponential in equation
(\ref{eq:reduced_Kompaneets}) around $n(x,0)$ for small $y$. 

The final distortion can then be written as

\be
{{\Delta n}\o{n}} \, = \,
{{n(x, y) - n(x, 0)}\o{n(x, 0)}} \,= \, y \, {{xe^{x}}\o{(e^{x}-1)}} \,
\left[ x\coth(x/2) \, - \, 4 \right]~~~.
\ee

Since the change in radiation spectrum $\Delta I(x)$ at frequency $x$
is given by $\Delta I(x) = x^3 \Delta n(x) I_0$, where $I_0 =
{{2h_{\rm pl}}\over{c^2}}\,{\left({k_{\rm B} T_{\rm CMB}}\over{h_{\rm
pl}}\right)}^3$, we obtain the distinct spectral signature of the TSZ
effect:

\bea
\Delta I(x) &=& I_0\,y\,{{x^4\,e^x}\o{(e^x - 1)}^2} \left[ x\coth(x/2) \, - 
\, 4 \right] \;\;\;  .
\eea
This signature assumes an incident Planckian spectrum and is
valid in the single-scattering approximation. The temperature
anisotropy due to inverse Compton scattering of CMB photons is given by:

\be
{{\Delta T}\over{T}} \, = \, 
{{\Delta I(x)}\over{I(x)}} \, {{d\mbox{ln}\, I(x)}\over{d\mbox{ln}T}} 
{{d\mbox{ln}\, T}\over{d\mbox{ln}I(x)}}
 \;=\;  y \, \left[ x\coth(x/2) \, - \, 4 \right] \; . 
\label{eqn_tSZ}
\ee

\begin{figure}  
\epsfxsize=10.cm
\epsfysize=9.cm
\hspace{2cm}
\epsffile{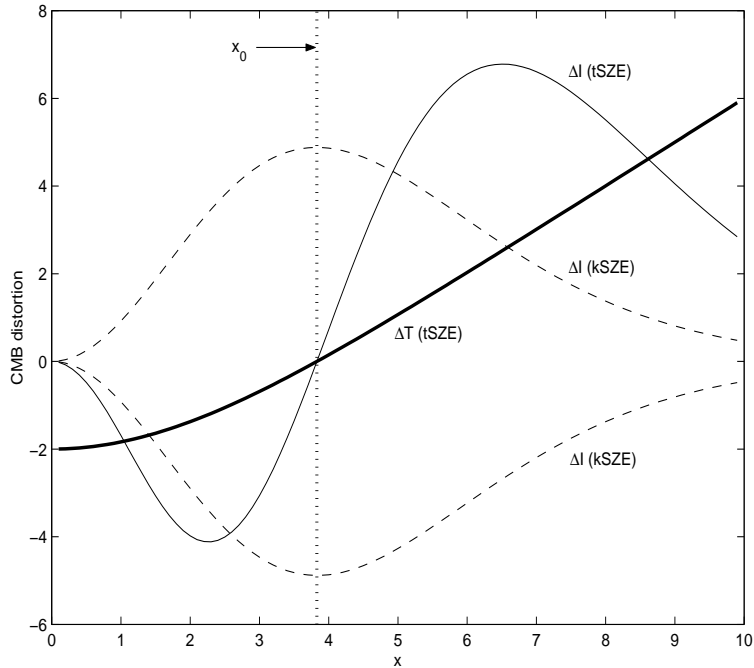}
\caption{Frequency dependence of TSZ and KSZ effects. The
thick line shows the frequency dependence of ${{\Delta T}/{\tcmb}}$
from TSZ effect, whereas the thin solid line shows the same for the
change in spectral intensity $\Delta I (x)$. The thin dashed lines
show the change in spectral intensity for KSZ effect, the upper one
for an approaching source and the lower one for a receding source. The
vertical dotted line shows the scaled frequency at which TSZ is zero
and KSZ effect is maximum. In here, $y,\,I_0$ and $\tcmb$ are all
scaled to unity.}
\label{fig_szspectra}
\end{figure}

In the non-relativistic limit, the frequency dependence of the
distortion, shown in figure (\ref{fig_szspectra}), is characterized
by three distinct frequencies : $x_{0} = 3.83$, where TSZ effect vanishes;
$x_{\rm min}=2.26$ which gives the minimum decrement of the CMB intensity
and $x_{\rm max}=6.51$ which gives the maximum distortion. In the
Rayleigh-Jeans (R-J) limit (\ie, when $x \rightarrow 0$) and in the
Wien region we have ${{\Delta \tcmb}\over{\tcmb}} \; = \; -\, 2\, y$
and $x^2\, y$ respectively. Thus at low frequencies we would see an
apparent {\it decrease} in the sky brightness of the CMB sky sometimes
referred to as ``holes in the sky'' (Birkinshaw \& Gull 1978).

\subsection{The kinetic SZ effect}
The KSZ effect occurs, along with TSZ effect, if the scattering plasma
has a bulk motion relative to the CMB.  In that case, the CMB photons
appear anisotropic in the reference frame of the scatterer and KSZ
effect tends to isotropise the radiation. This, however, makes the
radiation anisotropic in the reference frame of the observer, and
there is a distortion towards the scatterer with amplitude
proportional to the radial peculiar velocity $v_{\rm r}$ of the
scattering gas (Sunyaev \& Zel'dovich 1972, Rephaeli 1991).  To derive
the expression for the CMB temperature distortion due to KSZ effect,
one can either start with the Boltzmann equation (for example, see
Nozawa, Itoh \& Kohyama 1998) or use the radiative transfer equation
(see Birkinshaw 1999). In the limit of non-relativistic plasma moving
with $v_{\rm r} << c$, the change in the flux and temperature of the
CMB in the direction of an object giving rise to KSZ effect is given
by

\bea
{\Delta I(x) \over I(x) } \,&=&\, -\tau_{\rm clus} \, \frac{v_{\rm r}}{c} \, 
{x \, e^x \over e^x - 1} \;\; \\ \nonumber
\frac{\Delta T}{T} \,&=&\, -\frac{v_{\rm r}}{c} \, \tau_{\rm clus} \;,
\label{eqn_kSZ}
\eea
where $\tau_{\rm clus}$ is the optical depth of the intra-cluster
medium.  Unlike TSZ effect, the spectral distribution of the kinetic
SZE, figure
(\ref{fig_szspectra}), is Planckian making it impossible to
disentangle from CMB at overlapping angular scales.  The ratio of the
change in brightness temperatures caused by TSZ effect and KSZ is, at
a typical frequency for large galaxy cluster is

\bea {\Delta T_{\rm kinetic} \over \Delta
 T_{\rm thermal}} &=& {1 \over 2} \, {v_{\rm r} \over c} \, \left(
 {\boltz \te \over {m_{\rm e}} c^2} \right)^{-1} \nonumber \\ &\approx& 0.09
 \, (v_{\rm r} / 1000 \ {\rm km \, s^{-1}}) \, 
(\boltz \te /10 \ {\rm keV})^{-1} .\eea

Since typical peculiar velocities are around a few hundred kilometers
per second and typical temperatures a few keV, the kinetic effect
comes out to be at least an order of magnitude less than the thermal
effect. However, there can be cases when the kinetic distortion is
larger than the corresponding thermal distortion. This is so when the
plasma is either too tenuous or relatively cool or both and the
peculiar velocity is large. We discuss such scenarios in section
\ref{sec:szsource}.  For clusters of galaxies, which are the main
source of SZ distortion, it is very difficult to measure the KSZ
effect in the presence of TSZ effect. In CDM universes, $v_{\rm r}=300
- 400$ km s$^{-1}$, and hence the signal due to the presence of KSZ
effect is generally less than $5\,\%$.  However, the two effects
differ in their spectral shape and so can be, in principle,
separated. In fact, for non-relativistic cases, the kinetic effect
attains the maximum distortion at the frequency where the thermal
effect is zero (Figure \ref{fig_szspectra}).  However,
to be precise, one has to take the exact value of the cross-over
frequency which depends on both the plasma temperature, the optical
depth and the nature of the scattering medium which could include
non-thermal population (e.g. Aghanim et al. (2003)).

\subsection{Corrections to the SZ effect}

\begin{figure} 
\epsfxsize=7.cm
\hspace{1.5cm}
\epsffile{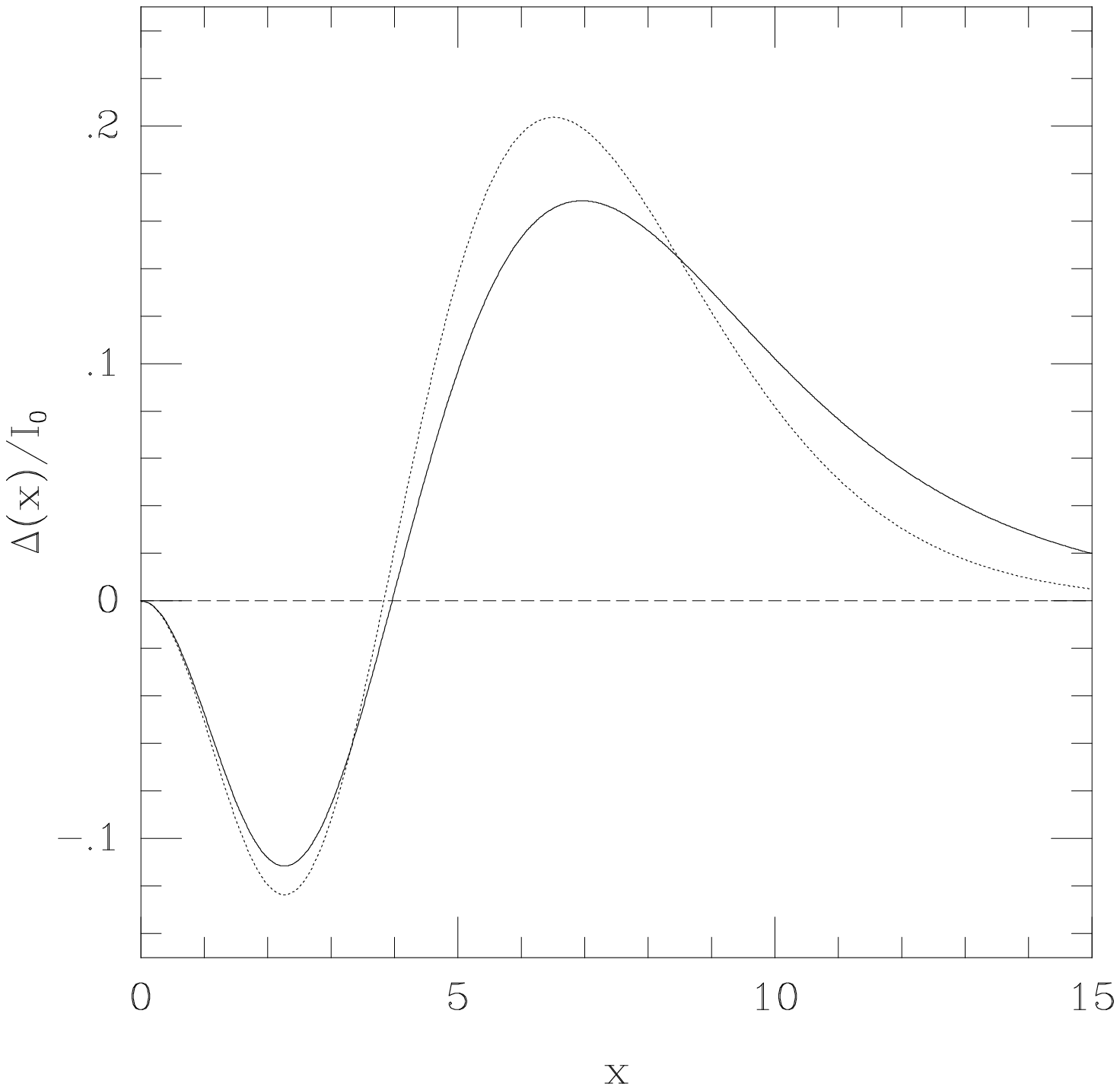}
\epsfxsize=7.cm
\epsfysize=6.5cm
\epsffile{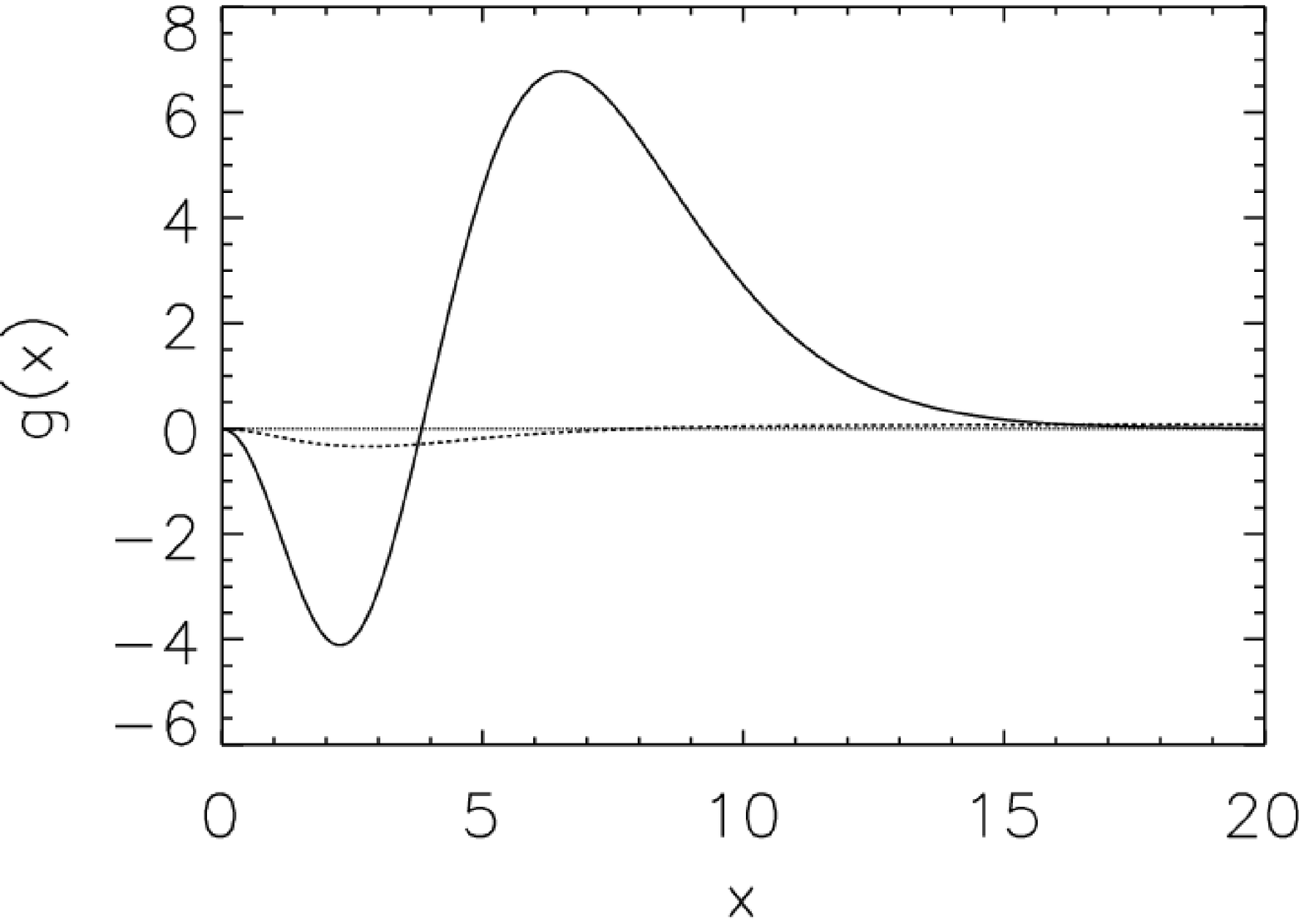}
\caption{Left Panel: From Birkinshaw (1999), $\Delta I/ I_0$ is
plotted against $x = h_{\rm pl}\nu/k_{\rm B} T$ for the relativistic
correction to SZ effect, from direct numerical integration shown in
solid line, and obtained from the corresponding Kompaneets kernel in
dotted line. The results are computed for a temperature $k_{\rm B}
T_{\rm e}\,=\,15\,$keV.  Right Panel: From Colafrancesco, Marchegiani
\& Palladino (2003): Comparison of frequency dependence $g(x)$ in the
thermal distribution (solid line) versus frequency dependence 
$g'(x)$ for a single power law non-thermal population of
electrons (dotted line).}
\label{fig:SZcorrection}
\end{figure}

When one deals with hotter and denser scattering media there are
corrections to the simple expressions of TSZ effect (equation
(\ref{eqn_tSZ})) and KSZ effect (equation(\ref{eqn_kSZ})) derived in
the previous sections, which become important. These issues have been
addressed in detail in many studies (Itoh, Kohyama \& Nozawa 1998,
Itoh et al. 2001, Nozawa, Itoh \& Kohyama 1998, Nozawa et al. 2000,
Challinor \& Lasenby 1998, 1999, Molnar \& Birkinshaw 1999,
Colafrancesco, Marchegiani \& Palladino 2003, Shimon \& Rephaeli
2004). The first step in most procedures, to calculate such corrections,
is to expand the Kompaneets equation in a power series in $\theta_{\rm
e} = k_{\rm B} T_{\rm e} / m_{\rm e} c^2$. This can be done for many
choices of parameters such as $p/m, v\equiv E/p$ etc. The convergence
of such expansions which are, in general, asymptotic expansions in
nature and converge slowly, is then an important issue.  For
scattering media having high temperatures, corrections up to 3-5
orders in $\theta_e$ are sufficient and match fully relativistic
numerical calculations well. The relativistic corrections modify the
frequency dependence of the SZE (see Figure \ref{fig:SZcorrection} left
panel).  There is an associated correction to the cross-over frequency
(Itoh, Kohyama \& Nozawa 1998) well approximated by a linear function
in $\theta_{\rm e}$ for $k_{\rm B}T_{\rm e} <20$ keV, and a quadratic
function in $\theta_{\rm e}$ up to 50 keV. The numerical fit is given
by:

\be
x_{0} \, = \, 3.830 \, \left( \, 1 + \, 1.1674 \theta_{\rm e} \, - \, 
0.8533 \theta_{\rm e}^{2} \, \right)  \, .
\label{eq:corr_to_tszcross}
\ee

The relativistic corrections to the KSZ effect can be obtained by
starting again from a generalized Kompaneets equation and applying a
Lorentz boost to the direction of the peculiar velocity. The electron
distribution functions are connected between the cluster frame and the
CMB frame by Lorentz transformations. One expands the Kompaneets
solution in powers of $\theta_{\rm e}$ including cross terms like
$\beta\theta_{\rm e}, \beta\theta_{\rm e}^2$ etc.  The
$\beta\theta_{\rm e}$ term can give rise to a correction of the order
of $10$ \% for a typical electron temperature of $10$ keV
(Figure ~\ref{fig:SZcorrection}). The other higher order terms
lead to negligible corrections for temperatures of interest. The
relativistic correction in the R-J limit is written:

\be \frac{\Delta n(X)}{n_{0}(X)} \rightarrow - 2 y \, \theta_{e} \,
\left[ \, 1 - \frac{17}{10} \theta_{e} +
\frac{123}{40}\theta_{e}^{2}\, \right] \, + \, \, y \, \beta \, \left[
\, 1 - \frac{2}{5} \theta_{e} + \frac{13}{5} \theta_{e}^{2} \, \right]
\,\, . \label{eq:corr_to_ksz1}, \ee 
where we considered cluster moving
along the the line of sight such that $\beta = v_{\rm r}/c $, and
neglected all $\beta^2$ and higher order terms.  Similarly, there is a
correction to the cross-over frequency which is very small.

Finally, one can relax the assumption of low optical depth and look at
multiple scatterings of the incident photon spectrum. In general, the
multiple scattering contribution is found out to be rather small
compared to single scatterings. As an example, for a 15 keV cluster,
the multiple scattering affects the final result by $-0.3$ \% in the
Wien region and $-0.03$ \% in the R-J region.  One can think of other
effects that will add further corrections to the SZ distortion. The
presence of magnetic fields would give rise to magnetic pressure which
will add to the gas pressure in determining the hydrostatic
equilibrium of the gas. Simple calculations that incorporates this
effect show a net decrease in the SZ effect distortion (Koch, Jetzer
\& Puy 2003, Zhang 2004). It has been proposed that the presence of
magnetic fields would lead to an anisotropic velocity distribution
such that one ends up with a two-temperature relativistic Maxwellian
distribution of the thermal electrons. This can lead to a net
enhancement of the SZ effect. Finally, it has been shown that the
presence of a temperature gradient in the cluster temperature would
lead to corrections to the electron momentum distribution thereby
leading to corrections of the TSZ spectrum (Hattori \& Okabe
2004). Unfortunately, the expected amplitude of the corrections is
almost two orders of magnitude smaller than that of TSZ effect.
Presence of a significant amount of non-thermal population of
electrons can also lead to deviations from the thermal SZ spectrum.  A
self consistent treatment of several corrections to the thermal SZ
effect in the presence of both thermal and non-thermal populations of
electrons is given in Colafrancesco, Marchegiani \& Palladino (2003)
(see Figure \ref{fig:SZcorrection} right panel).

\subsection{SZ observations}

\begin{figure}  
\epsfxsize=12.cm
\epsfysize=12.cm
\hspace{2cm}
\epsffile{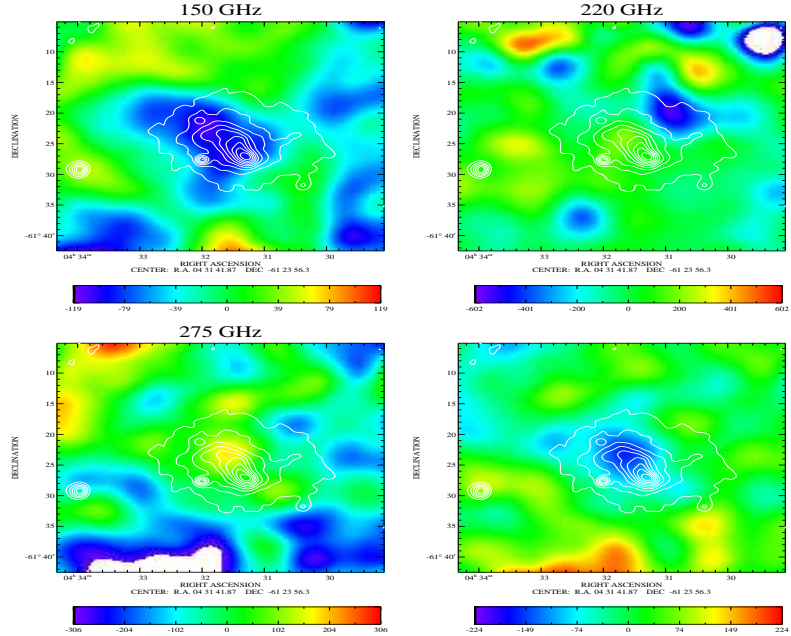}
\vspace{-1cm}
\caption[]{From Gomez et al. (2003): Mosaic of the 150 GHz, 220 GHz, 275
GHz, and CMB spectrally subtracted colorscale images of Abell 3266
(convolved with a Gaussian with FWHM $\sim 4.5$ arc minutes) overlaid onto
the ROSAT contours. The rms noise level of $\sim 25 \mu$K/beam. Most
of the CMB present in the 150 GHz, 220 GHz, and 275 GHz channels has
been minimized in the CMB subtracted map. As expected, the SZ
signal at 220 GHz is minimum.  }
\label{fig:sz_im}
\end{figure}

The first observations of SZ effect were targetted in nature and
looked at specific X-ray selected clusters. The SZ flux from these
observations were used along with X-ray modeling of the clusters to
estimate, in particular, the value of the Hubble Constant
(Sect. \ref{sec:h0}). The major instruments responsible for such
measurements were OVRO 5 meter telescope at 32 GHz, IRAM 30 meter
telescope at 140 GHz, the Nobeyama 45 metre telescope at 21 GHZ, 43
GHz and, 150 GHz, the SuZIE array at 140 GHz and the BOLOCAM 151
element array. Additionally, interferometers like the BIMA array at 30
GHz, the Ryle Telescope at 15 GHz, CBI working between 25--36 GHz,
ACBAR and AMI at 15 GHz have also been used (see Carlstrom, Holder \& Reese (2002), 
Birkinshaw \& Lancaster (2004) for a recent review).  In figure \ref{fig:sz_im}, the
SZ image of the cluster A3266 taken by ACBAR, with a beam of $\sim
4.5$ arc minutes, is shown for three frequencies 150 GHz, 220 and 275 GHz
including the crossover frequency of 217 GHz. Note that for 150 GHZ,
the cluster SZ effect appears as a decrement while for 275 GHz it is
an increment.  Note also that the combination of the three frequencies
permits to subtract the CMB contamination which remains important at
the beam scale.

\subsection{Hubble Constant from SZ effect} \label{sec:h0}

After the first TSZ observations of clusters started in the seventies, 
it was pointed out by Silk \& White (1978) that the
distance to a cluster can be estimated from the SZ and X-ray cluster
observations. If we put in plausible values for the matter and energy
budget of the universe, then one can estimate the value of
$H_\circ$ from this distance. This has been attempted or performed using SZ effect
observations: Single dish at radio wavelengths (Birkinshaw \& Hughes
1994, Hughes \& Birkinshaw 1998) millimeter wavelengths (Holzapfel et
al 1997, Pointecouteau et al.1999), submillimeter wavelengths (Komatsu
et al.1999) and also using interferometers (Jones at al.1993, Grego et
al 2001, Reese et al. 2002, Bonamente et al. 2006).

The gist of the method can be understood simply: the SZ temperature
decrement $\Delta T/T$ and the X-Ray surface brightness $S_X(r)$
depend on the cluster gas structure differently; $\Delta T/T \propto
n_{\rm e} T_{\rm e} L_{\rm cluster}$ and $S_X(r) \propto n_{\rm e}^2
T_{\rm e}^{1/2} L_{\rm cluster}$, where $n_{\rm e}, T_{\rm e}$ and
$L_{\rm cluster}$ are the characteristic density, temperature and
extent of the cluster gas. Eliminating the gas density, one can obtain
the cluster size in terms of the SZ and X-ray observables and the gas
temperature. Once the angular size of the cluster $\theta_{\rm
cluster}$ is measured, we are able to obtain the cosmologically
sensitive angular diameter distance $d_{\rm A} = L_{\rm
cluster}/\theta_{\rm cluster}$. For nearby ($z \ll 1$) clusters,
$d_{\rm A}$ can be approximated in terms of the deceleration parameter
$q_0 = {{\Omega_{0}}/{2}}\,-\,\Omega_\Lambda$ as \be d_{\rm A} =
{{c}\over{H_{0}\,(\,1\,+\,z\,)}}\left
[\,z\,-\,{{1\,+\,q_0}\o{2}}\,z^2\,\right] .
\label{eq:da_h0}
\ee The derived value of the Hubble constant depends on the other
cosmological parameters, $\Omega_0$ and $\Omega_\Lambda$. As long
as the redshift is less that 0.2, $d_{\rm A}$ does not change
significantly with small variation in the presently acceptable values
of the the cosmological parameters. For example, changing $q_0$ from
$0$ to $0.5$ for clusters A665 or A2218 (having $z \,\sim\, 0.17 -
0.18$) leads to a change in \ho by $\sim\, 3\,\%$. For a high redshift
cluster, the changes in \ho due to different cosmology can be higher
by $\sim 5 - 10 \,\%$ (Kobayashi, Sasaki \& Suto 1996, Reese etal 2002). More
generally, the combination of SZ effect and X-ray observations can be
used to probe dark energy. This is done especially when searching for
violations of the duality relation between the angular diameter
distance and the luminosity distance. The test of the reciprocity
relation (between the source angular distance and the observer area
distance) and the distance duality relation that derives from it was
proposed as an additional test of dark energy (Bassett \& Kunz 2004).
While the reciprocity relation holds when photons follow null geodesic
and that the geodesics deviation equation is valid, the distance
duality relation will hold if the reciprocity relation is valid and
the number of photons is conserved. Violations can thus occur if the
number of photons is not conserved (e.g. in the case of absorption by
dust) or if gravity is not described by a metric theory, i.e. photons
do not follow null geodesic. Uzan, Aghanim \& Mellier (2004) tested
for the duality relation. Using a data set of SZ effect
and X-ray clusters, they found no significant departure from the
reciprocity.

The procedure described here both for $H_0$ determinations or for
distance duality tests, in general, interprets the SZ effect and X-ray
observations with simple modeling of the cluster gas as spherical,
unclumped and isothermal distribution such as the $\beta$-model
(Cavaliere \& Fusco-Femiano 1978).  All clusters, however, show
departures from these simplistic assumptions (Evrard 1990, Navarro,
Frenk \& White 1997 ,Makino \& Sasaki 1998, Mohr et al.1995, Mohr,
Mathiesen \& Evrard 1999, Mathiesen, Evrard \& Mohr 1999). As a
result, sources of error in the determination of \ho as well as
limitations to the distance duality test are observational in nature
and come from the uncertainty in the cluster parameters such as its
core radius and temperature; the intracluster parameters such as the
central electron density $n_{{\rm e},0}$ and the central values of the
SZ effect and X-ray measurement. Errors can also be due to
contamination of SZ effect from point sources or a poor knowledge of
their spectra (e.g. Holder 2002, Aghanim, Hansen \& Lagache 2005).
There can also be systematic errors due to overall flux and brightness
temperature calibration uncertainties and from improper subtraction of
a zero level offset to the SZ data.

Without proper accounting for the many systematics present in the
observations, the estimates made from this method are biased
(Birkinshaw, Hughes \& Arnaud 1991, Inagaki, Suginohara \& Suto
1995,Majumdar \& Nath 2000, Puy et al. 2002, Reese et al. 2002). They
seem, in particular, to favour a low value of $H_0$ compared to
other methods. However, with a more careful treatment of the
systematics the discrepancies appear to be much less (Reese et al
2002, Ameglio et al. 2006). Combining recent Chandra X-ray data with
BIMA/OVRO SZ effect data on a large sample of clusters at $0.14 < z <
0.89$ yields $H_0=75\pm 10 $ km/s/Mpc for the standard LCDM model
(Bonamente et al. 2006). This result holds whether or not virial
equilibrium is assumed, and is consistent with the HST determination
of $H_0.$

\section{SZ cluster counts} 
\label{sec:SZcounts}

Counts of galaxy clusters, detected through their SZ
effect, can be used as major probe of cosmological as well as cluster
properties. The frequency dependence of the SZ effect can be used to
extract the clusters from a radio survey of the sky, making SZ cluster
catalogs possible. Once the clusters are detected, follow-up redshift
measurements can be carried out to get the cluster redshift
counts. The abundance of the clusters $N_{\rm tot}$, their redshift
distribution $dN/dz$, as well as their clustering $\xi(r)$, are governed 
by the geometry of the Universe and
the power spectrum of the initial density perturbations. Gas physics
related to cluster structure and evolution also enters through mapping
of the cluster SZ flux relative to the true mass of the cluster.

\subsection{Cluster mass and cluster mass--function}
\label{subsec:massfn}
The fundamental quantity that goes into calculating the observed
cluster counts is the cluster mass function, $dn/dM$, which
predicts the multiplicity function of clusters having mass in the
range $[M,M+\Delta M]$ at a given redshift for a choice of
cosmology. One starts by calculating the variance of the linear
density field, extrapolated to the redshift $z$ at which halos
(i.e. clusters) are identified, after smoothing the mean density field
with a spherical top-hat filter so as to enclose the mass
$M$. This variance can be expressed in terms of the power spectrum
$P(k)$ of the linear density field extrapolated to redshift zero as:
\begin{equation}
\sigma^2(M,z)  =  {D^2(z)\over2\pi^2}\int_0^\infty k^2P(k)W^2(k,M){\rm d}k,
\end{equation} 
where $D(z)$ is the growth factor of linear perturbations normalised to unity
 at $z=0$, and $W(k,M)$ is the
Fourier transform of a real-space top-hat filter 
\be \label{eq:tophat}
W(k)=\frac{3}{(kR_{\rm h})^3}[\sin(kR_{\rm h})-(kR_{\rm
 h})\cos(kR_{\rm h}).
\ee
In equation \ref{eq:tophat}, the mass $M$ is enclosed within a
 comoving radius
$R_{\rm h}$. An important cosmological parameter, related to the amplitude
of fluctuations, is the mass variance at $R_{\rm h} = 8h^{-1}$Mpc
denoted by $\sigma_8$.

One can define the mass function for a particular cosmological model
in terms of the quantity $\ln\sigma^{-1}(M,z)$ instead of $M$ as given
by Jenkins et al.(2001) : 
\be \label{eq:massfn} f(\sigma, z) \equiv
\frac{M}{\bar{\rho}}{{\rm d}n(M, z)\over{\rm d}\ln\sigma^{-1}}, 
\ee
where $n(M, z)$ is the abundance of halos with mass less than $M$ at
redshift $z$, and $\bar{\rho}(z)$ is the mean density of the universe
at that time. This implies that the mass function depends only on
$\sigma(M,z)$ which in turn depends on the background
cosmology. Further, the mass function is normalized to have
$\int_{-\infty}^\infty f(\sigma){\rm d}\ln\sigma^{-1} = 1$. In the
following paragraphs we list the four most commonly used mass
functions.

The first mass function was based on theoretical considerations (Press
\& Schecter 1974).  The number density of clusters is derived by
applying the statistics of peaks in a Gaussian random field (Bond et
al 1991, Lacey \& Cole 1993, Sheth, Mo \& Tormen 2001) to the initial
density perturbations.  It assumed that the fraction of matter
residing in objects of a mass $M$ can be traced to a portion of the
initial density lying at an overdensity over a critical threshold
value, $\delta_c$. This mass function is given by 
\be
\label{eq:PSmassfn} f(\sigma)_{\rm PS} = \sqrt{2\over\pi}
{\delta_c\over\sigma}\exp\bigg[-{\delta_c^2\over2\sigma^2}\bigg], \ee
where $\delta_c$ is the extrapolated linear overdensity of a spherical
perturbation at the time of collapses and is a weak function of
$\Omega_{\rm m}$ and $\Omega_\Lambda$ (Eke, Cole \& Frenk
1996). Notice, that the abundance of objects is exponentially
sensitive to their masses at a particular redshift. Recently, other
mass functions, mainly from fits to dark matter simulations have been
proposed in the literature. For example, the mass function of Sheth
and Tormen (1999) can be written as \be
\label{eq:STmassfn} f(\sigma)_{\rm ST} = A\sqrt{{2a\over\pi}}
\bigg[1+\big({\sigma^2\over a\delta_c^2}\big)^p\bigg]
{\delta_c\over\sigma}\exp\bigg[-{a\delta_c^2\over2\sigma^2}\bigg], \ee
with A=0.3222, $a=0.707$ and $p=0.3$ and the masses were
estimated with a spherical overdensity algorithm, by computing
the mass within the radius encompassing a mean overdensity equal to
the virial one. Jenkins et al. (2001), using a much larger
simulation and a friend-of-friend algorithm for cluster finding,
proposed 
\be f(\sigma)_{\rm Jenkins}=0.315\exp(-|\ln
\sigma^{-1}+0.61|^{3.8})\,.
\label{eq:Jenkinsmassfn}
\ee
This fit, which is widely used, has a fractional
accuracy better than 20\% for $-1.2\le\ln\sigma^{-1}\le1$. Recently,
Warren et al.(2006) have come up with an improved fit given
by 
\be \label{eq:Warrenmassfn} f(\sigma)_{\rm Warren}= A (\sigma^{-a} +b)
\exp{(-c/\sigma^2)} \ee with $A=0.7234, a=1.625, b=0.2538$ and
$c=1.1982$.

In spite of the progress in obtaining the mass function from dark
matter simulations, there have been considerable differences between
different simulation fits, especially at high redshifts.  Precision
cosmology with clusters may, ultimately, be limited by our
understanding of these differences. The first step in this direction
has already been taken recently by Lukic et al.(2007).

\subsection{Cluster abundance and redshift distribution}
\label{subsec:dndz}

\begin{figure}[!ht]  
\epsfxsize=11.0cm
\epsfysize=11.0cm
\hspace{2.2cm}
\vspace{-0.8cm}
\epsfbox{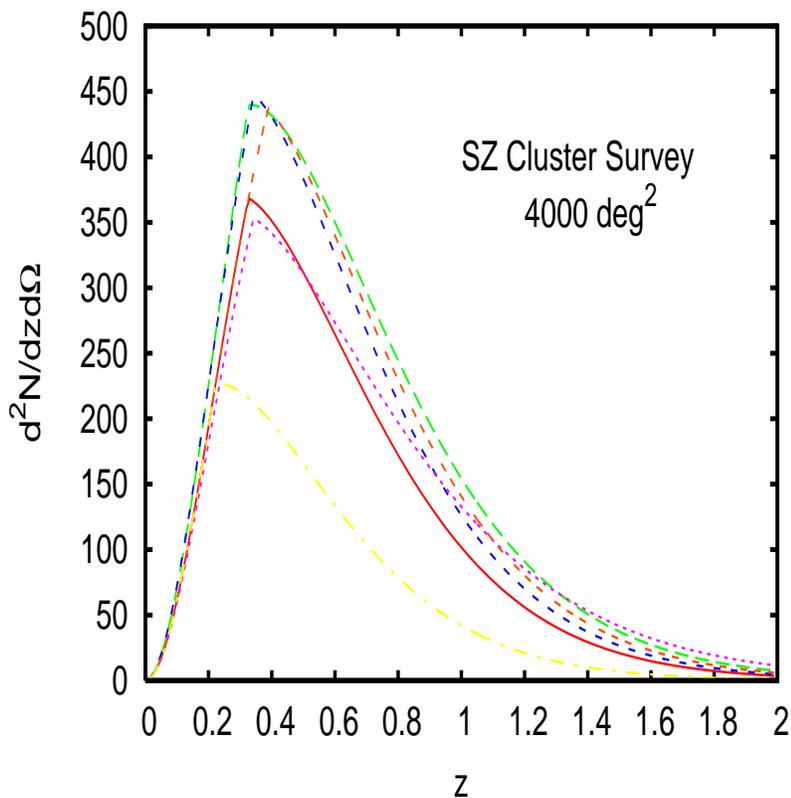}
\vspace{0.5cm}
\caption{Cluster redshift counts for a 4000 deg$^2$ SZ survey having a
  flux limit of 5 mJy. The solid red line is for fiducial $\Lambda$CDM
  model with WMAP-3yrs cosmological parameters. The green long-dashed
  line is for 10\% increase in the value of $\sigma_8$ and blue
  short-dashed line is for similar increase in the value of
  $\Omega_{\rm m}$. The purple dotted line is for a dark energy model
  with the equation state given by $w(a) = -0.8 + 0.3a$. The brown
  triple-dashed line shows the effect due 20\% change in the
  normalisation SZ flux to cluster mass. Finally, the yellow
  dot-dashed line shows the expected number counts for a flux limit of
  8 mJy. }
\label{fig:dndz}
\end{figure}

Once the halo mass function and its evolution are quantified, the
cluster redshift distribution, $d^2N/dzd\Omega$ (i.e the number of
clusters per unit redshift per unit solid angle), can be estimated by
multiplying the number density or abundance of clusters $n(z)$ with
the volume $d^2V/dzd\Omega$ surveyed. This volume depends on the
angular diameter distance $d_{\rm A}(z)$ and the Hubble parameter $H(z)$ at
that redshift. The abundance can then be easily computed by integrating the
mass function of over the limiting mass of a survey which depends on
the selection function $f_{\rm survey}(M,z)$ of the survey, 
\be
{d^2N\over dz\,d\Omega}={c\over H(z)} d^2_{\rm A}(z)\left(1+z\right)^2 \,\,
\int_0^\infty f_{\rm survey}(M,z) \, {dn\over dM}(M,z)\,dM\,, \ee 
where
$dn\over dM$ is calculated using equation \ref{eq:massfn}.  The
cosmological information contained in the observed cluster counts
comes through its dependence on the expansion history of the Universe
and on the growth rate of structures (Haiman, Mohr \& Holder 2001).

Once the mass function is written in the universal form (Eq.~\ref{eq:massfn}), its evolution is
completely governed by the growth factor $D(z)$. The difference in the cluster counts with
varying energy density of different components are explained as follows: small-amplitude density
perturbations grow as $D(z) = (1+z)^{-1}$ when $\Omega_{\rm m} (z) \approx 1$, but perturbation
growth stalls at around $z\sim \frac{1}{\Omega_{\rm m}} -1$ when  $\Omega_{\rm m} (z) \ll 1$. For
a fixed $\Omega_{\rm m}$ at $z=0$, its behaviour at a higher redshift depends on the Hubble
expansion factor $H(z)$ which inturn depends on the different energy densities including
parameters for the amount of dark energy, $\Omega_\Lambda$, and its equation of state $w$. Dark
energy starts to dominate the universe at a later time for  large $\Omega_\Lambda$ and a more
negative value of $w$. The different growth histories are manifest most strongly in high-mass
clusters where the exponential dependence of the mass functon on $\sigma(M,z) = D(z) \sigma(M,0)$
has a dramatic effect on the  abundance of clusters.
These cosmological sensitivities of cluster redshift
distribution (see Figure \ref{fig:dndz}) has led clusters to be
considered as probes of precision cosmology.

\subsection{Precision cosmology with cluster counts}
\label{subsec:cosmo_with_counts}

Galaxy cluster surveys of the nearby universe (Abell (1958)) have been
done for many decades. However, with the beginning of SZ surveys,
ambitious plans to do detect clusters at faraway universe has started
to take place. In recent past, a 12~deg$^{2}$ interferometric SZ
survey of the high redshift universe (Holder et al. 2000) has been
carried out. Future surveys covering many hundreds to thousands of
degrees capable of detecting tens of thousands of clusters are already
being attempted. The goal of all these surveys is to use the
sensitivity of the cluster redshift distribution to the cosmological
parameter as cosmological tools. Especially, it has been demonstrated
by many that a suitably large cluster survey can be used as a strong
discriminator of dark energy models (Haiman, Mohr \& Holder 2001,
Weller, Battye \& Kneissl 2002, Levine, Schulz \& White 2002, Majumdar
\& Mohr 2003, 2004). All of these authors forecast few percent level
constraints on cosmological parameters, including those of dark
energy.

At this point, let us stress the fact that the exponential sensitivity
of the cluster mass function to the cluster mass (equations
\ref{eq:PSmassfn}, \ref{eq:STmassfn}, \ref{eq:Jenkinsmassfn},
\ref{eq:Warrenmassfn}) is both the boon and bane for cosmological
studies with clusters. Any systematic error in the estimation of
cluster mass, including conversion between one definition of mass to
the other, are exponentially magnified by the steep slope of the mass
function. Numerous techniques have been proposed in the last few years
to tackle this complication, as described below.

\begin{figure} 
\epsfxsize=8.cm
\epsffile{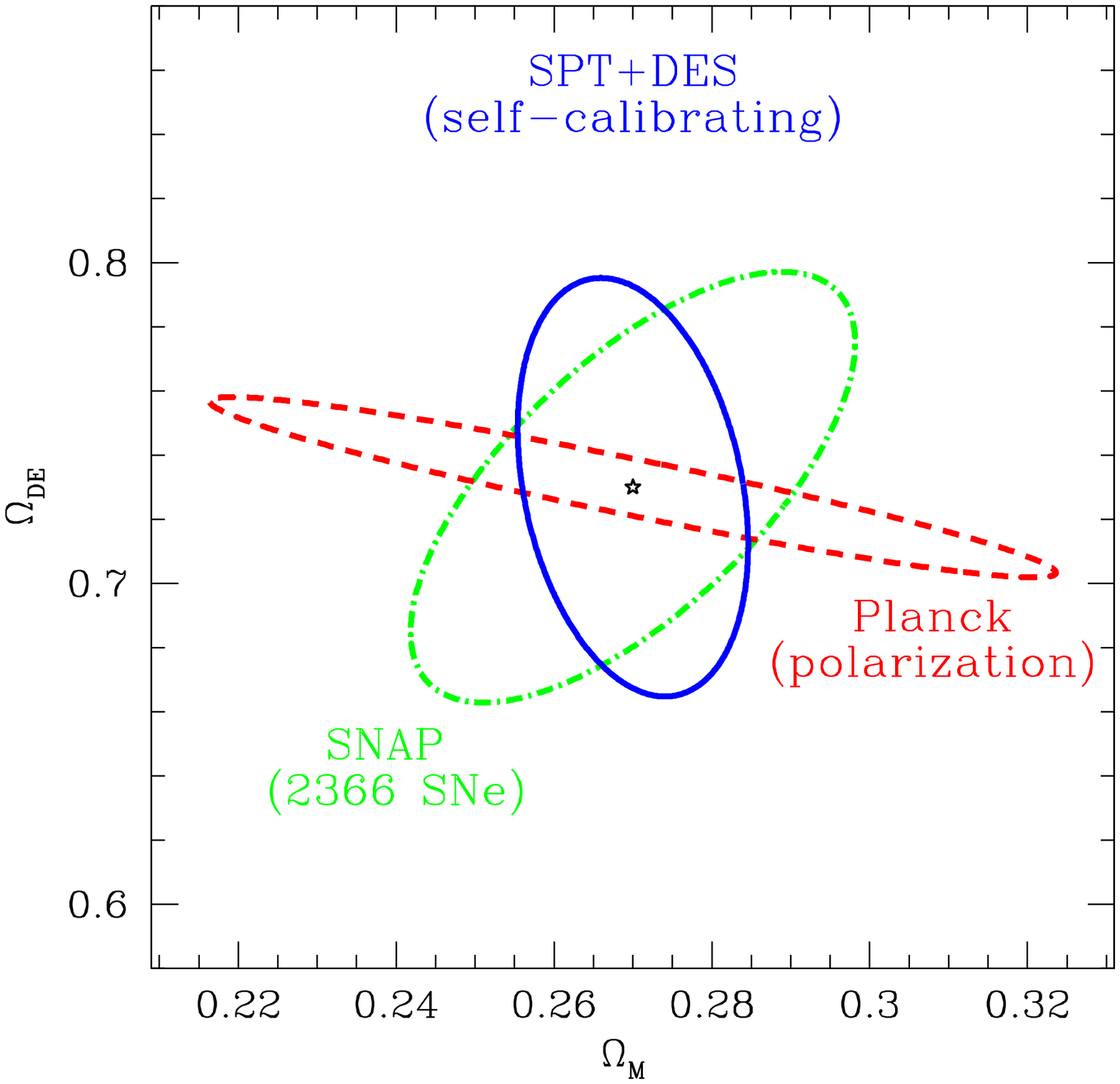}
\epsfxsize=8.cm
\epsffile{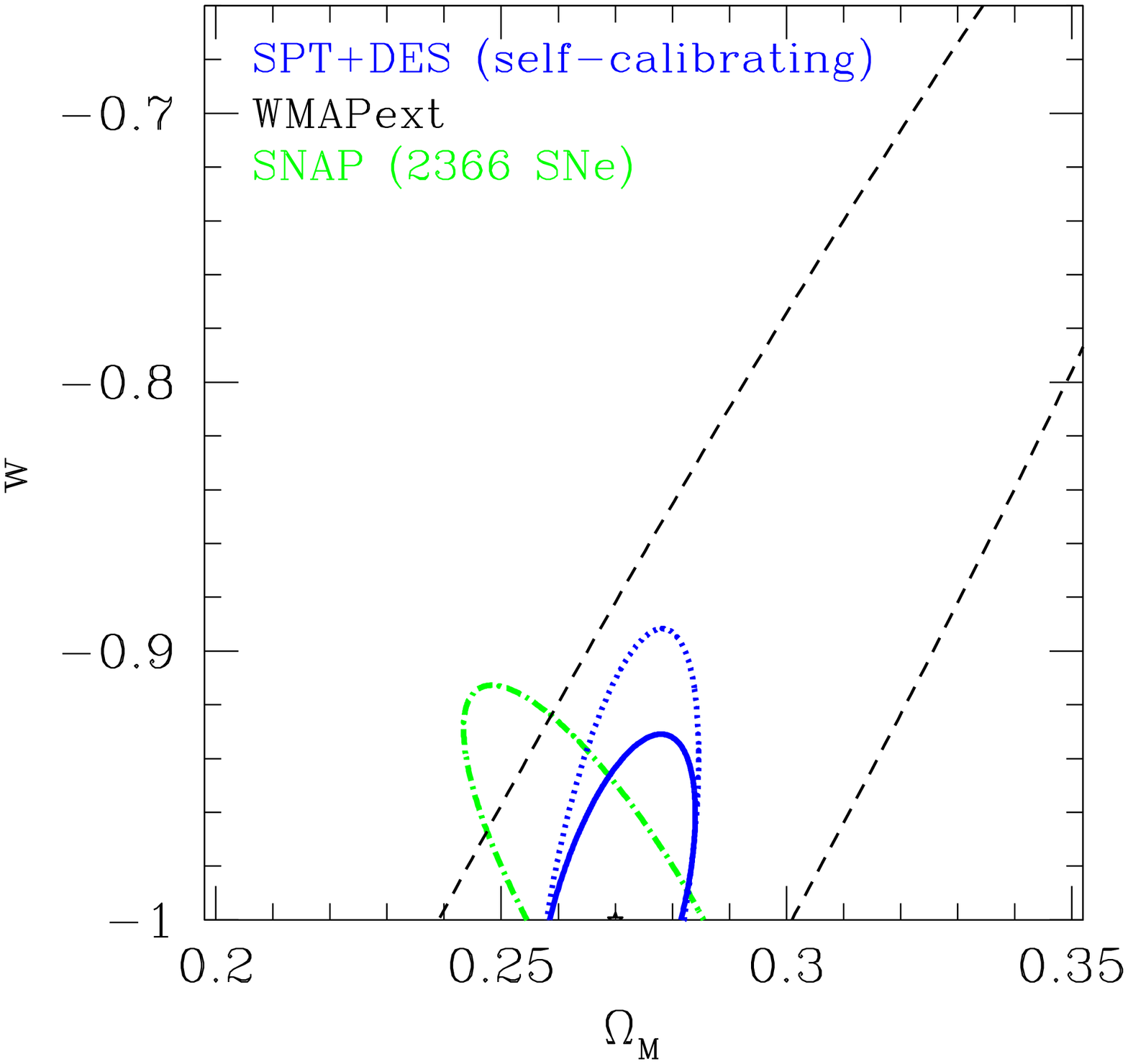}
\caption{Left Panel: Forecasts for $\Omega_{\rm m} -
\Omega_\Lambda$ constraints from the SPT galaxy cluster survey, the
SNAP SNe Ia mission, and the {\it Planck} CMB anisotropy
mission. Right Panel: Same, but for joint $\Omega_{\rm m}-w$
constraints from the WMAP CMB anisotropy mission, SNAP and SPT. Figure
courtesy J. Mohr.}
\label{fig:SPT_cosmo}
\end{figure}

In spite of the cosmological usefulness of cluster number counts,
there are several theoretical and observational requirements needed to
achieve precise cosmological constraints. These include advances in
understanding the formation and evolution of cluster size halos (in
practice, a tighter fitting for the mass function), a well
understood/controlled cluster selection function as well as a robust
observational proxy for the cluster mass. Additionally, one would need
a follow-up program to estimate the redshifts of these
clusters\footnote{Only photometric redshifts are plausible for a
sample of tens of thousands of clusters}. Finally, one needs to
calibrate and control any instrumental systematics. For a SZ survey
detecting 10,000-20,000 Poisson distributed clusters in roughly 10
redshift bins with equal weights, the statistical uncertainty is $\sim
10\%-7\%$. This gives a ballpark number at which systematic
uncertainties need to be controlled.

The first requirement depends on our ability to perform large
simulations. This becomes more feasible with the increase in
computational power. The second requirement translates into
understanding cluster selection function (the limiting mass and the
completeness level of the survey) from, as realistic as possible, mock
cluster catalogs.  For telescope beam larger than the cluster, a
survey is limited by SZ flux. Moreover, since for SZ fluxes, the
redshift dependence enters through the angular diameter distance
rather than the luminosity distance, the mass selection function is
more uniform than that of X-Ray surveys, except at nearby redshifts
($z<0.2$) where the clusters will be partially resolved. The
mass-selection function is directly linked with the cluster SZ-flux
measurements through cluster scaling relations (Kaiser 1982, Borgani
2006).  Finally, it has been pointed out (da Silva et al. 2001, Motl
et al. 2005, Pfrommer et al. 2006) that the SZ-flux of a galaxy cluster
is a good proxy for its virial mass with a tight scatter in the
scaling relation. The relation between the virial mass $M$ and the
SZ-flux $f_{\rm SZ}$ can be written as 
\be f_{\rm SZ}(z,\nu) d_{\rm
A}^2\,=\, f(\nu)\, f_{\rm gas}\, A_{\rm SZ} M_{\rm vir}^{\beta_{\rm
SZ}}\, E(z)^{2/3}\, {\cal F}(\gamma, z)\, , \ee 
where $H(z)=H_0\,E(z)^2$, $f(\nu)$ is the
SZ frequency dependence, $f_{\rm gas}$ is the gas fraction of the
cluster out to the virial radius, $A_{\rm SZ}$ and $\beta_{\rm SZ}$
are the amplitude and slope of the scaling relation and ${\cal
F}(\gamma, z)$ denotes any deviation from the standard evolution. For
simplicity, we usually set ${\cal F}(\gamma, z)=(1+z)^\gamma$. The
complexity of cluster structure is then encoded in the three
parameters of the SZ flux--mass relation $A_{\rm SZ}, \beta_{\rm SZ}$
and $\gamma$. Uncertainties in the mass--observable relation can in
principle be reduced by the use of ``self-calibration'' techniques
(Majumdar \& Mohr 2004)\footnote{For application to actual data, see
Gladders et al.(2007).}  where one uses additional observables such as
the power spectrum of galaxy clusters (Majumdar \& Mohr 2004,Lima \&
Hu 2004). Moreover, the distribution of clusters in observed flux at
each redshift provides additional mass information (Hu 2003). Finally,
direct mass measurements-- through X-ray observations, optical
spectroscopy or weak lensing-- provide important additional leverage
on cluster masses and hence on cosmology (Majumdar \& Mohr 2003,
2004). Once parameter degeneracies are broken through use of multiple
cluster information, it is possible to achieve strong cosmological
constraints from upcoming cluster surveys. As as example, in figure
\ref{fig:SPT_cosmo}, we show forecasts for $\Omega_{\rm m},
\Omega_\Lambda$ and $w$ constraint from South Pole Telescope (SPT)
survey\footnote{The Dark Energy Survey (DES) would be used to
follow-up SPT clusters to get their redshifts}.  It also shows that SZ
clusters as probes are highly complementary to other experiments since
each experiment constrains a different combination of cosmological
parameters and is subject to different systematics.

However, due to the very nature of the entanglement between gas physics
and cosmology in using clusters as cosmological probes, there must be
cosmology-cluster physics degeneracies (Majumdar \& Mohr 2003,
Majumdar \& Cox 2007). These degeneracies can be broken by adding constraints
from complementary information within cluster surveys (such as $dN/dz
+ P_{\rm cluster}(k)$) or external information (e.g. mass
follow-up). 

\section{The SZ power spectrum} 
\label{sec:SZpowspec}

The SZ effect from galaxy clusters is one of the major sources of
secondary temperature anisotropies. A convenient way of describing its
effect on the CMB is by computing its angular power spectrum.  The
fluctuations in the temperature background due to SZ effect from
clusters of galaxies can be expressed in terms of correlations between
the fluctuations along two lines of sight separated by an angle. The
rms distortion can be quantified by the spherical harmonic
coefficients $a_{\ell m}$, which is defined as $\Delta T({\bf n}) =
T_0^{-1} \sum_{\ell m} a_{\ell m} Y_{\ell m}({\bf n})$. The angular
power spectrum of the SZ effect is then given by $C_\ell=<{|a_{\ell
m}|}^2>$, the brackets denoting an ensemble average.

\subsection{Modeling the SZ power spectrum} 
\label{subsec:SZ_cl}

The SZ power spectrum can be derived from numerical simulation of
structure formation and evolution or from analytical computations.  In
the first approach, hydrodynamical numerical simulations are the most
appropriate way to describe both the dark matter of which the
gravitational potential wells are made of, and the baryonic gas which
is responsible of the SZ effect. Various groups (Scaramella, Cen \&
Ostriker 1993, da Silva et al. 2000, Refregier et al. 2000, Refregier \&
Teyssier 2002, Seljak, Burwell \& Pen 2001, Zhang, Pen \& Wang 2002)
have performed such kind of simulations and computed the associated
power spectra. A compilation of the predictions from the different
groups can be found in Springel, White \& Herquist (2001). The SZ
power spectra computed from numerical simulations globally agree
within a factor of two. However, the results are quite sensitive to
the resolution of the simulations which acts as an artificial damping
effect at small angular scales, and to the size of simulation which if
not large enough underestimates the number of massive clusters and
thus the power at large angular scales.

In the second approach, the SZ power spectrum can be computed
analytically (Cole \& Kaiser 1988, Makino \& Suto 1993,
Atrio-Barandela \& Mucket 1999, Molnar \& Birkinshaw 2000, Komatsu \&
Kitayama 1999, Cooray 2000, 2001 , Majumdar 2001, Komatsu \& Seljak
2002) The computation is based on two quantities:
\begin{itemize}
\item The cluster number counts or mass function $dn/dM$ which
provides us with the number of clusters of a given mass $M$ present at
a redshift $z$ (see Section \ref{sec:SZcounts}).
\item The cluster model or mass--SZ flux relation, i.e. its
temperature and density profiles which give the spatial form factor
of the associated SZ effect.
\end{itemize}

To begin with, let us assume that the cluster cross-correlation
function can be known (for details see Peebles 1980, Cole \& Kaiser
1988).  The pattern of temperature anisotropy on the sky, induced by a
population of clusters, is found by the convolution of the temperature
anisotropy due a single ``template'' cluster of mass $M$ at redshift
$z$ with the angular distribution of the clusters, and then
integrating over their mass and redshift distributions. If one takes
an ensemble average and further assumes that $n(M,z)$ is constant over
the range of comoving separations for which the cross-correlation
function $\xi (M_1, M_2, z, \delta r)$ is non zero, then the angular
temperature power spectrum $C_\ell$ can be written as the sum of two
terms, the ``1-halo" or the Poisson term and ``2-halo" or the clustering
term, i.e

\be
C_\ell^{\rm total} = C_\ell^{\rm Poisson} \,+\,C_\ell^{\rm clustering} ~~~.
\ee
The power spectrum for the Poisson distribution of objects can then be
written as (Cole \& Kaiser 1988) 

\be C_\ell^{\rm Poisson} =
\int_0^{z_{\rm max}} dz {{dV(z)}\over{dz}} \int_{M_{\rm min}}^{M_{\rm max}} dM
{{dn(M,z)}\over{dM}} {|y_\ell(M,z)|}^2 , \ee 
where $dV(z)/dz$ is the differential comoving volume and $dn/dM$ is
the number density of objects and $y_\ell$ is the 2D Fourier transform of
the projected Compton $y$-parameter. The mass range is chosen so as to
cover from group scale to the largest cluster scales.  Since these
fluctuations occur at small angular scales, we can use the small
angle approximation of the Legendre transformation and write $y_\ell$ as the
angular Fourier transform of $y(\theta)$ as $y_\ell = 2\pi \int y(\theta)
J_0 [(\ell+1/2])\theta]\theta d\theta $ (Peebles 1980, Molnar \& Birkinshaw 2000), where
$J_0$ is the Bessel function of the first kind and zero order.

The clustering power spectra depends on lines of sight passing though
an ensemble of correlated clusters. It can be estimated (Komatsu \&
Kitayama 1999) as

\be
C_\ell^{\rm clustering} = \int_0^{z_{\rm max}} dz {{dV(z)}\over{dz}} P(k) \times 
 {\left[\int_{M_{\rm min}}^{M_{\rm max}} dM {{dn(M,z)}\over{dM}} 
b(M,z) y_\ell(M,z)\right]} ^2 ,
\ee
where $b(M,z)$ is the time dependent linear bias factor.  The matter
power spectrum, $P(k,z)$, is related to the power spectrum of cluster
correlation function $P_{\rm cluster}(k,M_1,M_2,z)$ through the bias, \ie,
$P_{\rm cluster}(k,M_1,M_2,z)=b(M_1,z)b(M_2,z)D^2(z)P(k,z=0)$. Convenient
expressions for the bias at cluster scales are given by 
Sheth \& Tormen (1999) and Jing (1999).
 
When one calculates the variance in beams of fixed size, the
Poissonian model is a good approximation if the probability that a
cluster has a neighbour is small inside the beam.  This probability is
the product of the number density and the volume integral of the
cross-correlation function over the region probed by the beam. It can
be shown that for beams comparable to the size of rich clusters ($R
\sim 1.5 h^{-1}$ Mpc), the Poissonian approach is a valid
approximation. Only for very large beams, the variance will increase
due to positive correlation of the clusters. It can be shown that the
Poisson power spectrum dominates at all $\ell$ values greater than
100. However, by subtracting, X-ray selected clusters of galaxies over
a certain flux ($S_{X} > 10^{-13}$ erg cm$^{-2}$ s$^{-1}$), from both
power spectra, one can make the clustering part of the spectrum
dominates around $\ell \sim 700$ (Komatsu \& Kitayama 1999, Majumdar 2001b).

To calculate the 2D profile of each cluster in the cluster ensemble,
one needs a cluster gas density and temperature model.  This can be
either the empirical truncated $\beta$-profile (e.g. in Molnar \&
Birkinshaw 2000), or it can be derived by solving the hydrostatic
equilibrium equation of a gas within a NFW dark matter potential
(e.g. in Komatsu \& Seljak 2001), or it can simply be obtained from
fits to simulated cluster profiles as in Diego \& Majumdar (2004).

\begin{figure} 
\epsfxsize=8.cm
\epsfysize=9.cm
\epsffile{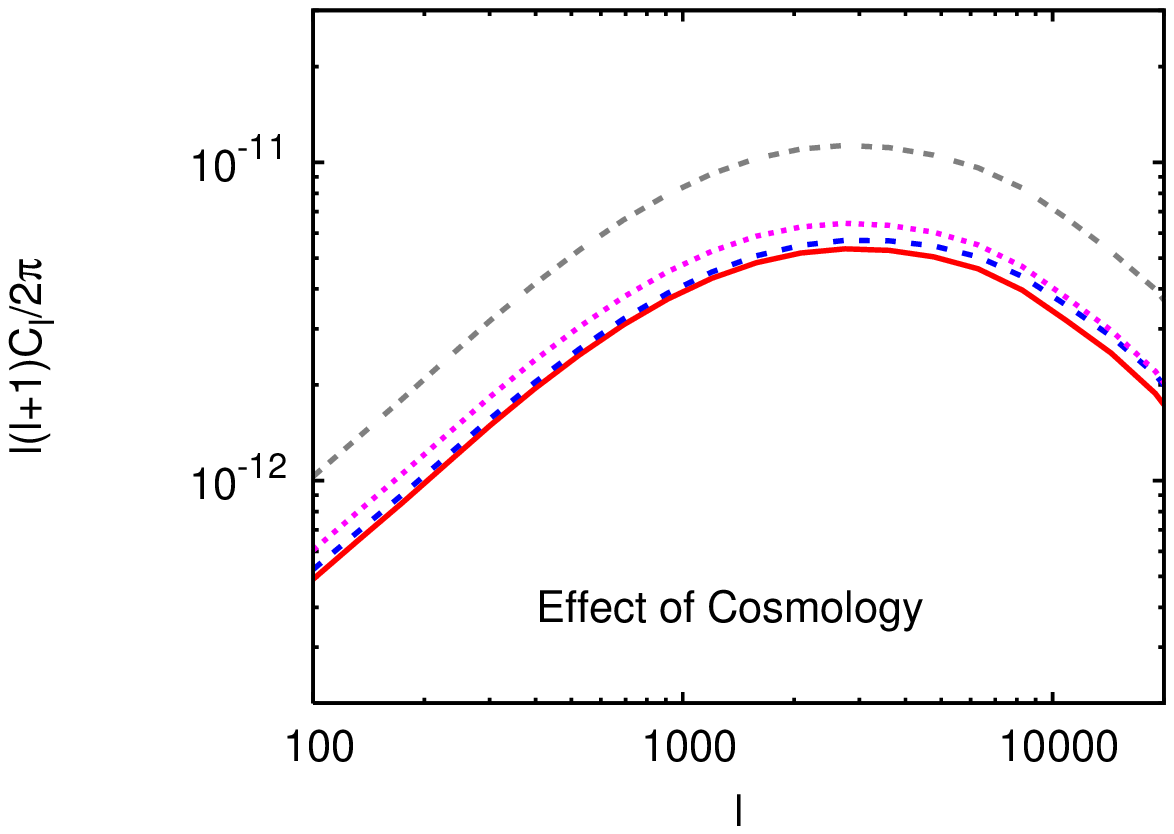}
\epsfxsize=8.cm
\epsfysize=9.cm
\epsffile{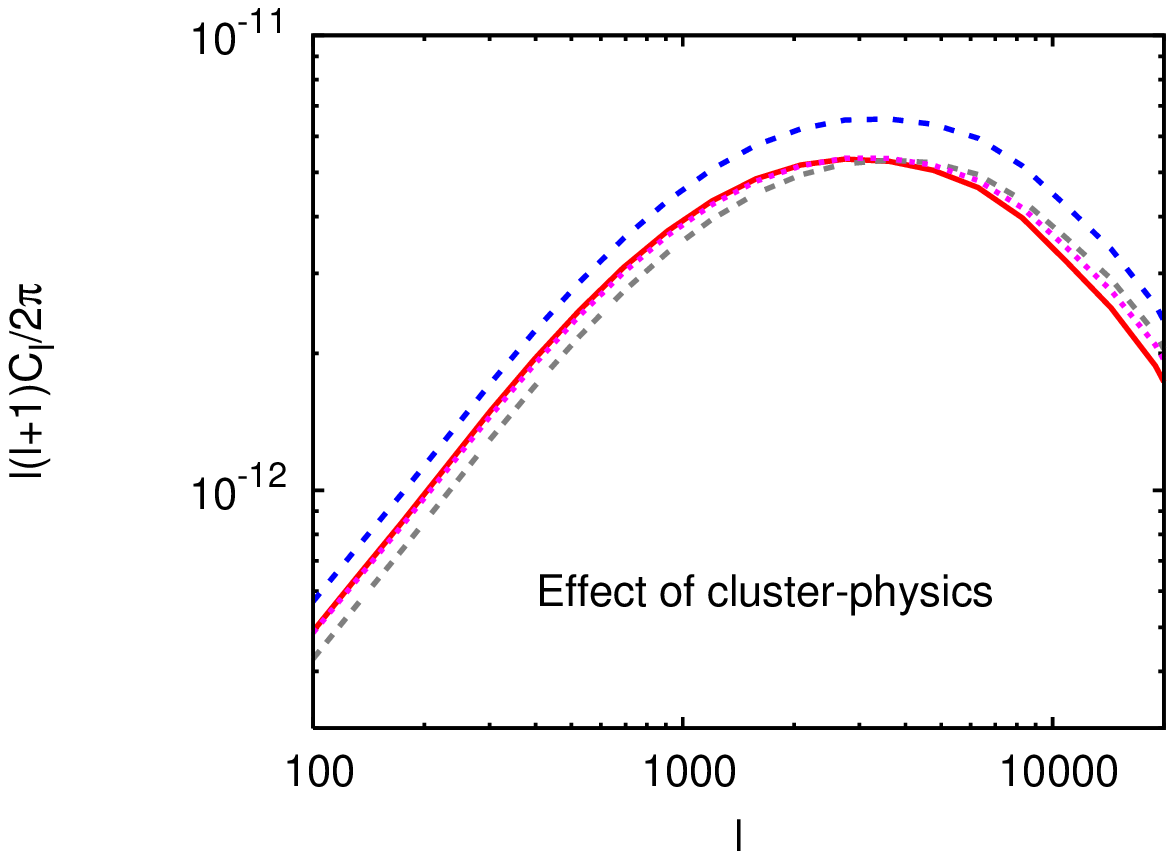}
\caption{Left Panel: The variation of the SZ power spectrum with the
cosmological parameters. The solid red line is for the fiducial model
for WMAP-3yrs cosmological parameters and cluster model from Komatsu
\& Seljak (2001). The triple dashed grey line shows the effect of 10\%
increase in $\sigma_8$, the purple dotted line for 10\% increase in
$\Omega_{\rm b}$ and blue dashed line for 10\% increase in
$\Omega_{\rm m}$.  The sensitivity of SZ C$_\ell$ to $\sigma_8$ is
maximum. Right Panel: The same as in the left panel but with changes
in the parameters relating to cluster physics while fixing the
cosmology. The blue dashed and the grey triple-dashed lines are for
cluster $M_{500}-T$ relation normalised to two different observational
results (Arnaud, Pointecouteau \& Pratt 2005) for an ensemble of
clusters. In these cases the normalisation and the slope of the
$M_{500}-T$ are both different than that obtained in the Komatsu \&
Seljak (2001) cluster model. The purple dotted line is for a 10\%
difference on the cluster halo central concentration. The red solid
line is for the fiducial model. The calculations are done for R-J
band. }
\label{fig:szpowspec}
\end{figure}

\subsection{Cosmological studies with SZ power spectrum}
\label{Subsec:cl_cosmo}

Both the volume element and the abundance of clusters depend on the
cosmological model. As a consequence the power spectrum will also vary
with varying cosmological parameters (Komatsu \& Seljak 2002, see also
Figure \ref{fig:szpowspec} left panel). For example, SZ power spectrum is
sensitive to the density parameter $\Omega_{\rm m}$ which mainly
affects the number of low and moderate redshift clusters, and the
equation of state of the dark energy $w=P/\rho_{\rm DE}$ which mainly
affects the number of high redshift clusters. However in the range of
allowed values for $\Omega_{\rm m}$ and $w$ the effects are rather
small. As for the other parameters the effects are quite negligible,
except that of $\sigma_8$. The SZ power spectrum is strongly sensitive
to the normalisation of mass fluctuations at cluster scales, i.e
$\sigma_8$. Numerical simulations of SZ clusters show
$C_\ell\propto\sigma_8^7$. Using simple scaling analysis, one can show
that $C_\ell\propto\sigma_8^{14/(3+n)}$ where $n$ is the effective
spectral index of mass fluctuations at cluster scales. If the highest
mass halos contribute to the power spectrum then $n\sim-1$ and we
recover back the simulation results. Note that when smaller mass hales
contribute significantly to the rms\footnote{For example, observed
smaller mass clusters show increased entropy over simple self-similar
predictions.}, then the effective $n$ is greater than $-1$ and
$C_\ell$ can have a stronger dependence on $\sigma_8$. The cluster
power spectrum depends, just like the mass function, on the cluster
physics more specifically the mass--observable relation affects the
power spectrum (Figure \ref{fig:szpowspec}, right panel).

Recently, the estimated power spectrum from the CBI-ACBAR-BIMA data
interpreted as an SZ signal has been used to constrain the value of
$\sigma_8$. The resulting value $\sigma_8 = 0.9$ was found higher than
other estimates of $\sigma_8$ which are now converging to the range
[0.7--0.8]. By forcing the clusters in the simulations or in the
analytic calculations of SZ-$C_\ell$ to have scaling properties
compatible with observed clusters, one can fit the CBI excess with a
$\sigma_8$ closer to WMAP 3 years value.  Note, however, that the
excess $\sigma_8$ only appears if we believe that the resulting excess
CMB fluctuations at the CBI scales is due to SZ effect from galaxy
clusters.  It has indeed been pointed out (Toffolatti et al. 2005,
Douspis, Aghanim \& Langer 2006) that SZ from clusters are only mildly
needed if there are unremoved point sources below the detection limit.
On a a more speculative note, non-Gaussianity in the primordial power
spectrum can boost the SZ power spectrum at cluster scales up to CBI
excess (Mathis, Diego \& Silk 2004, Sadeh, Rephaeli \& Silk 2006).

\subsection{Extraction of SZ effect from CMB data}
\label{subsec:SZextract}

The SZ contribution to the CMB power spectrum is dominant as compared
with other sources of secondary anisotropies. It was shown in Douspis,
Aghanim \& Langer (2006) that the SZ contribution, if coherently taken
into account, affects the determination of cosmological parameters
such as the normalisation $\sigma_8$, the optical depth $\tau$ and the
initial power spectrum index $n$. However, the TSZ contribution should
in principle be removed down to a given cluster mass from the measured
power spectrum provided multi-frequency observations are conducted. To
this aim many methods have been proposed and developed especially in
the context of Planck experiment (Sanz, Herranz \& Martinez-Gonzalez
2001, Vielva et al. 2001, Diego et al. 2002, Schaefer et al. 2006a,
2006b, Pierpaoli et al. 2006, Pires et al. 2006).  All of them take
benefit from the specific spectral signature of TSZ signal discussed
in Sect.\ref{sec_SZ}. These methods also use additional spatial
constraint based on adapted or matched filters, wavelets, etc. They
are mainly aimed at providing us with SZ cluster catalogues that will
be further used as cosmological probes. Consequently they help in
cleaning out the primary CMB signal.  In practice the sensitivity
limits of the experiments, their frequency coverages as well as their
finite beams prevent us from a complete cleaning of the TSZ effect.
The TSZ effect is not the only source of power at small scales. One
can also probe KSZ by its effect on the power spectrum at high
$\ell$. From amplitude arguments, it is easy to find that the
amplitude of the KSZ power spectrum is much smaller that that of the
TSZ. However, patchy reionisation has interesting implications for the
KSZ effect power spectrum (Iliev et al. 2007b). At fixed optical depth,
patchy reionisation approximately doubles the total KSZ power above
$\ell=3000$ by up to an order of magnitude compared to a uniform
reionisation scenario.  The KSZ effect has the same dependence as the
CMB anisotropies and the multi-frequency observations do not serve in
removing this contribution. Optimised methods to extract a map of the
KSZ temperature fluctuations from the CMB anisotropies can be
developed (Forni \& Aghanim 2004).  Assuming that a map of Compton
parameters for the TSZ effect can be obtained by multi-frequency
separation, one can take benefit from the spatial correlation between
KSZ and TSZ effects which are due to the same galaxy clusters. This
correlation allows us to use the TSZ map as a spatial template in
order to mask, in the temperature anisotropy map encompassing both CMB
and KSZ signal, the regions where clusters must have imprinted an SZ
fluctuation. By further using the statistical properties of KSZ, which
is a non-Gaussian, one can perform achieve good separation of the KSZ
signal out of the primary CMB.

\section{The SZ effect from other astrophysical sources}\label{sec:szsource}

The SZ effect, as proposed originally by Sunyaev \& Zel'dovich,
represents the shift experienced by the CMB photons when they undergo
inverse Compton interactions with the free electrons of the hot
ionised intra-cluster gas. The SZ effect is thus historically
associated with the galaxy clusters. However and more generally,
inverse Compton scattering can take place in all astrophysical
environments where both conditions of ionisation and high temperature
are fulfilled. As a result, the SZ effect was studied in a variety of
redshift domains and astrophysical sources from early proto-galaxies
and galaxies, to local galaxies like M31, as suggested by Taylor,
Moodley \& Diego (2003).

In the standard scenario of structure formation, baryonic matter is
believed to lay in the potential wells formed by the DM. The baryonic
matter in dynamical equilibrium with the DM can thus reach high
temperatures at virialisation. The baryonic gas can be heated by
additional means (photoionisation, mechanical heating,
...). Therefore, induced SZ anisotropies are expected to span a large
range amplitudes and angular scales.

At intermediate and large angular scales, a warm-hot gas
($T_e=10^5-10^7$ keV) is likely to exist in the large
structures of the cosmic web. This gas might account for a fraction of
the missing baryons (e.g. Fukugita, Hogan \& Peebles 1998). Due to its
relatively large temperature, this warm medium is expected to exhibit
an SZ signal but also an X-ray emission. Observing the SZ signal from
this warm medium, which would contribute to the CMB signal at
large scales, is certainly important from the CMB point of view for
disentangling between primary and secondary anisotropies.  Observing
the warm medium may be also a unique way to seek for, and find, the
missing baryons in the universe. Several studies have aimed at
studying and describing this contribution. Since it is associated with
non-linear structures the ideal tool is numerical simulations
(e.g. Springel, White \& Hernquist 2001, da Silva et al. 2001, Zhang,
Pen \& Wang 2002).  The expected SZ signal was found to have quite low
amplitudes ,thus, making it difficult to detect directly.  One way
around the problem might to target the correlations between SZ
signal and X-ray emission. Future experiments will tell us to what
extent this will be possible.

At smaller angular scale scales, when structures collapse to form
galaxies the temperature of the baryonic gas increases to larger
values due to shock heating. Moreover, the thermal content of the
collapsed structures increases due to feed-back processes (star
formation, AGN activity, ...).  In both cases, an SZ signal is
expected at the galaxy scale. Its detailed amplitude depends on the
efficiency of the heating mechanisms and of the galactic environment
(mainly the gas density). The predicted SZ signal has been computed in
both cases for shock heating and feed-back. In the first case, Valageas,
Balbi \& Silk (2001) found that the contribution from collapsed
objects dominates only at very small scales $\ell > 10^4$. The SZ
signal from galaxies can be even larger if they host central
supermassive black-holes (BH). In that case the galactic outflows
powered by the mechanical energy of accreting matter onto the BH
induce an important SZ signal as large as the COBE limit
(e.g. Natarajan \& Sigurdsson 1999, Aghanim, Balland \& Silk 2000,
Lapi, Cavaliere \& De Zotti 2003). SZ effect from quasar feedback
(Chatterjee \& Kosowsky 2007) has been predicted at $1 \mu$K level
which is potentially detectable by ALMA. The feed-back from stars and
its associated SZ effect was also studied. Rosa-Gonzalez et al.(2004)
calculated the signal expected from star-formation activity during the
formation of the most luminous bulges of normal galaxies. They found
that the temperatures and densities were high enough to produce $y$
parameters comparable to those of galaxy clusters. The supernova
driven galactic winds during the early stages of evolution of normal
galaxies can also cause the distortion of the CMB radiation as
proposed by (Majumdar \& Nath 2001).  Finally, SZ effect can not
only arise from forming or early formed objects but it can also be
associated with relic objects such as hot regions, ``cocoons'', around
radio galaxies as proposed by Yamada, Sugiyama \& Silk (1999). In that
case, the Compton parameter associated with the ensemble of cocoons
was found to be of the same order as the COBE constraint. Radio
galaxies in galaxy clusters can eject large quantities of energy which
is either thermalised or remains as relativistic radio plasma (radio
ghost) in the galaxy clusters. Ensslin \& Kaiser (2000) estimated the
Compton parameter from these two phases and found it too small to be
detected. However a statistical estimate of the relativistic
population in clusters can be envisaged by stacking the SZ signal from
all the cluster detected by future Planck satellite (Ensslin \& Hansen
2004).

In some cases the ionised regions have too low temperatures or
densities, or both, to exhibit significant Compton distortions. In
these cases, the kinetic SZ effect (if the gas moves with respect to
the CMB) becomes the dominant source of secondary anisotropies. This
is, particularly, the case for the patchy reionisation
(Aghanim et al.1996). Here, the temperature of the ionised bubbles
generated by emitting and ionising sources is low (typically $\sim
10^4$K) implying a negligible $y$ distortion, but the proper motion of
the ionised bubbles causes significant KSZ fluctuations. The secondary
anisotropies due to Doppler effect have been the subject of quite a
large number of studies, in the context of the reionisation problem,
especially in view of the first year WMAP constraints on optical depth
(see Sect. \ref{sec:reion} and references therein for details). A
large TSZ signal from the sources responsible for the reionisation is
however not excluded yet. It might on the contrary contribute to the
excess power measured by CBI and BIMA. An example for this is the case
in which early massive stars have played an important role in the
reionisation history of the universe (Oh, Cooray \& Kamionkowski 2003).

\section{Polarisation from galaxy clusters}\label{sec:polclus}

As shown by Sunyaev \& Zel'dovich (1980) not only CMB intensity is altered by the
presence of clusters, through the TSZ and KSZ effects, the CMB
polarisation is affected by the presence of galaxy clusters along the
photon lines of sight. Polarisation anisotropies are generated when
the photons scatter off free electrons in the intracluster medium. The
cluster-induced polarisation can be of a different origin: due to the
CMB quadrupole itself, to the cluster transverse motion or rotation
(Chluba \& Mannheim 2002), to its finite optical depth or to the presence of
cluster magnetic fields. In the following we briefly review each of
these processes

\begin{figure}
\epsfxsize=12.cm
\epsfysize=8cm
\hspace{2cm}
\epsfbox{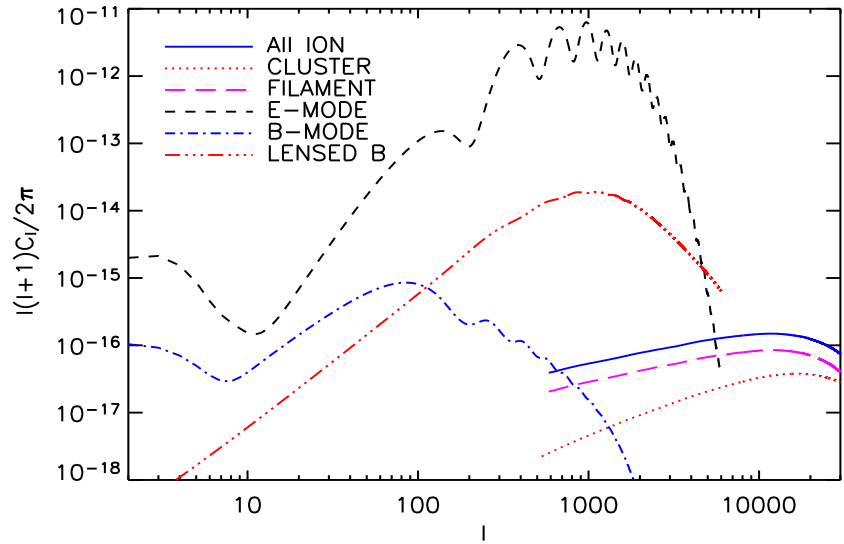}
\caption{From Liu, da Silva \& Aghanim (2005): The power spectra of
the polarised $E$ and $B$ modes from primary and secondary
anisotropies due to quadrupole-induced interactions.  }
\label{fig:secpol}
\end{figure}

\begin{enumerate}
\item{Transverse motion-induced polarisation: }\\ \noindent
The proper motion of the galaxy cluster relative to the CMB produces a
polarisation (Sunyaev \& Zel'dovich 1980, Itoh, Kohyama \& Nozawa 1998, Audit \& Simmons 1999, Sazonov \& Sunyaev 1999, Shimon et al. 2006). Keeping only the
first and second order terms in velocity, the polarisation towards a
cluster moving with a transverse velocity $v_{\mathrm t}$ is given by:
$$ P_\nu=0.1\frac{x^2{\mathrm e}^x({\mathrm e}^x+1)}{2({\mathrm
e}^x-1)^2}\,\left(\frac{v_{\mathrm t}}{c}\right)^2\tau
$$ 
The amplitude depends on the observed frequency. It is higher in the
Wien part of the spectrum. The frequency-integrated polarisation is
simply proportional to $\left(\frac{v_{\mathrm t}}{c}\right)^2\tau$.
The polarisation vector is perpendicular to the plane formed by the
velocity vector and the observing direction.

\item{Double scattering-induced polarisation:}\\ \noindent This
process is also called the finite optical depth effect. When the CMB
photons scatter off free electrons in the intra-cluster gas they
acquire anisotropies due to TSZ ($\propto\tau\left(\frac{k_{\rm
B}T_{\rm e}}{m_{\rm e}c^2}\right)$) and KSZ
($\propto\tau\left(\frac{v_{\mathrm r}}{c}\right)$) effects.  A second
scattering within the cluster induces polarisation of the order of
$\left(\frac{k_{\rm B}T_{\rm e}}{m_{\rm e}c^2}\right)\tau^2$ and
$\left(\frac{v}{c}\right)\tau^2$ without modifying the frequency
dependences. This effect can be generalised to any other source of
local anisotropy such as gravitational effects (moving gravitational
lens effects were computed by Gibilisco (1997), bulk motions of moving
gas clouds in the inner part of clusters (Diego, Mazzotta \& Silk
2003), collapse or expansion effects).  The amplitude of the
polarisation depends on the gas distribution $\rho(r)$.  For a
homogeneous spherical cloud with gas density $\rho_0$, (Sazonov \&
Sunyaev 1999) found that the maximal polarisation degrees are $0.025
\left(\frac{v_{\mathrm t}}{c}\right)\tau_0^2\,g(x)$ and $0.014
\left(\frac{k_{\rm B}T_{\rm e}}{m_{\rm e}c^2}\right)\tau_0^2\,f(x)$,
with $\tau_0=2\sigma_{\rm T}\rho_0$ and $f(x)$ and $g(x)$ the spectral
dependences of the TSZ and KSZ respectively. This results in a unique
spectral signature displayed in Figure \ref{fig:polfreq}).

\begin{figure}
\epsfxsize=12.cm 
\epsfysize=8.cm 
\hspace{2cm}
\epsfbox{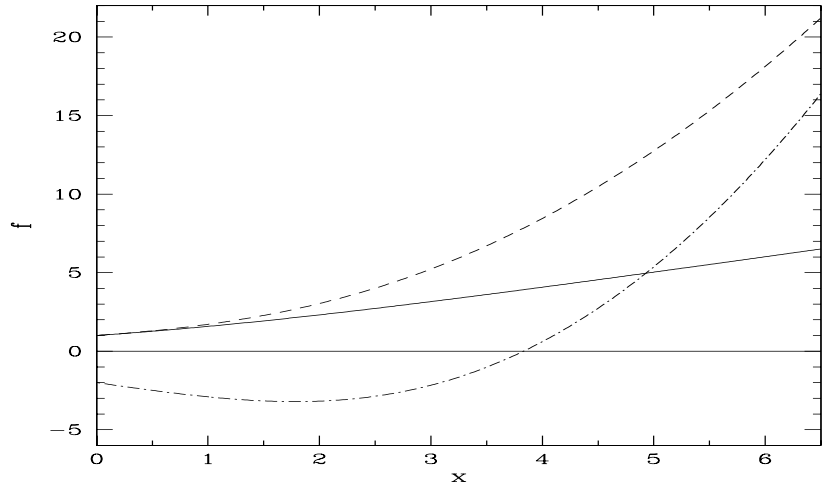}
\caption{From Sazonov \& Sunyaev (1999): The frequency dependencies of
the cluster-induced polarisation effects: CMB quadrupole-induced and
the $\left(\frac{v_{\mathrm t}}{c}\right)^2\tau$ effects (solid line),
$\left(\frac{v_{\mathrm t}}{c}\right)\tau^2$ effect (dashed line), and
the $\left(\frac{k_{\rm B}T_{\rm e}}{m_{\rm e}c^2}\right)\tau^2$ effect (dash-dotted
line). The last effect changes its sign at $x=3.83$.}
\label{fig:polfreq}
\end{figure}\hspace{2cm}

\item{Faraday rotation in magnetised intra-cluster medium:}\\
\noindent A radiation of frequency $\nu$ propagating through a plasma
in presence of a magnetic field ${\mathbf B}$ along direction
$\mathbf{n}$ sees its linear polarisation vector rotated by an angle
$\Delta \varphi$. This effect is the Faraday rotation (FR). Clusters
show evidence for magnetic fields (e.g. Murgia et al. (2004), and
references there in). The CMB polarised radiation passing through
magnetised galaxy clusters undergoes FR. This mixes the Stokes
parameters $Q$ and $U$ and thus generates $B$ modes out of $E$
polarisation.  The $B$-mode power spectrum depends on the details of
the electron density distribution per individual cluster, on the mass
function of clusters, as well as on the magnetic field distribution
and evolution. Such a contribution was computed by Takada, Ohno \&
Sugiyama (2001) and recently revisited by Tashiro, Aghanim \& Langer
(2007). It is proportional to the product $B_0^2\nu_{\rm obs}^{-4}$.
At the frequencies typically used for CMB observations, the amplitude
of the FR-induced $B$ polarisation is small. However, the polarisation
observed at the cluster scale could be a powerful tool for probing the
gas distribution (Ohno et al. 2003).

\item{CMB Quadrupole-induced polarisation:}\\ \noindent
The presence of a quadrupole component produces a polarisation signal
proportional to the cluster optical depth $\tau$. It has a maximum
amplitude of $P_{\rm max}\sim 2.\,10^{-6}g(x)\tau$ which changes with
frequency ($x={h_{\rm pl}}\nu/{k_{\rm B}}T$) following $g(x)=x{\mathrm
e}^x/({\mathrm e}^x-1)$.  This effect should be the dominant source of
polarisation related to clusters. The primary quadrupole-induced
polarisation due to galaxy clusters and to warm gas in filamentary
structures has been investigated using hydrodynamical simulations
(Liu, da Silva \& Aghanim 2005). As shown in Figure \ref{fig:secpol}, this effect
dominates at very small angular scales. On the larger scales the
signal is dominated by the contributions from the filamentary
structures. At the smallest scales it is the galaxy cluster
contribution which dominates.
\end{enumerate}
\noindent

Galaxy clusters produce both $E$ and $B$ polarisation but at a level
that is much smaller than the primary signal. However, despite the
relatively low signal amplitudes, the study of cluster
induced-polarisation has gained a new interest since it appears to be
a potentially interesting cosmological probe
(see e.g. Cooray \& Baumann 2003). Cluster polarisation measurement
were proposed to probe the large scale velocity fields through
measuring the cluster transverse motions, as well as the galaxy
cluster dynamics. However, the most promising application of cluster
polarisation measurement is associated with the dominant effect, the
quadrupole induced polarisation. The polarised signal from numerous
galaxy clusters has been suggested by Kamionkowski \& Loeb (1997) as a
method to probe CMB quadrupole by reducing the cosmic variance
uncertainty (see also Portsmouth 2004). This in turn provides a
new way to obtain accurate ISW measurements and probe the dark energy
content of the universe (e.g. Cooray, Huterer \& Baumann 2004, Bunn 2006).

\section{High-order statistics of secondary anisotropies}

Most, and indeed the simplest, inflationary scenarios
(e.g. Guth 1981, Sato 1981) predict that the temperature anisotropy
field obeys Gaussian statistics to first order.  In this case, the
statistical distribution of the CMB anisotropies is fully described by
its second moment, the power spectrum, given by:
$$
C_\ell=\sum^\ell_{m=\ell}|a_{\ell m}|^2,
$$ where the $a_{\ell m}$ are the multipole coefficients in the
spherical harmonic expansion $\Delta T(\theta)/T=\sum_{\ell,m}a_{\ell
m}Y_\ell^m(\theta)$. Nevertheless, other cosmological scenarios such
as topological defects (e.g. Vilenkin \& Shellard 1994, Landriau \&
Shellard 2003 and references therein) and multi-field inflation
(Gangui et al.1994, Bernardeau \& Uzan 2002) suggest departures from
the Gaussian hypothesis. One example of a Non-Gaussian model that
lends itself to specific predictions is the so-called $\chi_m^2$ model
for the multi-field inflaton potential (Koyama, Soda \& Taruya 1999)
implemented for $m=2$ (2 fields) by Sadeh, Rephaeli \& Silk (2006) to
study SZ observables. The issue of testing, through higher order
statistics, assumptions about the early universe is quite important
and is becoming feasible in the context of present and future CMB
experiments. Therefore, a battery of non-Gaussian (NG) estimators have
been recently developed and tested.  Among the most commonly used,
there are the three and four-point functions and their harmonic
analogues the bi- ($T_3$) and trispectrum ($T_4$) (e.g. Hu 2001,
Komatsu \& Spergel 2001, Kunz et al. 2001) respectively given by:
\begin{equation}
<T({\bf {\ell_1}})T({\bf \ell_2})T({\bf \ell_3})>_c=(2\pi)^2
\delta({\bf \ell_{123}})T_3({\bf \ell_1,\ell_2,\ell_3})
\end{equation}
and 
\begin{equation}
<T({\mathbf \ell_1})T({\mathbf \ell_2})T({\mathbf \ell_3})T({\mathbf
\ell_4})>_c=(2\pi)^2 \delta({\mathbf \ell_{1234}})T_4({\mathbf
\ell_1,\ell_2,\ell_3,\ell_4}),
\end{equation}
where ${\mathbf \ell_{123}}={\mathbf
\ell_1+\ell_2+\ell_3}$ and ${\mathbf \ell_{1234}}={\mathbf
\ell_1+\ell_2+\ell_3+\ell_4}$). Also widely used are the higher order
moments of the wavelet coefficients (skewness and excess kurtosis)
(e.g. Pando, Valls-Gabaud \& Fanf 1998, Forni \& Aghanim 1999, Hobson,
Jones \& Lasenby 1999, Barreiro \& Hobson 2001). The wavelet analysis,
in the dyadic wavelet transform scheme, decomposes a signal $s$ in a
series of the form :
\begin{eqnarray}
 s(l) = \sum_{k} c_{J,k} (\phi_{\mathrm A})_{J,l}(k)
       +  \sum_{k} \sum_{j=1}^J (\psi_{\mathrm A})_{j,l}(k) w_{j,k}
\end{eqnarray}
where $J$ is the number of decomposition levels, $w_{j,k}$ the wavelet
(or detail) coefficients at position $k$ and scale $j$ (the indexing
is such that $j = 1$ corresponds to the finest scale, i.e. highest
frequencies), and $c_{J}$ is a coarse or smooth version of the
original signal $s$. Other tests of non-Gaussianity are the global
Minkowski functionals like the total area of excursion regions
enclosed by isotemperature contours or total contour length and genus
(e.g. Gott et al.1990, Schmalzing \& Gorski 1998, Novikov, Schmalzing
\& Mukhanov 2000, Shandarin 2002), the harmonic space analysis
(Hansen, Pastor \& Semikoz 2002), the peak statistics (e.g. Bond \&
Efstathiou 1987, Vittorio \& Juszkiewicz 1987).

Not only departures from the simplest inflation model can generate
non-Gaussian signatures. Systematic effects, point source and
foreground-induced non-Gaussianities will inevitably
arise at small scales from the secondary anisotropies, either through
the non-linear growth of fluctuations or through the interactions of
CMB photons with the potential wells or ionised matter along their
lines of sight. Besides, the study of secondary non-Gaussianities is
very interesting on its own, since it is related to the cosmic
structures, their evolution and spatial distribution; it is also of
great importance in order to go beyond the information provided by the
power spectrum of the CMB primary anisotropies. In this context, the
higher order statistics of the secondary anisotropies are used to
predict the NG signatures of non-primordial origin in the CMB, and to
better detect and understand the structures themselves.

The NG signatures are of particular importance in the case of
gravitational lensing since they allow us in theory to reconstruct the
mass distribution of the lenses. As a matter of fact, the deflection
angles are small compared to the scale of structures and the lensing
effect is hardly seen directly in a CMB map. The effects on the power
spectrum are generally small and sub-dominant, and the two
point-statistics is thus not sufficient to allow for the
reconstruction of the mass distribution of lenses. To better identify
the effects of gravitational lensing on the CMB, one has therefore to
consider the induced NG signatures, naturally arising from the second
order effects in the anisotropies (correlations between large scale
gradients and small scale generated power), through higher-order
statistics. The week lensing of primary anisotropies produces a
four-point signature (e.g. Bernardeau 1997, Zaldarriaga 2000, Kesden,
Cooray \& Kamionkowski 2003). Quadratic statistics (such as the power
spectrum of the squared temperature maps) permit us to recover the
information in the four-point function about the mass distribution of
the lens field (e.g. Zaldarriaga \& Seljak 1999, Hu 2001, Takada 2001,
Hu \& Okamoto 2002, Cooray \& Kesden 2003).  These methods use lensed
anisotropy maps only, or combine them with the polarisation field
(especially the $B$ field) which is less contaminated by the primary
signal. In all cases, mapping the lens, and thus dark matter,
distribution requires high resolution, high signal-to-noise maps of
the CMB temperature fluctuations and polarisation fields.

\begin{figure} 
\epsfxsize=12cm
\epsfysize=10cm
\hspace{2cm}
\epsffile{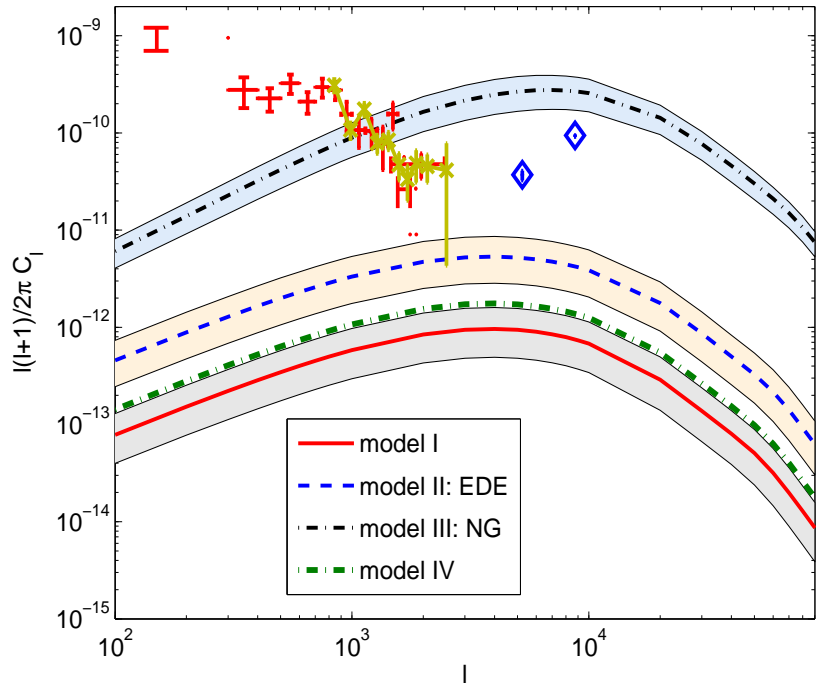}
\caption{From Sadeh, Rephaeli \& Silk (2007): SZ power spectrum
obtained for a $\Lambda$CDM model with $\sigma_8=0.74$ (solid line)
and 0.8 (thick dash-dotted line), an early dark energy model (dashed
line) and a non-Gaussian $\chi_m^2$ model (dash-dotted line). Shaded
area correspond to WMAP 1$\sigma$ error on $\sigma_8$. The data points
are those of BIMA (diamonds), ACBAR (x symbols) and CBI (crosses).}
\label{fig:nong}
\end{figure}

The interactions of CMB photons with the free electrons along their
lines of sight, through Compton or Doppler effects, also produce
secondary NG signatures. These sources of secondary anisotropies are
expected to be important; it was therefore necessary to forecast their
NG signal and study its detectability.  This was done mainly for the
SZ effect and the inhomogeneous reionisation through the trispectrum
(Cooray 2001) and through the high-order moments of the wavelet
coefficients (Aghanim \& Forni 1999). For the SZ thermal effect,
Cooray (2001) gave the expression for the trispectrum of the TSZ
effect in the flat sky approximation
\begin{equation}
<y({\mathbf \ell_1})y({\mathbf \ell_2})y({\mathbf \ell_3})y({\mathbf
\ell_4})>_c=(2\pi)^2 \delta({\mathbf \ell_{1234}})T^{\rm TSZ}({\mathbf
\ell_1,\ell_2,\ell_3,\ell_4}),
\end{equation}
where $_c$ designate the connected part and $T^{\rm TSZ}$ is given by 
\begin{equation}
T^{\rm TSZ}=\int{\rm d}r \frac{W^{\rm TSZ}(r)^4}{d_{\rm A}^6}T_\Pi\left(
\frac{{\mathbf \ell_1}}{d_{\rm A}},\frac{{\mathbf \ell_2}}{d_{\rm A}},
\frac{{\mathbf \ell_3}}{d_{\rm A}},\frac{{\mathbf \ell_4}}{d_{\rm A}};r\right).
\end{equation}
where $T_\Pi$ is the pressure trispectrum.
The weight function $W^{\rm TSZ}(r)=-2\frac{k_{\rm B}\sigma_{\rm T}
{\overline n_{\rm e}}}{a(r)^2m_{\rm e}c^2}$ is given in the
Rayleigh-Jeans regime.  In all cases, the signal from SZ effect
dominates at small angular scales. For all vector-like fields such as
the Ostriker-Vishniac effect, but also the mildly non-linear regime
probed by the KSZ effect for large scale structures, even moments were
shown to dominate over odd moments, making the trispectrum a more
sensitive estimator of non-Gaussianity than the bispectrum
(Castro 2004). As a result while the bispectrum is most likely
undetectable by future CMB experiments, the trispectrum of the OV
effect could be measured by Planck or by arc-minute scale
interferometric experiments.

The NG signatures associated with the secondary effects can be used to
probe and trace the matter distribution, they can also be use as
additional constraints to separate the secondary effects from the
primary CMB signal (e.g. Forni \& Aghanim 2004).  However in all these
cases, this signal at small angular scales is the sum of the CMB
anisotropies and all the secondary contributions. This makes it harder
to disentangle them and requires the use of the polarisation field
or cross-correlations and couplings between components.

Figure \ref{fig:nong} shows an example of SZ angular power spectra for
the galaxy cluster contribution, demonstrating that the non-Gaussian
$\chi_m^2$ model can have a substantial impact for $10^3< \ell <
10^4,$ especially in the case of WMAP-3 yrs normalisation
$(\sigma_8=0.74$). Also shown are examples of Gaussian models with
different normalisations $(\sigma_8=0.74, 0.8)$ and an early dark
energy model (Bartelmann, Doran \& Wetterich 2006).

\section{Discussion and conclusion}\label{sec:conclusion}

Secondary effects induce temperature and polarisation
anisotropies. These additional anisotropies contribute to the CMB
signal and modify (at certain scales) both its amplitude and its
statistical character. Such contribution was not actually important
within the context of first generation of CMB experiments (e.g. COBE).
Already now with WMAP, and even more so with future Planck
satellite, the aim of measuring the CMB signal with fundamental
instrumental noise limits forces us to investigate with extreme care
the effects of the secondary anisotropies. They might constitute in some
cases important limiting factor on the scientific objectives of future
CMB studies such that constraining the energy scale of inflation
through the $B$-mode polarisation induced by the stochastic
gravitational wave background, or constraining the inflationary field
through the statistical nature of the temperature
anisotropies. Present day CMB experiments are now reaching
sensitivities and angular resolutions such that secondary effects can
be no longer neglected. This is the case for lensing by large scale
structures which convert the $E$-mode primary polarisation into a
$B$-mode secondary contribution and is by far the largest
contaminant. This is also the case for the example of the SZ effect
from galaxy clusters which could explain the excess of power at
high multipoles measured by ACBAR, BIMA and CBI.

The SZ effect is more than a nuisance factor to cosmological parameter
extraction. It is a potentially powerful tool for cosmology. SZ
cluster counts can be used to probe the cosmological model and put
constraints on the nature of dark energy. In combination with other
observations, especially at X-ray energies, it allows us to measure
cosmological parameters such as the Hubble constant and the cluster
gas mass fraction (e.g. Grego et al. 2001). The SZ effect can also be
used to characterise the clusters themselves as it potentially can
measure their radial peculiar velocities
(Lamarre et al. 1998, Benson et al. 2003). The non-relativistic corrections to the
SZ effect can also be used to measure the gas temperature directly for
massive clusters. This might an important issue for future SZ surveys for
which X-ray counterparts will not be available. The spectral signature
of the SZ effect can in principle probe the electron gas distribution
and constrain any non-thermal electron population in the intracluster
medium. Moreover multi-frequency SZ measurements might provide a novel
way of constraining the CMB temperature and its evolution with
redshift (Battistelli et al. 2003, Horellou et al. 2005).

To achieve these goals, high precision measurements of the SZ effect
will be needed over large areas of the sky. This requires a new
generation of SZ telescopes that are already being built or
designed. Following OVRO and BIMA, the Sunyaev-Zel'dovich Array (SZA)
which consists of eight 3.5 metre telescopes is operating at 26--36
GHz and 85--115 GHz. The SZA along with BIMA/OVRO forms the Combined
ARray for Millimeter Astronomy (CARMA) telescope which aims at
providing high resolution, detailed imaging of SZ clusters. Several
other telescopes are being commissioned (e.g. AMI), in 2007, with the
aim of surveying large areas of the SZ sky for blind detection of
clusters. These deep SZ cluster surveys will be performed by the AMIBA
interferometer, the South Pole Telescope (SPT), the Atacama Cosmology
Telescope (ACT) and the Atacama Pathfinder Experiment
(APEX). Moreover, the Planck satellite scheduled to be launched in
2008 will detect thousands of SZ clusters over the whole sky.

\begin{figure} 
\epsfxsize=8.cm
\epsffile{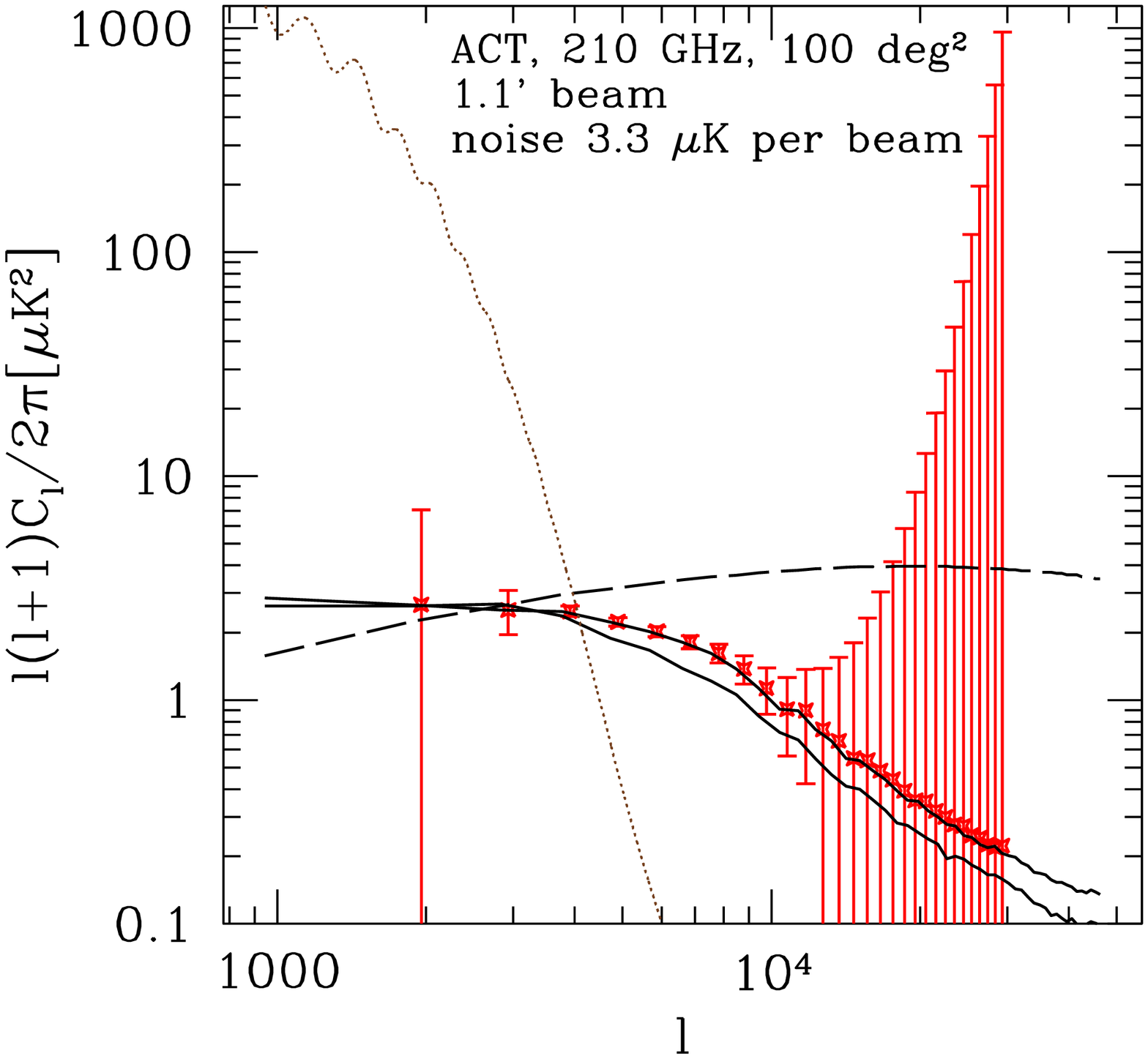}
\epsfxsize=8.cm
\epsffile{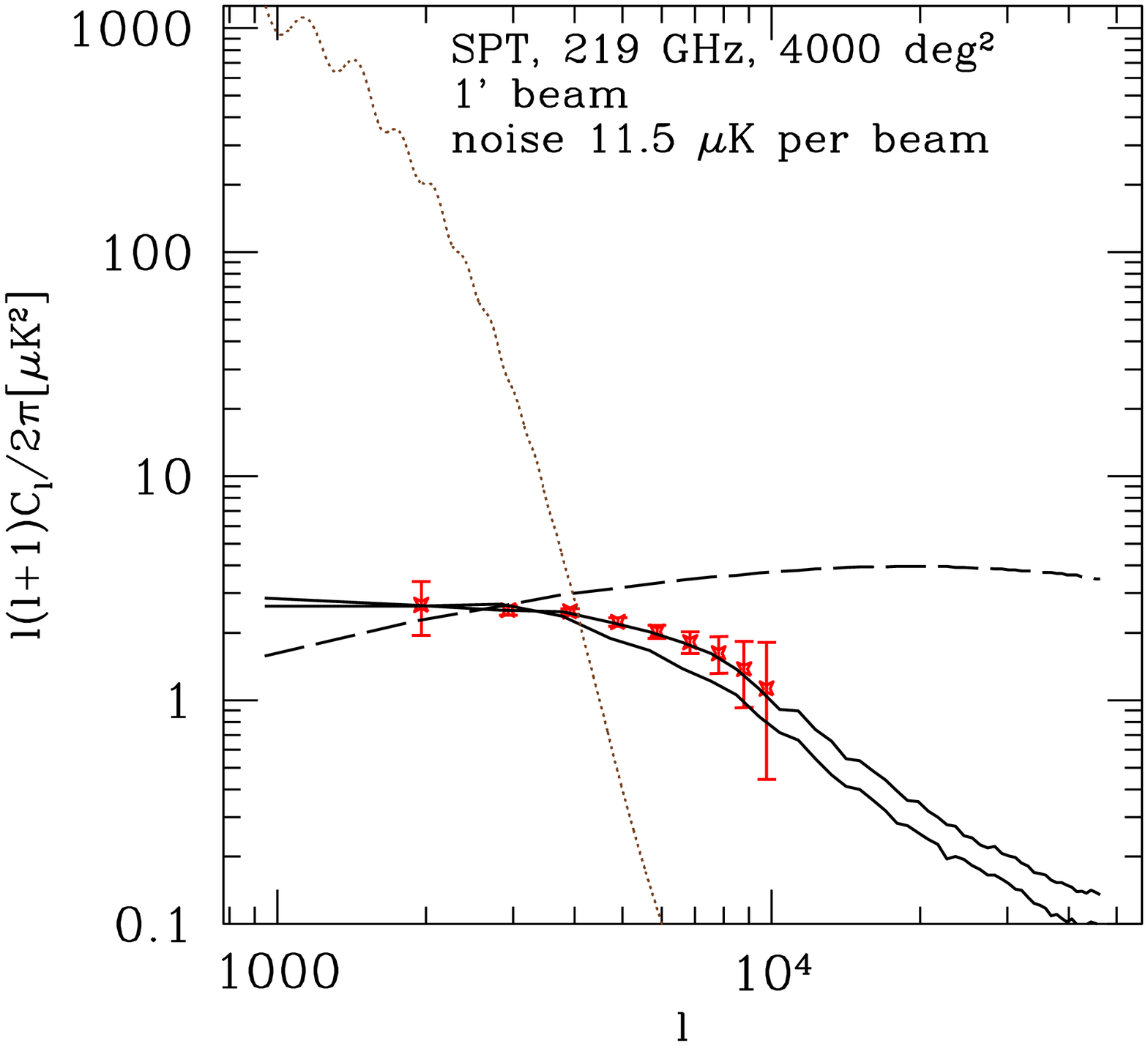}
\caption{From Iliev et al. (2007b): Observability of the Doppler induced
  temperature anisotropies: the sky power spectrum of the reionization
  signal (black, solid; from two simulations) with the forecast error
  bars for ACT (left) and SPT (right). The primary CMB anisotropy
  (dotted) and the post-reionization KSZ signal (dashed) are also
  shown and are added to the noise error bars for the reionization
  signal. The TSZ component is assumed to be completely separated.}
\label{fig:act-spt}
\end{figure}

Although both ACT and SPT are primarily designed for SZ cluster
detection, the predicted KSZ signal, at a few arc minute scales,
induced by the reionisation might be sufficiently strong to be
detected by these upcoming experiments (Figure \ref{fig:act-spt}). These
high $\ell$ measurements of the reionisation-induced temperature
anisotropies will however not suffice to unravel the ionisation
history. Polarisation measurements at low $\ell$ are the optimal CMB
tool to achieve this. In the near future, Planck will provide all sky
$E$-mode polarisation maps and will be sensitive to partial or double
reionisation models at the percent level. In principle this could help
discriminate between different models with identical optical depths
(Kaplinghat et al. 2003), subject to our being able to understand,
model and remove the relevant galactic foregrounds. In combination
with low frequency radio interferometer such as LOFAR and eventually
SKA it should be possible to probe the onset of the reionisation and
the end of the dark ages by anti-correlating 21~cm emission and CMB
temperature fluctuations (Alvarez et al. 2006).

The next generation of polarisation-optimised satellites, such as
B-POL or EPIC, is being designed to measure the primary $B$-modes from
inflation. These experiments will inevitably have high enough
sensitivity to actually reconstruct the ionisation history of the
universe. A new generation of moderate resolution ground-based and
balloon-born CMB polarisation (CLOVER, at 97, 150, and 220~GHz, QUIET
at 40 and 90~GHz, QUaD, EBEX, BICEP, SPIDER, BRAIN) are under
operation, construction or design. The principal aim is to measure
primordial $B$-modes. They will also measure weak lensing-induced
$B$-modes with resolution over multipoles $20<\ell< 1000$ and down to
$\sim$0.1$\mu$K precision. This is an essential prerequisite to
searching, at these scales, for the gravity-wave induced $B$-mode
background from inflation. For $20 <\ell <100$, current constraints on
the scalar to tensor ration should allow the primordial signal to
dominate lensing.

Secondary effects are not simply a ``foreground'' that adds noise and
limits our knowledge. They are by nature the best tools to probe
structure formation and evolution providing a complementary picture of
the late time universe to that obtained from traditional tools like
galaxy surveys.

\ack 
The authors would like to thank an anonymous referee and Matthias
Bartlemann for careful reading and commenting of the article. NA and
SM wish to thank Oxford University for hospitality. SM would like to
thank CITA where a large part of the work was done as well as IAS-Orsay  
for hospitality during the final stages of the review.

\References

\item[] 
{Abell}, G.O. 1958, \ApJS, {\bf 3}, 211

\item[]
Abel T., Bryan G.~L. \& Norman M.~L. 2000, \ApJ, {\bf 540}, 39
 
\item[]
Abel T., Bryan G.~L. \& Norman M.~L. 2002, {\it Science}, {\bf 295}, 93 

\item[]
{Abroe}, M.~E., et~al. 2004, \ApJ, {\bf 605}, 607

\item[]
Afshordi, N. 2004, \PhRvD, {\bf 70}, 083536

\item[]
Afshordi, N. {Lin}, Y.-T., \& {Sanderson}, A.~J.~R. 2005, \ApJ, {\bf 629}, 1

\item[]
{Aghanim}, N., {De Luca}, A., {Bouchet}, F.~R., {Gispert}, R. \& {Puget}, J.~L.
1997, \AaA, {\bf 325}, 9

\item[]
{Aghanim}, N., {Prunet}, S., {Forni}, O. \& {Bouchet}, F.~R.
1998, \AaA, {\bf 334}, 409

\item[]
{Aghanim}, N. \& {Forni}, O.
1999, \AaA, {\bf 347}, 409

\item[]
{Aghanim}, N., Balland, C. \& Silk J. 
2000, \AaA, {\bf 357}, 1

\item[]
{Aghanim}, N., {Hansen}, S.~H., {Pastor}, S., \& {Semikoz}, D.~V.\ 
2003, JCAP, 5, 7.

\item[]
Aghanim, N., Hansen, S.H., Lagache, G. 2005, \AaA, {\bf 439}, 901

\item[]
Alvarez, M.A., Shapiro, P.R., Ahn, K., Iliev, 
I.T. 2006, \ApJ, {\bf 644}, L101

\item[]
Alvarez, M. A., Komatsu, E., Dor\'e, O. \&
	Shapiro, P. R. 2006, \ApJ, {\bf 647}, 840

\item[]
Amblard, A., Vale, C., White, M. 2004, astr-ph/0403075

\item[]
Ameglio, S., Borgani, S., Diaferio, A., Dolag, K. 2006, \MNRAS,
{\bf 369}, 1459

\item[]
Arnaud, M.; Pointecouteau, E. \& Pratt, G. W., 
2005, \AaA, {\bf 441}, 893

\item[]
{Atrio-Barandela}, F. \& {Mucket}, J., 1999, \ApJ, {\bf 515}, 465

\item[]
Audit, E. \&  Simmons, J.~F.~L., 1999, \MNRAS, {\bf 305}, L27

\item[]
{Barreiro}, R.~B. \& {Hobson}, M.~P.
 2001, \MNRAS, {\bf 327}, 813
 
\item[]
Bartelmann, M., Doran, M., Wetterich, C. 2006, \AaA,
{\bf 454}, 27

\item[]
 Bassett, B.A. \&  Kunz, M. 2004, \ApJ, {\bf 607}, 661

\item[]
Battistelli, E.~S. et al. 2003, \ApJ, {\bf 598}, L75

\item[]
 Bean, R., Melchiorri, A., Silk, J.  2007, \PhRvD, {\bf 75}, 063505

\item[]
Benabed, K., Bernardeau, F. \& van Waerbeke, L. 2001, \PhRvD, {\bf 63}, 3501
 
\item[]
{Benson}, B.~A., {Church}, S.~E., {Ade}, P. A.~R., et al.
 2003, \ApJ, {\bf 592}, 674

\item[]
{Bernardeau}, F. 1997, \AaA, {\bf 324}, 15

\item[]
{Bernardeau}, F.,  1998, \AaA, {\bf 338}, 767

\item[]
{Bernardeau}, F. \& {Uzan}, J.-P.
 2002, \PhRvD, {\bf 66}, 103506
 
\item[]
{Birkinshaw}, M. \& {Gull}, S.~F. 1978, \Nat, {\bf 274}, 111

\item[]
{Birkinshaw}, M. \& {Gull}, S.~F. 1983, \Nat, {\bf 302}, 315

\item[]
Birkinshaw, M. 1989, in Moving Gravitational lenses, p.
59, eds. J. Moran, J. Hewitt \& K.Y. Lo; Springer-Verlag, Berlin

\item[]
Birkinshaw, M., Hughes, J.P., \& Arnaud, K. A. 1991, \ApJ, {\bf 379}, 466

\item[]
Birkinshaw, M. \& Hughes, J.P. 1994, \ApJ, {\bf 420}, 33

\item[]
{Birkinshaw}, M., 1999, {\it Phys. Rept.}, {\bf 310}, 97

\item[]
Birkinshaw, M. \& Lancaster, K., 2004, Background Microwave Radiation and Intracluster Cosmology, Proceedings of the International School of Physics "Enrico Fermi", {\it astro-ph/0410336}

\item[]
Blanchard, A. \& Schneider, J., 1987, \AaA, {\bf 184}, 1

\item[]
Bock, J., et al. 2006, astro-ph/0604101

\item[]
Bonamente, M., Joy, M., La Roque, S., Carlstrom, J., Reese, E.,
Dawson, K., 2006, \ApJ, {\bf 647}, 25 

\item[]
Bond, J. R. \&  Efstathiou, G., 1984, \ApJ, {\bf 285}, 45

\item[]
Bond, J. R.; Efstathiou, G., 1987, \MNRAS, {\bf 226}, 655

\item[]
{Bond}, J., {Kaiser}, N., {Cole}, S. \& {Efstathiou}, G., 1991, \ApJ, {\bf 379}, 440

\item[]
{Borgani}, 2006, Lectures for 2005 Guillermo Haro Summer School on Clusters, to appear in "Lecture notes in Physics", {\it astro-ph/0605575}
 
\item[]
Boughn, S. P.; Crittenden, R. G.; Turok, N. G., 1998, {\it New Astron.}, 
{\bf 3}, 275

\item[]
Boughn, S.P. \& Crittenden, R.G. 2002, \PhRvL, {\bf 88}, 021302

\item[]
Boughn, S.P. \& Crittenden, R.G. 2004, \Nat, {\bf 427}, 45

\item[]
Bromm, V., Kudritzki, R.~P. \& Loeb, A., 2001, \ApJ, {\bf 552}, 464

\item[]
Bromm, V., Coppi, P.~S. \& Larson, R.~B., 2002, \ApJ, {\bf564}, 23

\item[]
Bunn, E. F., 2006, \PhRvD, {\bf 73}, 3517

\item[]
{Carlstrom}, J.~E., {Holder}, G.~P. \& {Reese}, E.~D.
2002, \ARAA, {\bf 40}, 643

\item[]
{Carroll}, S.~M., {Press}, W.~H. \& {Turner}, E.~L., 1992, \ARAA, {\bf 30}, 499

\item[]
{Castro}, P.~G., 2004, \PhRvD, {\bf 67}, 044039
 
\item[]
Cavaliere, A. \& Fusco-Femiano, R., 1978, \AaA, {\bf 70}, 677

\item[]
Cayon, L., Martinez-Gonzalez, E., Sanz, J., 1993, \ApJ, {\bf 413}, 10

\item[]
Cen, R., 2003, \ApJ, {\bf 591}, 5

\item[]
Challinor, A. D. \& Lasenby, A. N., 1998, \ApJ, {\bf 499}, 1

\item[]
Challinor, A. D. \& Lasenby, A. N., 1999, \ApJ, {\bf 510}, 930

\item[]
Challinor, A. D. \& Lewis, A. , 2005, \PhRvD, {\bf 71}, 103010

\item[]
Chatterjee, S. \& Kosowsky, A., 2007, astrp-ph/0701759

\item[]
Chluba, J. \& Mannheim, K., 2002, \AaA, {\bf 396}, 419

\item[]
Chodorowski, M., 1992, \MNRAS, {\bf 259}, 218

\item[]
Chodorowski, M., 1994, \MNRAS, {\bf 266}, 897

\item[]
Ciardi, B. \& Madau, P., 2003, \ApJ, {\bf 596}, 1

\item[]
Ciardi, B., Ferrara, A. \& White, S.~D.~M., 2003, \MNRAS, {\bf 344}, L7

\item[]
Colafrancesco, S., Marchegiani, P. \& Palladino, E., 2003,
\AaA, {\bf397}, 27

\item[]
{Cole}, S. \& {Kaiser}, N., 1988, \MNRAS, {\bf 233}, 637

\item[]
Cooray A. 2001, \PhRvD, {\bf 64}, 0635 

\item[]
Cooray, A. 2002a, \PhRvD, {\bf 65}, 083518

\item[]
{Cooray}, A. 2002b, \PhRvD, {\bf 65}, 3510
 
\item[]
{Cooray}, A. 2002c, \PhRvD, {\bf 65}, 3512

\item[]
Cooray, A. \& Sheth, R., 2002, {\it Phys. Rep.}, {\bf 372}, 1

\item[]
{Cooray}, A. \& {Baumann}, D., 2003, \PhRvD, {\bf67}, 063505
 
\item[]
{Cooray}, A. \& {Kesden}, M.
2003, {\it New Astron.}, {\bf 8}, 231
 
\item[]
{Cooray}, A., Huterer, D., Baumann, D., 2004, \PhRvD, {\bf 69}, 027301
 
\item[]
{Cooray}, A. \& {Seto}, N., 2005, {\it Journal of Cosm. \& Astropart.},
 {\bf 12}, 004
 
\item[]
Corasaniti, P.-S., Gianantonio, T., Melchiorri, A., 2005,
\PhRvD, {\bf 71}, 123521
 
\item[]
Crittenden, R.G. \& Turok, N., 1996, \PhRvL, {\bf 76}, 575

\item[]
Dabrowski, Y., Hobson, M.P., Lasenby, A.N., Doran, C.J.L. 1999, 
  \MNRAS, {\bf 302}, 757

\item[]
 Daigne F., Olive K.~A., Vangioni-Flam E., Silk J., Audouze, J. 2004,
 \ApJ, {\bf 617}, 693

\item[]
{Dickinson}, C., et~al., 2004, \MNRAS, {\bf 353}, 732

\item[]
{Diego}, J.~M.,  {Vielva}, P., {Martinez-Gonz{\'a}lez}, E., 
{Silk}, J. \&{Sanz}, J.~L.,
2002, \MNRAS, {\bf 336}, 1351

\item[]
{Diego}, J.~M., {Hansen}, S.~H. \& {Silk}, J.,
2003, \MNRAS, {\bf338}, 796

\item[]
Diego, J. M.; Mazzotta, P.; Silk, J., 2003, \ApJ, {\bf 597}, 1

\item[]
 {Diego}, J.~M.~R., \& {Majumdar}, S., 2004, \MNRAS, {\bf 352}, 993
 
\item[]
{Dijkstra}, M., Haiman, Z. \& {Loeb}, A., 2004, \ApJ, {\bf 613}, 646

\item[]
Dijkstra, M., Haiman, Z., Loe,b A., 2004, \ApJ, {\bf 613}, 646

\item[]
{Dodelson}, S. \& {Jubas}, J.~M.
1995, \ApJ, {\bf 439}, 503

\item[]
{Dodelson}, S. 2004, \PhRvD, {\bf 70}, 3009

\item[]
 Dor\'e, O., Holder, G., Alvarez, M., Iliev, I.T., Mellema, G., Pen,
 U.-L., Shapiro, P.R., 2007, astro-ph/0701784.

\item[]
{Douspis}, M., {Aghanim}, N. \& {Langer}, M.,
2006, \AaA, {\bf 456}, 819

\item[]
Dyer, C. C. 1976, \MNRAS, {\bf 175}, 429

\item[]
Eke, V. R., Cole, S. \& Frenk, C. S., 1996,
\MNRAS,, {\bf 282} 263 
 
\item[]
{Ensslin}, T. A. \&  {Kaiser}, C. R.,
2000, \AaA, {\bf 360}, 417

\item[]
{Ensslin}, T. A. \&  {Hansen}, S. H.,
2004, preprint, {\it astro-ph/0401337}

\item[]
Evrard, A. E. 1990, ApJ, 363, 349.

\item[]
{Fan}, et al., 2003, {\it Astron. J.}, {\bf 125}, 1649 

\item[]
Fang, L.-Z. \& Wu, X.-P.,  1993, \ApJ, {\bf 408}, 25

\item[]
Ferrarese, L. \& Merritt, D., 2000, \ApJ, {\bf 539}, 9
 
\item[]
{Forni}, O. \& {Aghanim}, N., 1999, \AaAS, {\bf 137}, 553

\item[]
Forni, O. \& Aghanim, N. 2004, \AaA, {\bf 420}, 49  
 
\item[]
Fosalba, P. \& Gaztanaga, E., 2005, \MNRAS, {\bf 350}, 37
 
\item[]
Fukugita, M., Hogan, C. J., \&  Peebles, P. J. E., 1998, \ApJ, {\bf503}, 518

\item[]
{Gangui}, A., {Lucchin}, F., {Matarrese}, S. \& {Mollerach}, S.,
1994, \ApJ, {\bf 430}, 447
 
\item[]
Gaztanaga, E., Maneram, M., Multamaki, T. 2006, \MNRAS, {\bf 365}, 171
 
\item[]
Gebhardt et al,  2000, \ApJ, {\bf 543}, L5

\item[]
Gibilisco, M., 1997, {\it Astrop. \& Space Suppl.}, {\bf 249}, 189

\item[]
{Gladders}, M.~D., {Yee}, H.~K.~C., {Majumdar}, S., 
{Barrientos}, L.~F., {Hoekstra}, H., {Hall}, P.~B. \& 
{Infante}, L., 2007, \ApJ, {\bf 655}, 128

\item[]
{Gnedin}, N.~Y., 2000, \ApJ, {\bf 535}, 530

\item[]	
Gnedin, N. Y. \& Jaffe, A. H., 2001, \ApJ, {\bf 551}, 3

\item[]
{Gnedin}, N.~Y. \& Prada, F., 2004, \ApJ, {\bf 608}, 77

\item[]
Goldberg, D.M. \& Spergel, D.N., 1999, \PhRvD, {\bf 59}, 3002

\item[]
Gomez, P.L. et al., 2003, Contribution to "Matter and Energy in
Clusters of Galaxies", Taipei April 2002, astro-ph/0301024.

\item[]
Gott, J. R.; Park, C.; Juszkiewicz, R.;
	Bies, W. E.; Bennett, D. P.; Bouchet, F. R.;
	Stebbins, A., 1990, \ApJ, {\bf 352}, 1

\item[]
{Grego}, L., {Carlstrom}, J.~E., {Reese}, E.~D., 
{Holder}, G.~P., {Holzapfel},
  W.~L., {Joy}, M.~K., {Mohr}, J.~J. \& {Patel}, S.,
 2001, \ApJ, {\bf 552}, 2

\item[]
{Gruzinov}, A. \& {Hu}, W.,
1998, \ApJ, {\bf 508}, 435

\item[]
{Guth}, A., 1981, \PhRvD, {\bf 23}, 347

\item[]
Haiman, Z., Abel, T. \& Madau, P., 2001, \ApJ, {\bf 551}, 599

\item[]
Haiman, Z., Mohr, J. J. \& Holder, G. P., 2001, 
\ApJ, {\bf 553} , 545

\item[]
Haiman, Z., Holder, G. P., 2003, \ApJ,  {\bf 595}, 1

\item[]
Hansen, S. H., Pastor, S. \&  Semikoz, D. V.,
 2002, \ApJ, {\bf 573}, L69

\item[]
Hansen, S. \&  Haiman, Z., 2004, \ApJ,  {\bf 600}, 26

\item[]
Hattori, M. \& Okabe, N., 2004, JKAS, {\bf 37}, 543.

\item[]
Hern\`andez-Monteagudo, C. \&  Rubino-Martin, J. A., 2004, \MNRAS,
{\bf 347}, 403.

\item[]
Hirata, C.M. \&  Seljak, U., 2003, \PhRvD, {\bf 67}, 043001

\item[]
{Hobson} M.~P., {Jones} A.~W., {Lasenby} A.~N., {Bouchet} F.~R., 1998,
\MNRAS, {\bf 300}, 1

\item[]
{Holder}, G.~P., {Mohr}, J.~J., {Carlstrom}, J.~E., {Evrard}, A.~E. \&
  {Leitch}, E.~M.,
2000, \ApJ, {\bf 544}, 629

\item[]
{Holder}, G.~P., 2002, \ApJ, {\bf 580}, 36

\item[]
Holder G.P. et al. 2003, \ApJ, {\bf 595}, 13

\item[]
Holder, G. \& Kosowsky, A. 2004, \ApJ, {\bf 616}, 8

\item[]
{Holzapfel}, W.~L., {Ade}, P. A.~R., {Church}, S.~E., et al. 
{Mauskopf}, P.~D.,{Rephaeli}, Y., {Wilbanks}, T.~M. \& {Lange}, A.~E.,
1997, \ApJ, {\bf 481}, 35

\item[]
Horellou, C.; Nord, M.; Johansson, D.; L\'evy, A., 2005, \AaA, {\bf441}, 435

\item[]
Hu, W. \& White, M., 1996, \ApJ, {\bf 471}, 30

\item[]
Hu, W. \& White, M., 1997, \PhRvD, {\bf 56}, 596

\item[]
Hu, W., 2000, \ApJ, {\bf 529}, 12
 
\item[]
{Hu}, W., 2001, \PhRvD, {\bf 64}, 083005
 
\item[]	
Hu, W. \& Dodelson, S., 2002, \ARAA, {\bf 40}, 171

\item[]
{Hu}, W. \& Okamoto, T., 2002, \ApJ, {\bf 574}, 566

\item[]
{Hu}, W., 2003, \PhRvD, {\bf 67}, 081304

\item[]
{Hu}, W. \& Okamoto , T., 2004, \PhRvD, {\bf 69}, 043004
 
\item[]
{Hu}, W., {Scott}, D. \& {Silk}, J., 1994, \PhRvD, {\bf 49}, 648

\item[]
Hughes, J.P. \& Birkinshaw, M., 1998, \ApJ, {\bf501}, 1

\item[]
Iliev, I.T., Pen, U.-L., Bond, J.R., Mellema, G., Shapiro, P.R.,
 2007a, \ApJ, {\bf 660}, 933

\item[]
 Iliev, I.T., Mellema, G., Pen, U.-L., Bond, J.R., Shapiro, P.R. 2007b,
 preprint, {\it astro-ph/0702099}

\item[]
Inagaki, Y., Suginohara, T., Suto, Y., 1995, \PASJ, {\bf 47}, 411 

\item[]
Islam, R.~R., Taylor, J.~E. \& Silk J., 2003, \MNRAS, {\bf 340}, 647 

\item[]
Itoh, N., Kohyama, Y. \& Nozawa, S., 1998,  \ApJ, {\bf 502}, 7
 
\item[]
Itoh. N, Kawana. Y, Nozawa. S, \& Kohyama Y., 2001,
\MNRAS,  {\bf327}, 567
 
\item[]
{Jaffe}, A.~H. \& {Kamionkowski}, 
1998, \PhRvD, {\bf 58}, 043001

\item[]
Jenkins, A. et al., 2001, \MNRAS, {\bf 321}, 372

\item[]
{Jing}, Y.~P.,  1999, \ApJ, {\bf 515}, L45

\item[]
Jones, M. et al. 1993, \Nat, {\bf 365}, 320

\item[]
Kaiser, N. 1982, \MNRAS,  {\bf198}, 1033

\item[]
Kaiser, N., \& Stebbins, A., 1984, \Nat, {\bf310}, 391

\item[]
Kamionkowski, M. \& Spergel, D. N., 1994, \ApJ, {\bf 432}, 7

\item[]
Kamionkowski, M. 1996, \PhRvD, 54, 4169

\item[]
Kamionkowski, M. \& Loeb, A., 1997, \PhRvD, {\bf 56}, 4511

\item[]
Kaplinghat, M., Knox, L., Song, Y.-S., 2003, \PhRvL, {\bf 89}, 011303 

\item[]
Kaplinghat, M. et al., 2003, \ApJ, {\bf 583}, 24

\item[]
{Kashlinsky}, 1988, \ApJ, {\bf 331}, 1
 
\item[]
Kesden, M., Cooray, A., Kamionkowski, M. 2003, \PhRvD, {\bf 67}, 123507 

\item[] 
Kinkhabwala, A. \& Kamionkowski, M. 1999, \PhRvL, {\bf 83}, 4172

\item[]
{Knox}, L., {Scoccimarro}, R. \& {Dodelson}, S., 1998, \PhRvL, {\bf 81}, 2004

\item[]
Kobayashi, S., Sasaki, S. \& Suto, Y., 1996, \PASJ, {\bf 48}, 107

\item[]
Koch, P. M., Jetzer, P. \& Puy, D., 2003, {\it New Astron.} , {\bf8}, 1

\item[]{kofman85}
Kofman, L. A. \& Starobinskii, A. A. 1985, Sov. Astr. Let., 11,
271. 

\item[]
{Kogut}, A. et al., 2003, \ApJS, {\bf 148}, 161

\item[]
{Komatsu}, E. et al, 1999, \ApJ, {\bf 516}, L1
 
\item[]
{Komatsu}, E. \& {Kitayama}, T., 1999, \ApJ, {\bf 526}, L1

\item[]
{Komatsu}, E. \& {Seljak}, U., 2001 \MNRAS, {\bf 327}, 1353

\item[]
{Komatsu}, E. \& {Seljak}, U., 2002, \MNRAS, {\bf 336}, 1256

\item[]
{Komatsu}, E. \& {Spergel}, D.~N., 2001, \PhRvD, 63, 3002.

\item[]	
Koyama, K., Soda, J., Taruya, A. 1999, \MNRAS, {\bf 310}, 1111

\item[]
{Kunz}, M., {Banday}, A.~J.,
{Castro}, P.~G., {Ferreira}, P.~G., {G\'orski}, K.~M. , 
2001, \ApJ, {\bf 563}, L99
 
\item[]
{Kuo}, C.~L., et~al. 2004, \ApJ, {\bf 600}, 32

\item[]
{Lacey}, C. \& {Cole}, S., 1993, \MNRAS, {\bf 262}, 627

\item[]
{Lamarre}, J.~M., {Giard}, M., {Pointecouteau}, E., et al.
{Bernard}, J.~P., {Serra},
  G., {Pajot}, F., {D\'esert}, F.~X., {Ristorcelli}, I., {Torre}, J.~P.,
  {Church}, S., {Coron}, N., {Puget}, J.~L. \& {Bock}, J.~J.,
1998, \ApJ, {\bf 507}, L5.

\item[]
Landriau, M.; Shellard, E. P., 2003, \PhRvD, {\bf 67}, 3512

\item[]
Lapi, A.; Cavaliere, A.; De Zotti, G., 2003, \ApJ, {\bf 597}, 93

\item[]
Lasenby, A.N., Doran, C.J.L., Hobson, M.P., Dabrowski, Y., Challinor,
A.D., 1999, \MNRAS, {\bf 302}, 748

\item[]
Lesgourgues, J., Perotto, L., Pastor, S., Piat, M., 2006, \PhRvD,
{\bf 73}, 5021

\item[]
{Levine}, E.~S., {Schulz}, A.~E. \& {White}, M., 2002, \ApJ, {\bf 577}, 569

\item[]
Lewis, A. 2005, \PhRvD, {\bf 71}, 083008

\item[]
Lewis, A. \& Challinor, A. 2006, {\it Phys. Rept.}, {\bf 429}, 1

\item[]
{Lima}, M. \& {Hu}, W. 2004, \PhRvD, {\bf 70}, 043504

\item[]
Linder, E., 1988, \AaA, {\bf 206}, 199

\item[]
Liu, G., et al., 2001, \ApJ, {\bf 561}, 504

\item[]
Liu, G., da Silva, A. \& Aghanim, N. 2005, \ApJ, {\bf 561}, 504

\item[]
{Lukic}, Z., {Heitmann}, K., {Habib}, S., {Bashinsky}, S. \& {Ricker}, P. M.,
2007, preprint, {\it astro-ph/0702360}

\item[]
Ma, C.-P. \& Bertschinger, E., 1995, \ApJ, {\bf 455}, 7

\item[]
Ma, C.P. \& Fry, J.N., 2000, \ApJ, {\bf 543}, 503

\item[]
Ma, C.P. \& Fry, J.N., 2002, \PhRvD, {\bf 88}, 211301

\item[]
Madau, P., Meiksin, A. \& Rees, M.J., 1997, \ApJ, {\bf 475}, 429

\item[]
Madau, P., Rees, M.~J., Volonteri, M., Haardt, F. \& Oh, S.~P. 2004,
\ApJ, {\bf 604}, 484

\item[]
Majumdar, S. \& Nath, B. B., 2000, \ApJ, {\bf 542}, 597

\item[]
Majumdar, S. \& Nath, B. B., 2001, \MNRAS, {\bf 324}, 537

\item[]
{Majumdar}, S., 2001, \ApJ, {\bf 555}, L7

\item[]
{Majumdar}, S. , 2001b, PhD Thesis, Indian Institute of Science, Bangalore

\item[]
{Majumdar}, S. \& {Mohr}, J.~J., 2003, \ApJ, {\bf 585}, 603

\item[]
{Majumdar}, S. \& {Mohr}, J.~J., 2004, \ApJ, {\bf 613}, 41

\item[]
{Majumdar}, S. \& {Cox}, G., 2007, preprint in preparation

\item[]
{Makino}, N. \& {Suto}, Y., 1993, \ApJ, {\bf 405}, 1

\item[]
Makino, N., Sasaki, S. Suto, Y., 1998, \ApJ, {\bf 497}, 555

\item[]
Martinez-Gonz\'alez, E., Sanz, J.-L., \& Silk, J., 1990, \ApJ, {\bf355}, L5

\item[]
Mathiesen, B., Evrard, A. E., \& Mohr, J. J., 1999, \ApJ, {\bf 520}, 21

\item[]
Mathis, H., Diego, C. \& Silk, J., 2004, \MNRAS, {\bf 353}, 681

\item[]
Metcalf, R.B. \& Silk, J., 1997, \ApJ, {\bf 489}, 1

\item[]
Mohr, J. J., Evrard, A. E., Fabricant, D. G., \& Geller,
M. J., 1995, \ApJ, {\bf 447}, 8

\item[]
Mohr, J. J., Mathiesen, B., \& Evrard, A. E. 1999, \ApJ, {\bf 517}, 627

\item[]
Mollerach, S., Harary, D., \& Matarrese, S., 2004, \PhRvD, {\bf 69}, 3002
 
\item[]
{Molnar}, S.~M. \& {Birkinshaw}, M.,
1999, \ApJ, {\bf 523}, 78.
 
\item[]
Molnar, S.\& Birkinshaw, M., 2000, \ApJ, {\bf 537}, 542

\item[]
Mortonson, M.J. \& Hu, W., 2007, \ApJ, {\bf 657}, 1

\item[]
{Motl}, P. M., {Hallman}, E. J., {Burns}, J. O. \& {Norman}, M. L.,
2005 \ApJ, {\bf 623}, L63

\item[]
Mukhanov, V. F., Feldman, H. A. \& Brandenberger, R. H. ,1992,
\PhRvD, {\bf 215}, 203

\item[]
Murgia, M., Govoni, F., Ferretti, L., Dallacasa, D., Giovaninni, G., Fanti,
R., Taylor, G.B., Dolag, K., 2004, \AaA, {\bf 424}, 429

\item[]
Natarajan, P. \& Sigurdsson, S. 1999, \MNRAS, {\bf 302}, 288

\item[]
Navarro, J. F., Frenk, C. S. \& White, S. D. M., 1997, \ApJ, {\bf 490}, 493

\item[]
Ng, K. L. \& Ng, K.-W., 1996, \ApJ,  {\bf 456}, 413

\item[]
Nolta, M.R. et al. , 2004, \ApJ, {\bf 608}, 10

\item[]
Novikov, D., Schmalzing, J. \& Mukhanov, V. F., 2000, \AaA, {\bf 364}, 17

\item[]
Nozawa, S., Itoh, N. \& Kohyama, Y., 1998,
\ApJ, {\bf 508},  17
 
\item[]
Nozawa, S., Itoh, N., Kawana, Y. \& Kohyama, Y., 2000,
\ApJ, {\bf 536}, 31

\item[] 
Oh, S.~P., 2001, \ApJ, {\bf 553}, 499 

\item[]
Oh, S. P., Cooray, A., \& Kamionkowski, M., 2003, \MNRAS, {\bf 342}, 20

\item[]
Ohno, H., Takada, M., Dolag, K., Bartelmann, M. \& Sugiyama,
N., 2003, \ApJ, {\bf 584}, 599

\item[]
Okamoto, T. \& Hu, W., 2003, \PhRvD, {\bf 67}, 3002

\item[]
{Ostriker}, J.~P. \& {Vishniac}, E.~T.,
1986, \ApJ, {\bf 306}, L51

\item[]
Padmanabhan,  N. \& Finkbeiner, D. 2005, \PhRvD, {\bf 72}, 023508.

\item[]
Padmanabhan, N. et al. 2005, \PhRvD, {\bf 72}, 043525
 
\item[]
{Pando}, J., {Valls-Gabaud}, D. \& {Fang}, L.~Z.
1998, \PhRvL, {\bf 81}, 4568
 
\item[]
Panek, M. 1992, \ApJ, {\bf 388}, 225

\item[]	
Peebles, P. J. E. \& Yu, J. T. 1970, \ApJ, {\bf 162}, 815

\item[]
{Peebles} P.~J.~E., 1980, Large Scale Structure of the Universe., Princeton 
University Press, Princeton 

\item[]
{Pfrommer}, C., {Ensslin}, T. A., {Springel}, V, {Jubelgas}, M. \& {Dolag}, K.,
2006, preprint, {\it astro-ph/0611037}

\item[]
{Pierpaoli}, E., {Anthoine}, S., {Huffenberger}, K. \&	{Daubechies}, I.,
2006, \MNRAS, {\bf 359}, 261

\item[]
{Pires}, S., {Juin}, J.~B., {Yvon}, D., {Moudden}, Y., 
{Anthoine}, S. \& {Pierpaoli}, E.
2006, \AaA, {\bf 455}, 741

\item[]
Pogosian, L. 2006, {\it New Astr. Rev.}, {\bf 50}, 932

\item[]
Portsmouth, J. 2004, \PhRvD, {\bf 70}, 063504

\item[]
Pointecouteau, E. 1999, \ApJ, {\bf 519}, 115.

\item[]
{Press}, W.~H. \& {Schechter}, P.,  1974, \ApJ, {\bf 187}, 425

\item[]
Puy, D., Grenacher, L., Jetzer, Ph., Signore, M. 2000, \AaA, {\bf 363}, 415

\item[]
Rassat, A., Land, K, Lahav, O., Abdalla, F.B. 2006, astro-ph/0610911

\item[]
{Readhead}, A.~C.~S., et~al. 2004, \ApJ, {\bf 609}, 498

\item[]
{Rees}, M.~J. \& {Sciama}, D.~W.
1968, \Nat, {\bf 511}, 611

\item[]
{Reese}, E.~D., {Carlstrom}, J.~E., {Joy}, M., {Mohr}, J.~J., {Grego}, L. \&
  {Holzapfel}, W.~L.
2002, \ApJ, {\bf 581}, 53

\item[]
{Refregier}, A., {Komastsu}, E., {Spergel}, D.~N. \& {Pen}, U., 2000,
\PhRvD, {\bf 61}, 123001

\item[]
{Refregier}, A. \& {Teyssier}, R., 2002, \PhRvD, {\bf 66}, 043002

\item[]
{Rephaeli}, Y. 1995, \ARAA, {\bf 33}, 541

\item[]
Ricotti, M. \& Ostriker, J. 2004, \MNRAS, {\bf 352}, 547

\item[]
Ricotti, M., Ostriker, J. \& Gnedin, N. 2005, \MNRAS, {\bf 357}, 207

\item[]
Rosa-Gonzalez, D.; Terlevich, R.; Terlevich, E.; Friaca, A.;
Gaztanaga, E. 2004, \MNRAS, {\bf 348}, 669

\item[]
{Ruhl}, J.~E., et~al. 2003, \ApJ, {\bf 599}, 786

\item[]
Sachs, R.K., Wolfe, A.M. 1967, \ApJ, {\bf 147}, 73

\item[]
Sadeh, S., Rephaeli, Y., Silk, J. 2006, \MNRAS, {\bf 368}, 1583

\item[]
Sadeh, S., Rephaeli, Y., Silk, J. 2007, astro-ph/0706.1340

\item[]
{Sanz} J.~L.,  {Herranz} D. \& {Martinez-Gonzalez} E.,
2001, \ApJ, {\bf 552}, 484

\item[]
Santos, M. B., Cooray, A., Haiman, Z., Knox, L., Ma, C.-P. 2003, \ApJ,
{\bf 598}, 756 

\item[]
Sato, K. 1981, \MNRAS, {\bf 195}, 467

\item[]
Sazonov, S. Y.; Sunyaev, R. A.  1999, \MNRAS, {\bf 310}, 765.
 
\item[]
Scannapieco, E. 2000, \ApJ, {\bf 540}, 20

\item[]
{Scaramella}, R., {Cen}, R.,  \& {Ostriker}, J., 1993, \ApJ, {\bf 416}, 399

\item[]
Schaerer, D. 2002, \AaA, {\bf 24}, 337

\item[]
{Sch{\"a}fer}, B.~M.,  {Pfrommer}, C., {Bartelmann}, M., 
{Springel}, V. \& {Hernquist}, L.
2006a, \MNRAS, {\bf 370}, 1309

\item[]
{Sch{\"a}fer}, B.~M.,  {Pfrommer}, C., {Hell}, R.~M. \&
{Bartelmann}, M.
2006b, \MNRAS, {\bf 370}, 1713

\item[]
Schmalzing, J.; Gorski, K. M. 1998, \MNRAS, {\bf 297}, 355
 
\item[]
{Seljak}, U.  1996a, \ApJ, {\bf 460}, 549

\item[]
{Seljak}, U. 1996b, \ApJ, {\bf 463}, 1

\item[]
{Seljak}, U., {Burwell}, J.  \& {Pen}, U., 2001, \PhRvD, {\bf 63}, 063001.

\item[]
Seshadri, T. R. \& Subramanian, K. 1998, \PhRvD, {\bf 58}, 3002.

\item[]
Shandarin, S. F. 2002, \MNRAS, {\bf 331}, 865 

\item[]
{Sheth}, R.~K. \& {Tormen}, G.
1999, \MNRAS, {\bf 308}, 119.

\item[]
{Sheth}, R. K., {Mo}, H. J, \& {Tormen}, G., 2001, \MNRAS, {\bf 323}, 1
 
\item[]
Shimon, M. \& Rephaeli, Y. 2004, {\it New Astron.}, {\bf 9}, 69.

\item[]
Shimon, M., Rephaeli, Y., O'Shea, B.W., Norman, M.L. 2006, \MNRAS, {\bf 368}, 511

\item[]
Silk, J. 1967, \Nat, {\bf 215}, 1155

\item[]
Silk, J. \& White, S. D. M. 1978, \ApJ, {\bf 226}, 103

\item[]
{da Silva}, A.~C., {Barbosa}, D., {Liddle}, A.~R. \& {Thomas}, P.~A.
2000, \MNRAS, 317, 37.

\item[]
da Silva, A. C.; Kay, S. T.; Liddle, A. R.; Thomas, P. A.; Pearce,
F. R.; Barbosa, D 2001, \ApJ, {\bf 561}, 15

\item[]
Sokasian, A., Abel, T., Hernquist, L., \& Springel, V.
2003, \MNRAS, {\bf 344}, 607

\item[]
Sokasian, A. et al. 2004, \MNRAS, {\bf 350}, 47

\item[]
Somerville, R. \& Livio, M. 2003, \ApJ, {\bf 593}, 611

\item[]
{Spergel}, et al. 2003, \ApJS, {\bf 148}, 175

\item[]
{Spergel}, et al. 2007, \ApJS, 170, 377.

\item[]
{Springel}, V., {White}, M.  \& {Hernquist}, L., 2001, \ApJ, {\bf 549}, 681

\item[]
{Stebbins}, A.  1988, \ApJ, {\bf 327}, 584.

\item[]
Sugiyama, N., Zaroubi, S. \& Silk, J. 2004, \MNRAS, {\bf 354}, 543 

\item[]
Sunyaev, R. A. \& Zel'dovich, Ya. B.,  1970, 
{\it Astroph. \& Space Sc.}, {\bf 7}, 20

\item[]
Sunyaev, R. A. \& Zel'dovich, Ya. B. 1972, 
{\it Comments Astrop. Space Phys.}, {\bf 4}, 173

\item[]
{Sunyaev}, R. A. \& {Zel'dovich}, Ya. B.
  1980, \ARAA,  {\bf 18}, 537

\item[]
Takada, M. 2001, \ApJ, {\bf 558}, 29

\item[]
Takada M., Ohno H.,\& Sugiyama N. {\it Preprint} {\bf astro-ph/0112412}

\item[]
Tashiro H., Aghanim N., \& Langer M. 2007, astro-ph/0705.2861

\item[]
Taylor, J. E.; Moodley, K.; Diego, J. M. 2003, \MNRAS, {\bf 345}, 1127

\item[]
Theuns et al. 2002, \ApJ, {\bf 574}, 111

\item[]
Thompson, K.L. \& Vishniac, E.T. 1987, \ApJ, {\bf 313}, 517

\item[]
{Toffolatti}, L., {Negrello}, M., {Gonz\`alez-Nuevo}, J., {De Zotti},
G., {Silva}, L.,  {Granato}, G. L. \& {Argueso}, F.,  2005, \AaA, {\bf 438}, 475

\item[]
Tomita, K. 2005, \PhRvD, {\bf 72}, 3526

\item[]
Tomita, K. 2006, \PhRvD, {\bf 73}, 9901

\item[]
Tuluie, R., Laguna, P., Anninos, P. 1996, \ApJ, {\bf 463}, 15

\item[]
Uzan, J.-P., Aghanim, N., Mellier, Y. 2004, \PhRvD, {\bf 70}, 083533.

\item[]
Valageas, P.; Balbi, A.; Silk, J. 2001, \AaA, {\bf 367}, 1

\item[]
Vald\'es, M., Ciardi, B., Ferrara, A., Johnston-Hollitt, M.,
Rottgering, H. 2006, \MNRAS, {\bf 369}, 66.

\item[]
Vale, C. 2005, astro-ph/0509039.

\item[]
{Vielva}, P., {Barreiro}, R.~B., {Hobson}, M.~P., {Martinez-Gonz{\'a}lez}, E., 
{Lasenby}, A.~N., {Sanz}, J.~L. \& {Toffolatti}, L.
2001, \MNRAS, {\bf 328}, 1.

\item[]
Vielva, P., Martinez-Gonzalez, E., Tucci, M. 2006, \MNRAS, {\bf 365}, 891. 

\item[]
Vilenkin, A.; Shellard, E. P. S. 1995, {\it Science}, {\bf 267}, 1845

\item[]
{Vishniac}, E.~T. 1987, \ApJ,  {\bf 322}, 597.

\item[]
Vittorio, N.; Juszkiewicz, R. 1987, \ApJ, {\bf 314}, 29

\item[]
Volonteri, M., Haardt, F. \& Madau P. 2003, \ApJ,  {\bf 582}, 559

\item[]
Walter, et al. 2004, \ApJ, {\bf 615}, L17

\item[]
Warren, M. S., Abazajian, K., Holz, D. E., \& Teodoro, L.
2006, \ApJ, {\bf 646}, L881

\item[]
{Weller}, J., {Battye}, R.~A. \& {Kneissl}, R., 2002, \PhRvD , {\bf 88}, 231301

\item[]
Wyithe, \& Loeb, A. 2003, \ApJ, {\bf 588}, 69

\item[]
Yamada, M.; Sugiyama, N.; Silk, J. 1999, \ApJ, {\bf 522}, 66

\item[]
{Zhang}, P.J,  {Pen}, U.-L.  \& {Wang}, B., 2002, \ApJ, {\bf 577}, 555

\item[]
Zhang, P.J. 2004, \MNRAS, {\bf 348}, 1348

\item[]
Zhang, P.J., Pen,U.-L, Trac, H. 2004, \MNRAS, {\bf 347}, 1224

\item[]
Zaldarriaga, M. \&  Seljak, U. 1997, \PhRvD, {\bf 55}, 1830 

\item[]
Zaldarriaga, M. \&  Seljak, U. 1998, \PhRvD, {\bf 58}, 3003

\item[]
Zaldarriaga, M. \&  Seljak, U. 1999, \PhRvD, {\bf 59}, 3507

\item[]
Zaldarriaga, M. 2000, \PhRvD, {\bf 62}, 063510

\item[] Zaroubi, S. \& Silk, J. 2005, \MNRAS, {\bf 360}, 64

\endrefs

\end{document}